\documentclass{pasj00}

\def\commenta{$^*$}
\def\commentb{$^\dagger$}
\def\commentc{$^\ddagger$}
\def\commentd{$^\S$}
\def\commente{$^\|$}
\def\commentf{$^\#$}

\def\submitted{submitted}
\def\inpress{in press}

\def\arxiv#1{ (arXiv astro-ph/#1)}

\DeclareAbbreviation\AAHam{Astron. Abh. Hamburg. Sternw.}
\DeclareAbbreviation\AARv{Astron. Astrophys. Rev.}
\DeclareAbbreviation\AAS{American Astron. Soc. Meeting Abstracts}
\DeclareAbbreviation\AcA{Acta Astron.}
\DeclareAbbreviation\actaa{Acta Astron.}
\DeclareAbbreviation\Afz{Astrofizika}
\DeclareAbbreviation\AGAb{Astronomische Gesellschaft Abstract Ser.}
\DeclareAbbreviation\an{Astron. Nachr.}
\DeclareAbbreviation\AnAp{Annales d'Astrophysique}
\DeclareAbbreviation\AnTok{Tokyo Astron. Obs. Annals, Sec. Ser.}
\DeclareAbbreviation\Ap{Astrophysics}
\DeclareAbbreviation\ARep{Astron. Rep.}
\DeclareAbbreviation\ATel{Astron. Telegram}
\DeclareAbbreviation\ATsir{Astron. Tsirk.}
\DeclareAbbreviation\AcApS{Acta Astrophys. Sinica}
\DeclareAbbreviation\AstL{Astron. Lett.}
\DeclareAbbreviation\BaltA{Baltic Astron.}
\DeclareAbbreviation\BANS{Bull. of the Astron. Institutes of the Netherlands Suppl. Ser.}
\DeclareAbbreviation\BASI{Bull. Astron. Soc. India}
\DeclareAbbreviation\BeSN{Be Newslett.}
\DeclareAbbreviation\BHarO{Harvard Coll. Obs. Bull.}
\DeclareAbbreviation\CBET{Cent. Bur. Electron. Telegrams}
\DeclareAbbreviation\ChJAA{Chinese J. of Astron. and Astrophys.}
\DeclareAbbreviation\caa{Chinese J. of Astron. and Astrophys.}
\DeclareAbbreviation\CoAsi{Asiago Contr.}
\DeclareAbbreviation\CoSka{Contributions of the Astronomical Observatory Skalnat\'e Pleso}
\DeclareAbbreviation\GCN{GRB Coord. Netw. Circ.}
\DeclareAbbreviation\ErgAN{Erg. Astron. Nachr.}
\DeclareAbbreviation\ibvs{IBVS}
\DeclareAbbreviation\IEEEP{IEEE Proc.}
\DeclareAbbreviation\JAD{J. Astron. Data}
\DeclareAbbreviation\JAVSO{J. American Assoc. Variable Star Obs.}
\DeclareAbbreviation\JBAA{J. Br. Astron. Assoc.}
\DeclareAbbreviation\JPhCS{J. of Physics Conference Series}
\DeclareAbbreviation\JPSJ{J. Phys. Soc. Japan}
\DeclareAbbreviation\JSARA{J. of the Southeastern Assoc. for Research in Astron.}
\DeclareAbbreviation\LowOB{Lowell Obs. Bull.}
\DeclareAbbreviation\MitAG{Mitteil. der Astronom. Gesell. Hamburg}
\DeclareAbbreviation\MitVS{Mitteil. Ver\"{a}nderl. Sterne}
\DeclareAbbreviation\MmSAI{Mem. Soc. Astron. Ital.}
\DeclareAbbreviation\Msngr{Messenger}
\DeclareAbbreviation\NewA{New Astron.}
\DeclareAbbreviation\na{New Astron.}
\DeclareAbbreviation\NewAR{New Astron. Rev.}
\DeclareAbbreviation\NInfo{Nauchnye Informatsii}
\DeclareAbbreviation\OAP{Odessa Astron. Publ.}
\DeclareAbbreviation\Obs{Observatory}
\DeclareAbbreviation\OEJV{Open Eur. J. on Variable Stars}
\DeclareAbbreviation\PASA{Publ. Astron. Soc. Australia}
\DeclareAbbreviation\PASAu{Publ. Astron. Soc. Australia}
\DeclareAbbreviation\PCCP{Phys. Chem. Chem. Phys.}
\DeclareAbbreviation\PAZh{Pis'ma AZh}
\DeclareAbbreviation\PhR{Phys. Rep.}
\DeclareAbbreviation\PVSS{Publ. Variable Stars Sect. R. Astron. Soc. New Zealand}
\DeclareAbbreviation\PZ{Perem. Zvezdy}
\DeclareAbbreviation\PZP{Perem. Zvezdy, Prilozh.}
\DeclareAbbreviation\QJRAS{QJRAS}
\DeclareAbbreviation\RA{Ricerche Astronomiche}
\DeclareAbbreviation\RMxAA{Rev. Mexicana Astron. Astrof.}
\DeclareAbbreviation\RvMA{Reviews of Modern Astron.}
\DeclareAbbreviation\SASS{Society for Astronom. Sciences Ann. Symp.}
\DeclareAbbreviation\Sci{Science}
\DeclareAbbreviation\SPIE{SPIE Proc.}
\DeclareAbbreviation\SvA{Soviet Astronomy}
\DeclareAbbreviation\SvAL{Soviet Astronomy Letters}
\DeclareAbbreviation\VeSon{Ver\"{o}ff. Sternw. Sonneberg}
\DeclareAbbreviation\VSOLJBul{VSOLJ Variable Star Bull.}
\DeclareAbbreviation\yCat{VizieR Online Data Catalog}
\DeclareAbbreviation\ZA{Z. Astrophys.}

\def\ASPConf#1#2{ASP Conf. Ser. #1, #2}

\def\PublisherCambridge{Cambridge: Cambridge University Press}

\def\PublisherASP{San Francisco: ASP}

\def\PublisherSpringer{Berlin: Springer-Verlag}

\begin{document}
\SetRunningHead{T. Kato et al.}{Characterization of Dwarf Novae Using SDSS Colors}

\Received{201X/XX/XX}
\Accepted{201X/XX/XX}

\title{Characterization of Dwarf Novae Using SDSS Colors}

\author{Taichi \textsc{Kato}}
\affil{Department of Astronomy, Kyoto University,
       Sakyo-ku, Kyoto 606-8502}
\email{tkato@kusastro.kyoto-u.ac.jp}

\author{Hiroyuki \textsc{Maehara}}
\affil{Kwasan and Hida Observatories, Kyoto University, Yamashina,
       Kyoto 607-8471}

\and

\author{Makoto \textsc{Uemura}}
\affil{Astrophysical Science Center, Hiroshima University, Kagamiyama, 1-3-1
       Higashi-Hiroshima 739-8526}



\maketitle

\begin{abstract}
   We have developed a method for estimating the orbital periods 
of dwarf novae from the Sloan Digital Sky Survey (SDSS) colors 
in quiescence using an artificial neural network.
For typical objects below the period gap with sufficient 
photometric accuracy, we were able to estimate the orbital periods 
with an accuracy to a 1 $\sigma$ error of 22 \%.
The error of estimation is worse for systems with longer orbital periods.
We have also developed a neural-network-based method for categorical
classification.  This method has proven to be efficient in classifying
objects into three categories (WZ Sge type, SU UMa type and SS Cyg/Z Cam type)
and works for very faint objects to a limit of $g$=21.
Using this method, we have investigated the distribution of the orbital
periods of dwarf novae from a modern transient survey
(Catalina Real-Time Survey).  Using Bayesian analysis developed by
\citet{uem10shortPCV}, we have found that the present sample tends to 
give a flatter distribution toward the shortest period and 
a shorter estimate of the period minimum, which may have resulted 
from the uncertainties in the neural network analysis and photometric errors.
We also provide estimated orbital periods, estimated classifications and
supplementary information on known dwarf novae with quiescent SDSS
photometry.
\end{abstract}

\section{Introduction}

   Cataclysmic variables (CVs) are close binary systems consisting of
a white dwarf (WD) and a red-dwarf secondary transferring matter via
the Roche-lobe overflow
[for reviews, see \citet{war95book}; \citet{hel01book}].
Dwarf novae (DNe) are a class of CVs characterized by the presence of
outbursts, which are generally believed to be a result of
thermal instabilities in the accretion disk.  Dwarf novae are
classified into three major classes:
SS Cyg-type, Z Cam-type and SU UMa-type dwarf novae.
Among them SU UMa-type dwarf novae show superhumps during their long-lasting
superoutbursts.  The superhumps are generally believed to arise from
the precessing eccentric accretion disk whose eccentricity is excited by
the 3:1 resonance \citep{osa89suuma}.

   According to the standard scenario of CV evolution, CVs with orbital
periods ($P_{\rm orb}$) longer than $\sim$3 hr evolve towards shorter
$P_{\rm orb}$ through the loss of angular momentum by magnetic braking
(\cite{ver81magneticbraking}; \cite{rap83CVevolution}) and gravitational
wave radiation.
When the systems reach certain periods around $P_{\rm orb} \sim 3$ hr,
the secondary becomes fully convective and the effect of the magnetic braking
is believed to decrease dramatically, followed by the shrinkage of the
secondary and reduction of the CV activity.  This state lasts till
the secondary again fills the Roche lobe at around $P_{\rm orb} \sim 2$ hr
and forms the famous ``period gap'' in the $P_{\rm orb}$ distribution of CVs.
After crossing the period gap, $P_{\rm orb}$ further decreases mainly
through the loss of angular momentum by gravitational wave radiation
until the secondary becomes degenerate.  Around the time when this point
is reached, the $P_{\rm orb}$ increases due to two reasons:
the thermal time-scale of the secondary exceeds the mass-transfer
time-scale and the mass-radius relation is reversed for degenerate dwarfs.
This mechanism leads to the existence of the minimum period for ordinary
CVs \citep{pac81CVGWR}.\footnote{
   We only showed the outline for the reason of the existence of
   the period minimum.  Modern works have suggested that the mechanism
   is more complex than this simplified picture
   (cf. \cite{kol99CVperiodminimum}; \cite{kni11CVdonor};
   \cite{ara05MCV}).
}
Those systems whose periods go past the period minimum are usually 
called period bouncers.

   The early model calculations yielded the minimum period of
60--80 min (\cite{pac81CVGWR}; \cite{rap82CVevolution};
\cite{pac83CVevolution}).  Later refined models yielded short periods
of 65--70 min (\cite{kol99CVperiodminimum}; \cite{how01periodgap}),
which is significantly shorter than the observed value
(e.g. \cite{kol93CVpopulation}).  This disagreement is called
``period minimum problem''.  Furthermore, population synthesis studies
expect that most systems have already reached the period minimum and
that there is a heavy accumulation of systems around the period minimum
(period spike).  Since such a spike was not observationally evident until
very recently, this disagreement was called ``period spike problem''
(\cite{kol99CVperiodminimum}; \cite{ren02CVminimum}).  The observational
evidence for the period spike has only recently been shown
(cf. \cite{gan09SDSSCVs}; \cite{uem10shortPCV}).

   Some of non-magnetic CVs above the period gap and most of CVs below
the period gap exhibit dwarf nova-type outbursts.  In addition to
color-based surveys such as Palomer Green (PG) survey
\citep{PGsurvey} and Hamburg Quasar Survey (HQS; \cite{HQS}), and
color-selected spectroscopic survey such as Sloan Digital Sky Survey
(SDSS; \cite{SDSS}), dwarf nova-type outbursts have been both traditionally,
and in modern times, playing an important role in discovering CVs.

   In variability-based surveys of CVs, SU UMa-type dwarf novae are
a particularly important group for two reasons:
(1) SU UMa-type dwarf novae show superhumps and we can estimate
the orbital periods using only photometric observations.
(2) Most of dwarf novae below the period gap are SU UMa-type dwarf novae,
and WZ Sge-type dwarf novae (a subgroup of SU UMa-type dwarf novae;
cf. \cite{bai79wzsge}; \cite{odo91wzsge}; \cite{kat01hvvir}) are
considered to occupy the terminal stage of the CV evolution.
Indeed, the recent increase in discoveries of new SU UMa-type dwarf novae,
helped by massive transient surveys such as Catalina Real-Time Survey
(CRTS, \cite{CRTS})\footnote{
   $<$http://nesssi.cacr.caltech.edu/catalina/$>$.
   For the information of the individual Catalina CVs, see
   $<$http://nesssi.cacr.caltech.edu/catalina/AllCV.html$>$.
}, the All Sku Automated Survey-3 (ASAS-3, \cite{ASAS3}) and 
by amateur observers (notably K. Itagaki),
have a significant impact on the distribution of orbital periods in
CVs (\cite{gan09SDSSCVs}; \cite{uem10shortPCV}).  Both works have
indicated that the period minimum spike, which has only become
observationally apparent, are heavily composed of newly identified
SU UMa-type dwarf novae or faint SDSS CVs.  Using Bayesian statistical
analysis, \citet{uem10shortPCV} suggested a possibility that
the true period minimum is even shorter than what is observed,
considering the effect of low detectability of very low mass-transfer systems.

   These surveys and statistics, however, unavoidably suffer from various
kinds of biases.  There are a bias because essentially all surveys are
magnitude-limited, with bias being introduced by color selection criteria,
and bias resulting from follow-up strategies.  The first two biases
are more serious in color and spectroscopy-selected searches.  The last
bias is more important in variability-based searches, because
long-$P_{\rm orb}$ SS Cyg and Z Cam-type dwarf novae do not show superhumps,
and short-$P_{\rm orb}$ SU UMa-type ones show such $P_{\rm orb}$ as
are more easily determined.  Furthermore, it is known that large-amplitude
dwarf novae tend to receive popular attention, potentially leading to
a bias toward detecting a larger number of shorter-$P_{\rm orb}$ systems.

   Estimating $P_{\rm orb}$ from colors of dwarf novae in quiescence,
if it is feasible, could reduce the effect of the last bias.
There has been a long history of using colors in classifying CVs:
\citet{bru84CVUBV} compiled $UBV$ colors of CVs and tried to classify them
on the color-color diagram and \citet{bru94CVUBV} extended this work.
\citet{szk87shortPCV} used $UBVJHK$ photometry and spectroscopy to
characterize quiescent dwarf novae.  \citet{szk87shortPCV} showed that
($V-J$) and ($U-B$) colors become bluer in shorter-$P_{\rm orb}$.
\citet{spr96CVabsmag} used ($J-K$) color to characterize the secondary
and CV type.  \citet{hoa02CV2MASS} complied 2MASS colors of CVs
and extended the results by \citet{spr96CVabsmag}.  \citet{ima06j0137}
closely examined 2MASS color to characterize dwarf novae.
\citet{ak07CVabsmag} further used 2MASS color to estimate absolute
magnitudes of CVs and derived space distributions \citep{ak08CVdist}.
\citet{wil10newCVs} used SDSS color cuts, UV color and variability
for detecting new dwarf novae.

   In this study, we used the SDSS photometric catalog for exploring a new
way to estimate $P_{\rm orb}$, and we discuss the implication of the
resultant period distribution.

\section{The Sample}\label{sec:data}

   The sample we used are the known dwarf novae in the General Catalog
of Variable Stars (GCVS, \cite{GCVSelectronic2011}), Downes CV Catalog,
Archival Edition \citep{DownesCVatlas3}, the online version of
RKCat [\cite{RKCat}, (update RKcat7.15, 2011)], SDSS CVs
(\cite{szk02SDSSCVs}; \cite{szk03SDSSCV2}; \cite{szk04SDSSCV3};
\cite{szk05SDSSCV4}; \cite{szk06SDSSCV5}; \cite{szk07SDSSCV6};
\cite{szk09SDSSCV7}),
newly recognized dwarf novae by CRTS,
CRTS Mount Lemmon Survey (MLS)\footnote{
   $<$http://nesssi.cacr.caltech.edu/MLS/AllCV.html$>$.
}, CRTS Siding Spring Survey (SSS)\footnote{
   $<$http://nesssi.cacr.caltech.edu/SSS/AllCV.html$>$.
},
candidate dwarf novae selected by color and variability
\citep{wil10newCVs}, and newly discovered dwarf novae whose properties
have been investigated in \citet{Pdot}, \citet{Pdot2} and \citet{Pdot3}.
For CRTS objects, most of dwarf novae were easily recognized based on
their light curves (typical dwarf nova-type light curve and existence
of past outbursts).  Some CRTS dwarf novae were selected based on
single outburst detections and typical CV colors in quiescence.
For SDSS CVs, we selected objects that are recognized as dwarf novae
based on the presence of outbursts and objects that have dwarf nova-type
spectra.

   We selected the objects whose magnitudes are present in
the SDSS Photometric Catalog, Releases 8 (DR8).  Since some objects or
measurements are missing in SDSS DR8, we used SDSS DR7 instead. 

   We rejected SDSS magnitudes measured during outbursts or
standstills (in Z Cam-type dwarf novae) by comparing with other photometric
catalogs, known ranges of variability and typical quiescent magnitudes
in the CRTS data.  If there are two or measurements in SDSS and their
magnitudes differ by more than 1 mag, we rejected the measurements
whose magnitudes are brighter by more than 1 mag than the faintest 
one in order to minimize the contamination of outbursts.  
Even if multiple SDSS entries are present for the same object, 
we used the measurements individually
and did not use an averaged value of each object before analysis.
We rejected measurements that have saturated pixels
(shown as blanks in table \ref{tab:dnlist}).

   For estimating Galactic extinction, we used \citet{sch98reddening}
for a through-the-Galaxy estimate, and obtained the extinction at the
distance of the object using \citet{bah80extinction} assuming
a scale height of $H=100$ pc for the interstellar dust
(\cite{Spitzer78ISMbook}; \cite{bah80extinction}).
We employed an iterative process to obtain
self-consistent estimates of distance and extinction.
We examined the dependence of the results on the assumption of $H$,
using extreme set of $H=70$ pc (typical value for molecular coulds)
and $H=200$ pc (typical value for cold neutral medium).
The resultant periods of neural network analysis (section \ref{sec:nnet})
varied within 5 \% of the $H=100$ pc result for 68 \% of objects.

   We used distance estimates tabulated in \citet{pat11CVdistance} and
those of \citet{roe07amcvndistance} for GP Com,
\citet{sla95novaremnant} for GK Per and \citep{und08gd552} for GD 552.
For estimating the distance of the rest of the objects, we used Warner's
relations (\cite{war87CVabsmag}; \cite{war95book}).
If apparent magnitudes of maximum are available we estimated
the distances by assuming the absolute magnitudes ($M_V{\rm (max)}$)
using the updated Warner's relation \citep{pat11CVdistance} for objects
with $P_{\rm orb} < 0.5$ d, and our own calibration based on GK Per
[the interstellar extinction correction of $E(B-V)=0.3$ was taken
from \cite{wu89gkper}] for longer $P_{\rm orb}$:

\begin{equation}
M_V{\rm (max)} = \left\{
  \begin{array}{ll}
    5.70 - 17.2 P_{\rm orb}, & \mbox{($P_{\rm orb} < 0.5$)} \\
    2.79 - 1.05 P_{\rm orb}, & \mbox{($P_{\rm orb} \ge 0.5$)}
  \end{array}
  \right.
\end{equation}

If the apparent magnitudes of maximum were unavailable, we used the
following equation (corresponding to equation 3.3 in \cite{war95book}):

\begin{equation}
M_V{\rm (min)} = 7.1 + 1.64 log T_n {\rm (d)} - 6.24 P_{\rm orb},
\end{equation}

assuming $\log T_n$=2.5 to estimate minimum $M_V$ and estimated
observed apparent magnitudes of minimum to the distances.
When the orbital periods were not available, we used the results of
a neural network analysis (section \ref{sec:nnet}).
We used a mean maximum $M_V=4.95$ for objects without
measured and estimated orbital periods.

   Since Warner's relation for the maximum is not very
sensitive to the orbital period, the uncertainty introduced by the
uncertainty of the orbital period is estimated to be sufficiently small.
We used the coefficients in \citet{sch98reddening} to convert $V$-band
absorption $A(V)$ to extinctions in SDSS passbands.  Since the application
of Warner's relation requires orbital periods, the estimation of
Galactic extinction is dependent on the results of the neural network
analysis for objects without known orbital periods.  We therefore repeated
this process three times, using the results of the neural network analysis
for estimating the distances, to obtain self-consistent estimates
of distances and orbital periods from de-reddened colors.
These values and the methods of estimation are placed in a later table
(table \ref{tab:dnlist2}) in order to save space.

   Since there are different kinds of samples in this paper,
we summarized the samples in table \ref{tab:samples} for the
convenience of readers.

\begin{table}
\caption{Samples in this paper.}\label{tab:samples}
\begin{center}
\begin{tabular}{c|l}
\hline
Sample type & Objects \\
\hline
All samples        & Sample described in section \ref{sec:data}. \\
                   & Identical with objects in table \ref{tab:dnlist}. \\
\hline
WZ Sge-type DNe    & UZ Boo, EG Cnc, V592 Her, RZ Leo, \\
(subsection \ref{sec:twocolor}) & UW Tri, BC UMa, HV Vir, \\
                   & SDSS J080434.20$+$510349.2, \\
                   & SDSS J133941.11$+$484727.5, \\
                   & SDSS J160501.35$+$203056.9, \\
                   & OT J012059.6$+$325545, \\
                   & OT J074727.6$+$065050, \\
                   & OT J090239.7$+$052501, \\
                   & OT J104411.4$+$211307, \\
                   & SDSS J161027.61$+$090738.4 \\
\hline
Systems below the  & EI Psc, SDSS J150722.30$+$523039.8, \\
period minimum     & OT J112253.3$-$111037 \\
(subsection \ref{sec:belowperiodmin}) & \\
\hline
Training set       & Sample used for training the neural \\
(subsection \ref{sec:training}) & network.  Objects in table \ref{tab:dnlist} with \\
                   & $P_{\rm orb}$ entries, excluding two \\
                   & AM CVn stars (GP Com and SDSS \\
                   & J012940.05$+$384210.4), \\
                   & AR Cnc, GZ Cet, MT Com, EI Psc, \\
                   & QZ Ser and OT J231308.1$+$233702. \\
\hline
DNe used for study of & CRTS transients with known SDSS \\
period distribution & colors and $70 < P_{\rm orb} {\rm (min)} < 130$. \\
(section \ref{sec:porbdist}) & Listed in table \ref{tab:crtsotlist}. \\
\hline
\end{tabular}
\end{center}
\end{table}

\section{Color Analysis}

\subsection{Location of WZ Sge-Type Dwarf Novae on Color-Color Diagrams}\label{sec:twocolor}

   We attempted to separate WZ Sge-type dwarf novae based on locations
on color-color diagrams.  We selected WZ Sge-type dwarf novae using the
criteria described in \citet{Pdot}.  The selected WZ Sge-type dwarf novae
were UZ Boo, EG Cnc, V592 Her, RZ Leo, UW Tri, BC UMa, HV Vir,
SDSS J080434.20$+$510349.2, SDSS J133941.11$+$484727.5,
SDSS J160501.35$+$203056.9, OT J012059.6$+$325545, OT J074727.6$+$065050,
OT J090239.7$+$052501, OT J104411.4$+$211307, and the CV selected by
\citet{wil10newCVs} SDSS J161027.61$+$090738.4.
Among them, RZ Leo and BC UMa are ``borderline'' WZ Sge-type dwarf novae with
relatively frequent (super)outbursts, but with well-defined early superhumps.
Most of other objects are more extreme WZ Sge-type dwarf novae
with very infrequent (super)outbursts or with multiple rebrightenings.

\begin{figure*}
  \begin{center}
    \FigureFile(190mm,95mm){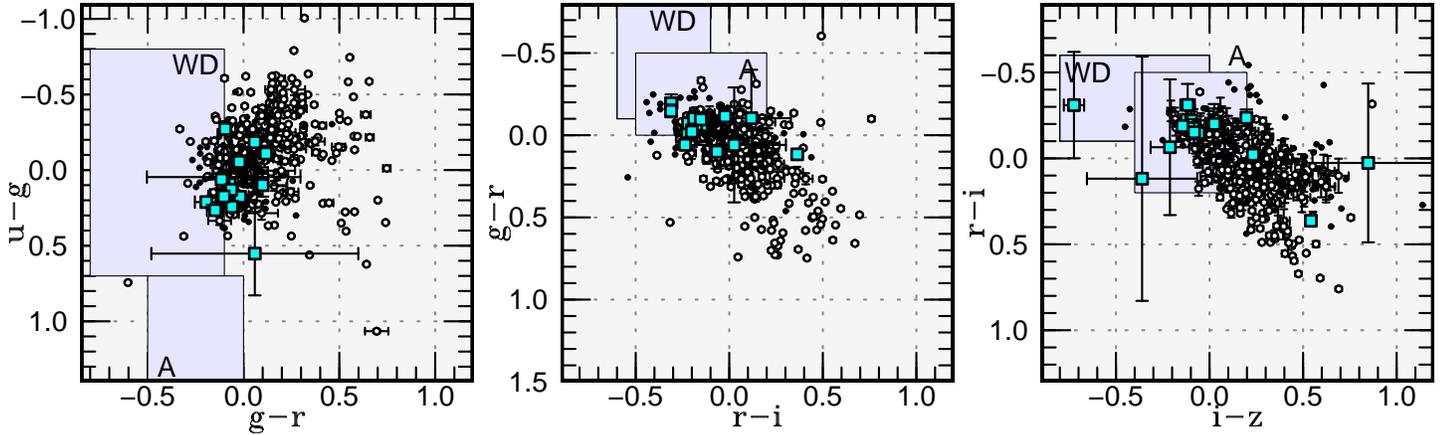}
  \end{center}
  \caption{Location of WZ Sge-type dwarf novae on color-color diagrams.
  Lightly filled squares represent well-characterized WZ Sge-type dwarf novae.
  The other open circles represent other classifications or objects without
  known outburst properties.  For better visualization, the plots of
  object other than WZ Sge-type dwarf novae are limited to those
  having errors smaller than 0.1 mag in all
  ($u-g$, $g-r$, $r-i$, $i-z$) colors.  The small dots represents
  objects other than WZ Sge-type dwarf novae with larger (up to 0.3 mag)
  photometric errors.  Two boxes in the figures are white dwarf (WD)
  exclusion box and A-type star (A) exclusion one in the spectroscopic 
  follow-up of SDSS quasar candidates.}
  \label{fig:wzloc}
\end{figure*}

   In figure \ref{fig:wzloc}, WZ Sge-type dwarf novae and dwarf novae
of other classes without known outburst properties are superimposed 
on the diagram.  It seems that WZ Sge-type dwarf novae tend to
cluster on the $(u-g,g-r)$ color-color diagram.  This trend can be
interpreted as the smaller contribution of
the Balmer continuum, arising from the accretion disk, to the $u$ light;
in WZ Sge-type dwarf novae, this spectral range is usually dominated
by the white dwarf and the level of disk-originated $u$ light is smaller.
The $(g-r,r-i)$ diagram is less diagnostic.
The $(r-i,i-z)$ diagram again shows some degree of clustering of
WZ Sge-type dwarf novae.  This trend can be interpreted as 
the contribution of the secondary to the $z$ light in longer-$P_{\rm orb}$
objects.  The two WZ Sge-type objects with large $i-z$ colors are
RZ Leo and OT J090239.7$+$052501.  The former has a long $P_{\rm orb}$
and the secondary significantly contributes to the $z$ band.
The unusual position of the latter was caused by large photometric
errors in SDSS due to its faintness ($g$=23.2).
The locations of WZ Sge-type dwarf novae, excluding these two objects,
can be used to discriminate WZ Sge-type candidates in the color-color space.
The average values and standard deviations for $u-g$, $g-r$, $r-i$ and $i-z$
colors for these samples are 0.03(0.17), $-$0.04(0.07), $-$0.12(0.18),
0.04(0.22), respectively.

\subsection{Dependence of Colors on Orbital Period}\label{sec:depporb}

   We studied the dependence of colors on the orbital period.
Whenever available, we used well-determined orbital periods from
radial-velocity studies or photometric observations of eclipses.\footnote{
   The values of orbital periods are taken from tables in our compilations
   (\cite{Pdot}, \cite{Pdot2} and \cite{Pdot3}; references
   are listed therein), and supplemented for
   RX And \citep{kai89rxand}, UU Aql (\cite{RKCat}, J. R. Thorstensen,
   private communication),
   VZ Aqr, V516 Cyg, V478 Her, VZ Sex, HS 1055$+$0939 \citep{tho10longPDN},
   CR Boo \citep{pro97crboo}, AM Cas \citep{tay96arandamcaspyper},
   V630 Cas \citep{oro01v630cas}, WW Cet \citep{tap97wwcet},
   EN Cet, SDSS J090103.93$+$480911.1, SDSS J124426.26$+$613514.6,
      SDSS J171145.08$+$301320.0 \citep{dil08SDSSPorb3},
   GY Cet, SDSS J005050.88$+$000912.6, SDSS J155531.99$-$001055.0,
      SDSS J205914.87$-$061220.5, SDSS J233325.92$+$152222.2
      \citep{sou07SDSSCV2},
   HP Cet, SDSS J091127.36$+$084140.7, SDSS J121607.03$+$052013.9
      \citep{sou06SDSSCV},
   AR Cnc \citep{how90faintCV3}, GY Cnc, IR Com \citep{fel05gycncircomhtcas},
   GP Com \citep{mar99gpcom}, MT Com \citep{pat05j1255},
   AB Dra \citep{tho85abdrawwcet}, DO Dra \citep{has97dodraHST},
   ES Dra \citep{rin11esdra},
   BF Eri, BI Ori, FO Per \citep{she07CVspec},
   LT Eri \citep{ak05j0407},
   AH Her \citep{hor86ahher}, X Leo \citep{sha86xleossaur},
   DO Leo \citep{abb90doleo}, HM Leo \citep{tho01CVperiod},
   CW Mon \citep{kat03cwmon}, HX Peg \citep{rin94hxpeg},
   IP Peg \citep{cop10ippeg}, V367 Peg \citep{wou05CVphot},
   V405 Peg \citep{tho09v405peg}, GK Per \citep{cra86gkperorbit},
   KT Per \citep{tho97ktper}, AY Psc \citep{dia90aypsc},
   X Ser \citep{tho00v533herv446herxser}, RY Ser,
      SDSSp J081321.91$+$452809.4 \citep{tho04longPDN},
   QZ Ser \citep{tho02qzser}, V386 Ser \citep{wou04CV4},
   EL UMa (\cite{RKCat}, derived from \cite{Pdot2}),
   VW Vul \citep{tho98cylyrtwtrivwvul}, GD 552 \citep{hes90gd552},
   1RXS J171456.2+585130 \citep{wil11j1714asas1509},
   HS 1016$+$3412, HS 1340$+$1524 \citep{aun06HSDN},
   HS 2205$+$0201 (\cite{RKCat}, A. Aungwerojwit, Ph.D.
   thesis, Univ. of Warwick), HS 2219$+$1824 \citep{rod05hs2219},
   RX J1715.6$+$6856, RX J1831.7$+$6511 \citep{pre07CVspacedensity},
   SDSS J003941.06$+$005427.5, SDSS J075059.97$+$141150.1 \citep{sou10j0039},
   SDSS J004335.14$-$003729.8, SDSS J165837.70$+$184727.4,
      SDSS J220553.98$+$115553.7 \citep{sou08CVperiod},
   SDSS J074531.92$+$453829.6, SDSS J155656.92$+$352336.6 \citep{szk06SDSSCV5},
   SDSS J075507.70$+$143547.6, SDSS J080534.49$+$072029.1,
      SDSS J143544.02$+$233638.7 \citep{szk07SDSSCV6},
   SDSS J080303.90$+$251627.0, SDSS J142955.86$+$414516.8 \citep{szk05SDSSCV4},
   SDSS J080846.19$+$313106.0, SDSS J153817.35$+$512338.0,
      SDSS J223439.93$+$004127.2, SDSSp J230351.64$+$010651.0
      (\cite{gan09SDSSCVs}, Dillon et al. in prep.),
   SDSS J084400.10$+$023919.3, SDSS J091945.11$+$085710.1
      (\cite{gan09SDSSCVs}, Thorstensen et al. in prep.),
   SDSS J090016.56$+$430118.2 \citep{szk04SDSSCV3},
   SDSS J090403.48$+$035501.2 \citep{wou05SDSSCVpuls},
   SDSS J100658.40$+$233724.4 \citep{sou09j1006},
   SDSS J103533.02$+$055158.3, SDSS J143317.78$+$101123.3,
      SDSS J150137.22$+$550123.4, SDSS J150722.30$+$523039.8
      \citep{lit08eclCV},
   SDSS J123813.73$-$033933.0 \citep{zha06j1238},
   SDSS J124058.03$-$015919.2 \citep{roe05j1240},
   SDSS J145758.21$+$514807.9 \citep{uth11CVthesis},
   SDSS J154453.60$+$255348.8 \citep{ski11j1544},
   SDSS J204448.92$-$045928.8 \citep{pet05CVlongP}.
}  For some WZ Sge-type dwarf novae, we could
use the periods of early superhumps when spectroscopic orbital periods
are not available.  We could estimate orbital periods fairly well
for SU UMa-type dwarf novae with known superhump periods ($P_{\rm SH}$).
For systems with $P_{\rm SH} < 0.084$ d (stage B superhumps, see
\cite{Pdot} for definition of stages) or $P_{\rm SH} < 0.076$ d
(stage C superhumps),
we employed the newly calibrated relations in \citet{Pdot3} for
deriving estimated $P_{\rm orb}$.  The 1 $\sigma$ error for the
estimated $P_{\rm orb}$ is 0.0003 d.  For systems with longer
$P_{\rm SH}$, we used the relation in \citet{sto84tumen}.

\begin{figure*}
  \begin{center}
    \FigureFile(190mm,95mm){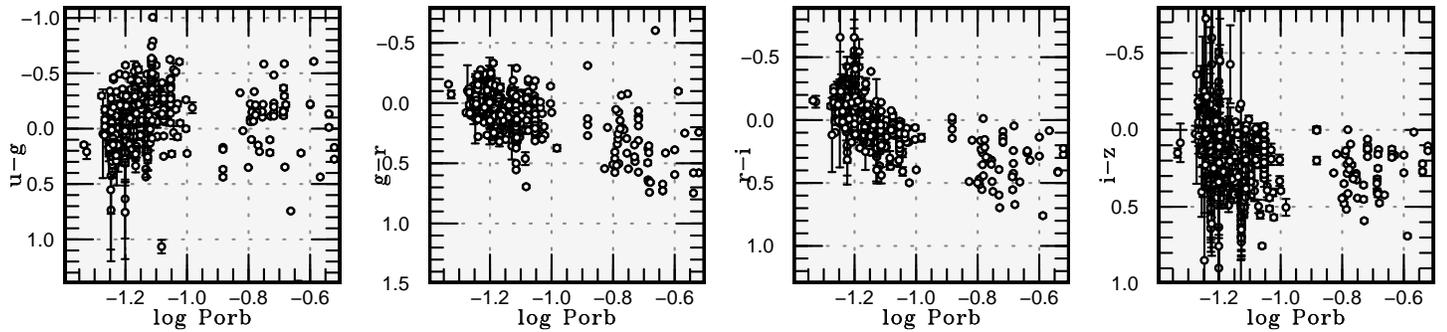}
  \end{center}
  \caption{Dependence of SDSS colors of dwarf novae on orbital period.}
  \label{fig:porbcol}
\end{figure*}

   The result is shown figure \ref{fig:porbcol}.  The figure includes
unusual short-$P_{\rm orb}$ systems EI Psc, and OT J112253.3$-$111037
but does not include AM CVn-type objects (for a review of AM CVn-type
stars, see \cite{sol10amcvnreview}) which show dwarf-nova outbursts
or which are similar to dwarf novae in quiescence
(GP Com and SDSS J012940.05$+$384210.4 in the present sample).
The only SDSS CV having $P_{\rm orb}$ longer than 1 d
(SDSS J204448.92$-$045928.8) is outside this figure.

   Excepting $u-g$, the colors of dwarf novae in quiescence become
redder in longer $P_{\rm orb}$.  The $u-g$ color is practically
constant for $\log P_{\rm orb} > -1.2$.  In shorter $P_{\rm orb}$,
the $u-g$ color becomes noticeably redder, reflecting the increasing
contribution of the white dwarf in systems with very low mass-transfer
rates.

   It is particularly noteworthy that all colors have significant
dependence on $P_{\rm orb}$ even below the period gap.
This dependence is expected to provide a way to estimate
$P_{\rm orb}$ only from SDSS photometric data.  The $i-z$ color
tends to become bluer in very short-$P_{\rm orb}$ systems.

\subsection{Period Bouncers and SDSS Fiber Exclusion Boxes}\label{sec:bouncers}

   SDSS spectroscopic follow-up observations used exclusion boxes
to locate the fibers in order to reduce the contamination from
white dwarfs, A-type stars and white dwarf-main sequence binaries
(cf. figure \ref{fig:wzloc} and figure 6 in \cite{gan09SDSSCVs}).
There has been a concern that SDSS-selected CVs may have a bias resulting
from these exclusion policies.

   As we have seen, there appears to be a tendency that the $i-z$
color sharply becomes bluer according as the $P_{\rm orb}$ shortens
(figure \ref{fig:porbcol}).  Most of the shortest-$P_{\rm orb}$
objects $i-z$ bluer than 0, and the spectroscopic detections these
objects and potential period bouncers are expected to be limited
by the exclusion criterion of $i-z \le -0.1$.  This might explain
the too small number of candidates for period bouncers in known
and SDSS-selected CVs compared to predictions by population models
\citep{gan09SDSSCVs}.

   Among CRTS and other transients, OT J010329.0$+$331822,
OT J010550.1$+$190317, OT J012059.6$+$325545, OT J020056.0$+$195727,
OT J035003.4$+$370052, OT J074419.7$+$325448, OT J074727.6$+$065050,
OT J091453.6$+$113402, OT J151037.4$+$084104, OT J164748.0$+$433845
and OT J213122.4$-$003937 were inside the WD exclusion box.
Two objects (OT J010329.0$+$331822 and OT J035003.4$+$370052)
were brighter than the limit ($i=19.1$) of the quasar target
selection algorithm up to SDSS DR7 \citep{ric02SDSSseletion}
and one object (OT J151037.4$+$084104) was brighter than the limit
($i=19.9$) for SDSS DR8 \citep{eis11SDSS3},\footnote{
   Although SDSS did not obtain spectra for the objects in the Sloan
   Extension for Galactic Understanding and Exploration (SEGUE)
   imaging footprint, none of the objects within the WD exclusion box
   are within the SEGUE field.
}, and there
seems to have been a possibility that by this exclusion criterion 
significant fraction of CVs have escaped detection by SDSS.
Notably, OT J074727.6$+$065050 is one of the renowned candidate
for a period bouncer with multiple rebrightenings \citep{Pdot}.
OT J012059.6$+$325545 is also a WZ Sge-type dwarf nova with multiple
rebrightenings \citep{Pdot3}.  Although the latter two objects
were below the limit of the target selection algorithm of SDSS,
these results suggest that the fraction of shortrst-$P_{\rm orb}$
objects and period bouncers was significantly biased in the SDSS CVs.

\subsection{Systems Below the Period Minimum}\label{sec:belowperiodmin}

   There are three objects (EI Psc, SDSS J150722.30$+$523039.8,
OT J112253.3$-$111037) having
$0.04 < P_{\rm orb} < 0.05$ (d) in our sample.
SDSS J150722.30$+$523039.8 is an unusual hydrogen-rich CV with
a very short $P_{\rm orb}$ (\cite{lit07j1507}; \cite{pat08j1507}).
It has been well-known that the region of this $P_{\rm orb}$ contains
objects with unusually massive and luminous secondaries:
EI Psc and V485 Cen (\cite{aug93v485cen}; \cite{aug96v485cen}).
The remaining one object, OT J112253.3-111037, has a blue $g-z$ color
($+0.01$) unlike EI Psc ($+0.74$).  The color is more similar to
SDSS J150722.30$+$523039.8 ($-0.13$).  A low resolution spectrum
of OT J112253.3$-$111037 in quiescence shows a stronger signature of
helium lines (vsnet-alert 12026).  The object might supply
an important missing link between the enigmatic object
SDSS J150722.30$+$523039.8, whose population and evolutionary
status are not yet unclear, and EI Psc-like objects, and detailed
follow-up observations are indispensable.

\section{Neural-Network Analysis}\label{sec:nnet}

\subsection{Method and Training Set}\label{sec:training}

   In order to estimate orbital periods from SDSS colors, we employed
a single-hidden-layer artificial neural network, a kind of multiple-layer
perceptrons (\cite{bis95nnet} and references therein),
using $\log P_{\rm orb}$ as the response, and colors as inputs.\footnote{
   We used {\bf nnet} package in R software
   (The R Foundation for Statistical Computing:
   $<$http://cran.r-project.org/$>$).
}  The procedure is practically the same
as in photometric redshift estimations (see \cite{ANNz} for the definition
of the activation function and numerical method for the training).
We used samples with known orbital periods in training the neurons.
The samples were restricted with photometric errors (in each SDSS band)
smaller than upper 90\% quantiles.  GZ Cet = SDSS J013701.06$-$091234.9
was rejected from the sample because this object has an unusually evolved
massive secondary (\cite{ima06j0137}; \cite{ish07CVIR}).
OT J231308.1$+$233702 was not included due to its unusually red color,
suggesting the possibility of similarity to GZ Cet.
AR Cnc was also rejected because this object is deeply eclipsing
\citep{how90faintCV3}, and the SDSS observation was likely obtained
during its eclipse.
We also rejected EI Psc (subsection \ref{sec:depporb}), QZ Ser
(system with an unusually luminous secondary, \cite{tho02qzser})
and MT Com [suggested to be a WZ Sge-like object \citep{pat05j1255},
but the true nature and the period are still poorly known].

   When there are multiple measurements for the same object, we used
weights reversely proportional to the number of measurements
in training the neural network.

\subsection{Optimization and Uncertainties}\label{sec:nnetoptim}

   We have optimized the number of units in the hidden layer
($N_{\rm hid}$) by the following cross-validation method.
We first randomly selected 70 \% of sample for the training data;
the remaining samples were used for estimating the prediction error.
The residuals for the training data were used for estimating
the training error.
We used 100 different sets of randomly selected samples and randomly
created and trained neural networks for each $N_{\rm hid}$ and averaged
resultant errors for each $N_{\rm hid}$.
Although the training error decreased as $N_{\rm hid}$ increases,
the decrease becomes slower for $N_{\rm hid} > 3$.  The prediction error
slowly increases as $N_{\rm hid}$ increases, especially $N_{\rm hid} > 5$.
We therefore adopted $N_{\rm hid} = 3$.  This adoption of $N_{\rm hid}$
was also supported by the following analysis of estimated uncertainties
and analysis of resultant correlation coefficients.

   We estimated uncertainties of period estimations using the
following process.  For each object (with a known orbital period),
a subsample excluding the target object was created and estimated the
orbital period of the target object using the parameters of
the neural network resulting from the subsample.

   The maximum correlation coefficient excluding outliers reached a maximum
for $N_{\rm hid} = 3$.  The number of outliers suggests that $\sim$2 \%
of ordinary objects will give totally unreliable $P_{\rm orb}$ in this
method.\footnote{
   Since our primary aim is to qualify dwarf novae, e.g. whether
   they are likely candidates for objects below the period gap,
   we used ``totally unreliable'' here for objects whose estimated
   periods are below the period gap while the true $P_{\rm orb}$
   are well above the period gap, or in the reverse cases.
   Since the members of these outliers are not fixed, but dependent on
   the neural networks, only the statistical description in fraction
   of outliers is meaningful.
   Some of these outliers can be properly characterized in the category
   analysis (subsection \ref{sec:category}).
}
   In a model with $N_{\rm hid} = 1$, there remained a strong tendency
of departure of estimated $P_{\rm orb}$ for long-$P_{\rm orb}$ systems,
and thus we did not adopt this model.  For $N_{\rm hid} \ge 5$, correlation
coefficient tends to decrease.
The correlation for objects below the period gap, however, was almost
always equally good regardless of $N_{\rm hid}$ tested.
The correlation was generally worse in long-$P_{\rm orb}$ systems than
in systems below the period gap.  This can be naturally explained by
the large spread of mass-transfer rate above the period gap
(cf. \cite{war87CVabsmag}).

\begin{figure}
  \begin{center}
    \FigureFile(70mm,130mm){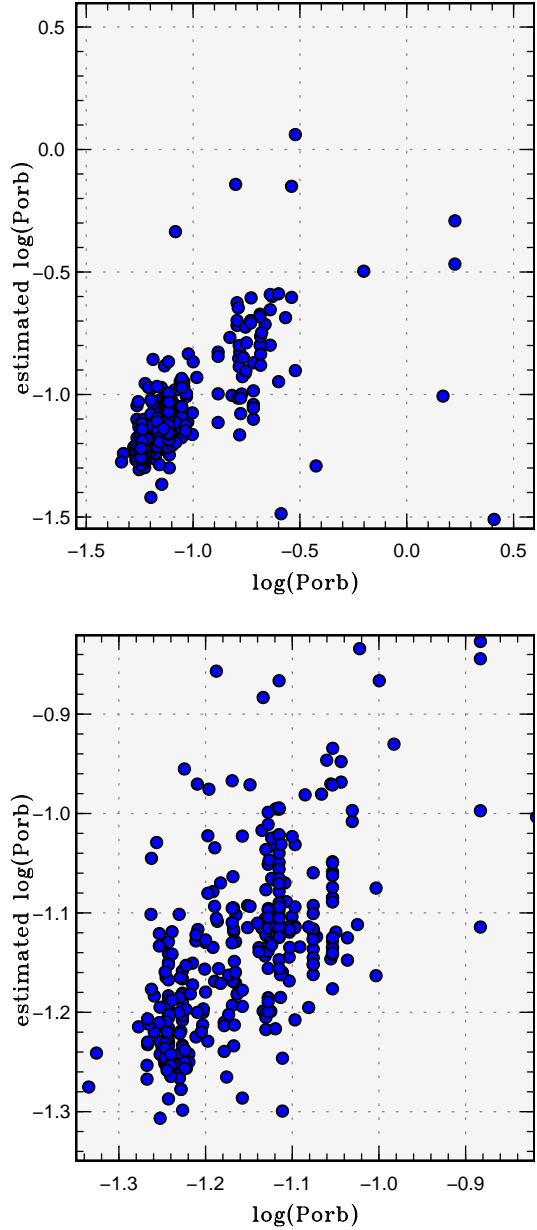}
  \end{center}
  \caption{Estimated versus true orbital periods.
  (Upper): All samples.
  (Lower): Estimations for objects below the period gap.  The upper object
  situated outside this panel is GZ Cet.
  }
  \label{fig:esterror}
\end{figure}

   Diagrams of the estimated orbital periods versus the true ones 
using this $N_{\rm hid}$
are shown in figure \ref{fig:esterror}.  The 1 $\sigma$ error for
estimating $P_{\rm orb}$ for systems below the gap (excluding outliers)
was 22 \%.  For systems with $0.12 < P_{\rm orb} < 0.4$ (d)
(excluding outliers), the 1 $\sigma$ error was 57 \%.

\subsection{Estimation of Orbital Period}

   In order to reflect the uncertainties of the neural network and
the photometric ones, we used different neural networks using all subsamples
described in subsection \ref{sec:nnetoptim}.  We also added random
errors to the observed data according to error estimates of the SDSS
photometry.  After adopting 90\% quantiles of the results,
the mean estimates (mean in its logarithmic value) of the resultant
$P_{\rm orb}$ and their standard deviations are given
in table \ref{tab:dnlist}.

   The columns of estimated orbital periods were left blank if 
the estimated periods were shorter than 0.04 d or longer than 1 d.  
The object with large estimated errors (larger than 0.4 times 
the estimated orbital periods) were also rejected.

\begin{figure}
  \begin{center}
    \FigureFile(70mm,70mm){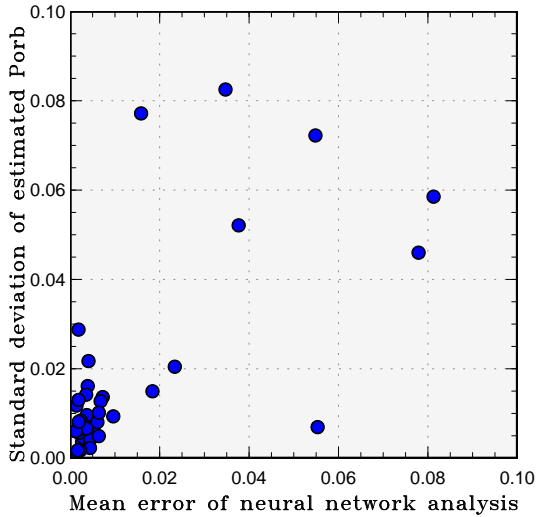}
  \end{center}
  \caption{Standard deviations of estimated $P_{\rm orb}$ and mean errors
    of the neural-network analysis for objects with more than four SDSS
    measurements.  The standard deviation of estimated $P_{\rm orb}$
    tends to be larger than the estimated errors by a factor of two.}
  \label{fig:dispers}
\end{figure}

   There are objects with a multiplicity of up to 31 SDSS scans.  For objects
with more than four measurements, we have obtained standard deviations of
estimated $P_{\rm orb}$ in order to assess the quality of estimated
errors of the neural-network analysis.  Although these estimates were
generally well-correlated ($r^2=0.86$, figure \ref{fig:dispers}
and table \ref{tab:aveper}), the estimated errors tend to be smaller
than the standard deviations of estimated $P_{\rm orb}$.  The ratio of
the standard deviations of estimated $P_{\rm orb}$ to the estimated errors
was 1.8 (median).  This difference may be partly interpreted as
the intrinsic variation of the objects.

\begin{table}
\caption{Averaged $P_{\rm orb}$ estimetes for objects with more than four SDSS scans.}\label{tab:aveper}
\begin{center}
\begin{tabular}{cccc}
\hline
Object & $N$\commenta & mean $P_{\rm est}$ & error of $P_{\rm est}$ \\
\hline
V496 Aur & 4 & 0.077(20) & 0.023 \\
EN Cet & 13 & 0.067(4) & 0.003 \\
GS Cet & 15 & 0.062(14) & 0.007 \\
BI Ori & 6 & 0.103(29) & 0.002 \\
QZ Ser & 4 & 0.210(7) & 0.055 \\
FSV J1722$+$2723 & 4 & 0.081(5) & 0.003 \\
OT J032839.9$-$010240 & 31 & 0.062(13) & 0.007 \\
OT J043829.1$+$004016 & 13 & 0.077(7) & 0.002 \\
OT J051419.9$+$011121 & 4 & 0.077(6) & 0.003 \\
OT J052033.9$-$000530 & 6 & 0.325(83) & 0.035 \\
OT J055842.8$+$000626 & 4 & 0.263(77) & 0.016 \\
OT J074222.5$+$172807 & 4 & 0.092(16) & 0.004 \\
OT J074419.7$+$325448 & 4 & 0.054(9) & 0.010 \\
OT J081418.9$-$005022 & 5 & 0.076(12) & 0.001 \\
OT J082123.7$+$454135 & 6 & 0.109(14) & 0.004 \\
OT J092839.3$+$005944 & 4 & 0.119(15) & 0.018 \\
OT J171223.1$+$362516 & 4 & 0.236(46) & 0.078 \\
OT J211550.9$-$000716 & 11 & 0.124(59) & 0.081 \\
OT J213122.4$-$003937 & 12 & 0.100(72) & 0.055 \\
OT J221128.7$-$030516 & 5 & 0.065(7) & 0.002 \\
OT J234440.5$-$001206 & 22 & 0.081(10) & 0.004 \\
ROTSE3 J221519.8$-$003257.2 & 16 & 0.080(7) & 0.005 \\
SDSS J003941.06$+$005427.5 & 14 & 0.051(4) & 0.004 \\
SDSS J004335.14$-$003729.8 & 21 & 0.064(6) & 0.002 \\
SDSS J053659.12$+$002215.1 & 6 & 0.042(2) & 0.004 \\
SDSS J075507.70$+$143547.6 & 4 & 0.056(1) & 0.001 \\
SDSS J075939.79$+$191417.3 & 4 & 0.114(22) & 0.004 \\
SDSS J085623.00$+$310834.0 & 4 & 0.055(2) & 0.002 \\
SDSS J091001.63$+$164820.0 & 4 & 0.078(6) & 0.002 \\
SDSS J094325.90$+$520128.8 & 4 & 0.081(5) & 0.006 \\
SDSS J113215.50$+$624900.4 & 4 & 0.076(8) & 0.002 \\
SDSS J155644.24$-$000950.2 & 4 & 0.075(6) & 0.001 \\
SDSS J155644.24$-$000950.2 & 4 & 0.075(6) & 0.001 \\
SDSS J160501.35$+$203056.9 & 4 & 0.057(2) & 0.002 \\
SDSS J204720.76$+$000007.7 & 10 & 0.073(7) & 0.004 \\
SDSS J210014.12$+$004446.0 & 13 & 0.076(6) & 0.001 \\
SDSS J210449.94$+$010545.8 & 13 & 0.074(8) & 0.006 \\
SDSS J211605.43$+$113407.5 & 4 & 0.087(52) & 0.038 \\
SDSS J223439.93$+$004127.2 & 17 & 0.082(8) & 0.002 \\
SDSSp J230351.64$+$010651.0 & 25 & 0.087(13) & 0.002 \\
SDSS J081408.42$+$090759.1\commentb & 4 & 0.122(10) & 0.006 \\
\hline
  \multicolumn{4}{l}{\commenta Number of SDSS scans.}\\
  \multicolumn{4}{l}{\commentb Dwarf nova proposed by \citet{wil10newCVs}.}\\
\end{tabular}
\end{center}
\end{table}

\subsection{Category Classification by Neural Network}\label{sec:category}

   Since a direct estimation of $P_{\rm orb}$ is sometimes difficult,
especially if the object is too faint (large photometric errors), or
if the object has an unusually red color (due to the relatively small number
of training samples for long-$P_{\rm orb}$ objects),
we provide an alternative way for distinguishing dwarf novae into 
three categories: {\it ultrashort} ($P_{\rm orb} < 0.06$),
{\it short} ($0.06 \le P_{\rm orb} < 0.10$) and {\it long}
($P_{\rm orb} \ge 0.10$), approximately corresponding to WZ Sge type,
SU UMa type and SS Cyg/Z Cam type.  We applied neural-network analysis
to these categories, and estimated the probability belonging to each
category (table \ref{tab:dnlist2}).

   Except for some objects with unusually bright and luminous secondaries,
and, and except for the objects that are unusual for dwarf novae 
(see appendix \ref{sec:individual}
for individual notes), this classification appears to be very effective,
and can be applied to very faint objects.  We can obtain reasonable
classifications for ordinary dwarf novae of $g=21$, and even $g=22$ if the
objects are dominated by the secondary.
In particular, objects with probabilities for {\it ultrashort} category
exceed 0.4 can be considered to be very likely candidates for WZ Sge-type
dwarf novae or related objects.  We infer that this probability index 
can be used to identify new WZ Sge-type dwarf novae or 
potential period bouncers.

\begin{table*}
\caption{List of dwarf novae}\label{tab:dnlist}
\begin{center}
\begin{tabular}{cccccccc}
\hline
Object & $u$ & $g$ & $r$ & $i$ & $z$ & $P_{\rm orb}$ & $P_{\rm est}$\commenta \\
\hline
RX And & 13.996(4) & 14.082(4) & 13.741(3) & -- & 13.108(5) & 0.2099 & -- \\
BV And & 16.893(7) & 16.366(4) & 16.319(5) & 16.222(5) & 16.115(7) & -- & 0.072(1) \\
FN And & 18.718(21) & 18.719(8) & 18.473(9) & 18.140(9) & 17.664(17) & -- & 0.105(3) \\
IZ And & 19.402(30) & 19.621(13) & 19.312(13) & 18.878(13) & 18.361(27) & -- & 0.108(5) \\
LL And & 20.182(47) & 20.073(17) & 19.993(20) & 20.079(28) & 20.165(105) & 0.0551 & 0.066(2) \\
LL And & 20.209(70) & 20.152(21) & 20.163(30) & 20.349(49) & 20.523(211) & 0.0551 & 0.057(2) \\
LT And & 19.767(31) & 19.505(11) & 19.281(12) & 19.214(15) & 19.041(45) & -- & 0.099(3) \\
\end{tabular}
\end{center}
\end{table*}

\addtocounter{table}{-1}
\begin{table*}
\caption{List of dwarf novae (continued)}
\begin{center}
\begin{tabular}{cccccccc}
\hline
Object & $u$ & $g$ & $r$ & $i$ & $z$ & $P_{\rm orb}$ & $P_{\rm est}$\commenta \\
\hline
GS Cet & 20.578(68) & 20.395(23) & 20.412(36) & 20.543(55) & 20.434(206) & -- & 0.060(2) \\
GS Cet & 20.600(65) & 20.427(22) & 20.412(27) & 20.489(40) & 20.768(217) & -- & 0.070(5) \\
GS Cet & 20.610(61) & 20.381(20) & 20.369(31) & 20.477(50) & 20.739(261) & -- & 0.066(4) \\
GS Cet & 20.847(80) & 20.508(26) & 20.563(34) & 20.662(54) & 20.977(330) & -- & 0.065(5) \\
GS Cet & 21.237(242) & 20.455(53) & 20.411(51) & 20.594(67) & 21.001(283) & -- & 0.114(45) \\
GY Cet & 18.267(16) & 18.294(7) & 18.329(8) & 18.525(12) & 18.458(33) & 0.0566 & 0.056(1) \\
GY Cet & 18.276(15) & 18.296(7) & 18.414(9) & 18.586(14) & 18.433(38) & 0.0566 & 0.058(2) \\
\end{tabular}
\end{center}
\end{table*}

\addtocounter{table}{-1}
\begin{table*}
\caption{List of dwarf novae (continued)}
\begin{center}
\begin{tabular}{cccccccc}
\hline
Object & $u$ & $g$ & $r$ & $i$ & $z$ & $P_{\rm orb}$ & $P_{\rm est}$\commenta \\
\hline
V592 Her & 21.648(115) & 21.421(36) & 21.413(48) & 21.541(82) & 21.426(335) & 0.0558\commentb & 0.062(4) \\
V592 Her & 21.483(114) & 21.301(35) & 21.272(45) & 21.387(73) & 21.402(334) & 0.0558\commentb & 0.062(3) \\
V610 Her & 21.552(101) & 21.222(39) & 20.918(38) & 20.630(44) & 20.438(145) & -- & 0.176(19) \\
V610 Her & 21.564(161) & 21.421(57) & 21.136(54) & 20.862(59) & 20.716(223) & -- & 0.137(14) \\
V610 Her & 21.649(241) & 21.582(79) & 21.262(86) & 20.827(84) & 20.578(296) & -- & 0.152(17) \\
V611 Her & 20.311(45) & 20.584(22) & 20.534(27) & 20.387(39) & 19.847(93) & -- & 0.076(3) \\
V611 Her & 20.510(53) & 20.659(25) & 20.687(35) & 20.595(47) & 20.058(108) & -- & 0.072(3) \\
\end{tabular}
\end{center}
\end{table*}

\addtocounter{table}{-1}
\begin{table*}
\caption{List of dwarf novae (continued)}
\begin{center}
\begin{tabular}{cccccccc}
\hline
Object & $u$ & $g$ & $r$ & $i$ & $z$ & $P_{\rm orb}$ & $P_{\rm est}$\commenta \\
\hline
GV Psc & 20.501(55) & 20.409(21) & 20.192(23) & 19.903(24) & 19.511(67) & 0.0904\commentb & 0.108(5) \\
GV Psc & 20.032(49) & 20.508(27) & 20.200(25) & 19.764(23) & 19.346(59) & 0.0904\commentb & 0.111(5) \\
X Ser & 17.388(10) & 17.298(5) & 16.821(5) & 16.604(5) & 16.523(8) & 1.478 & 0.110(3) \\
RY Ser & 16.759(8) & 16.091(3) & 15.470(3) & 14.953(3) & 14.626(4) & 0.3009 & 0.159(10) \\
RY Ser & 17.372(11) & 16.598(4) & 15.640(4) & 15.171(4) & 14.813(4) & 0.3009 & 0.505(34) \\
QW Ser & 17.795(11) & 17.805(6) & 17.823(7) & 17.717(7) & 17.408(14) & 0.0745 & 0.074(1) \\
QZ Ser & 17.822(12) & 16.275(3) & 15.721(3) & 15.516(4) & 15.427(5) & 0.0832 & 0.180(56) \\
\end{tabular}
\end{center}
\end{table*}

\addtocounter{table}{-1}
\begin{table*}
\caption{List of dwarf novae (continued)}
\begin{center}
\begin{tabular}{cccccccc}
\hline
Object & $u$ & $g$ & $r$ & $i$ & $z$ & $P_{\rm orb}$ & $P_{\rm est}$\commenta \\
\hline
FSV J1722$+$2723 & 20.340(46) & 20.902(28) & 20.605(29) & 20.410(35) & 20.320(117) & -- & 0.084(4) \\
GD 552 & 16.427(7) & 16.398(4) & 16.197(4) & 16.533(5) & 16.539(10) & 0.0713 & 0.057(3) \\
GSC 847.1021 & 16.639(6) & 15.138(3) & 14.620(3) & 14.412(3) & 14.349(4) & -- & 0.170(55) \\
GUVV J090904.4$+$091714.4 & 21.049(92) & 21.447(57) & 21.092(50) & 20.724(52) & 20.459(154) & -- & 0.102(7) \\
GUVV J090904.4$+$091714.4 & 22.028(170) & 22.360(91) & 22.514(169) & 21.656(135) & 21.265(227) & -- & 0.151(39) \\
GUVV J090904.4$+$091714.4 & 22.316(192) & 22.467(110) & 22.103(108) & 21.572(88) & 21.221(228) & -- & 0.148(16) \\
HS 1016$+$3412 & 17.845(12) & 18.391(7) & 18.202(7) & 18.072(8) & 17.904(19) & 0.0794 & 0.077(1) \\
\end{tabular}
\end{center}
\end{table*}

\addtocounter{table}{-1}
\begin{table*}
\caption{List of dwarf novae (continued)}
\begin{center}
\begin{tabular}{cccccccc}
\hline
Object & $u$ & $g$ & $r$ & $i$ & $z$ & $P_{\rm orb}$ & $P_{\rm est}$\commenta \\
\hline
OT J010411.6$-$031341 & 19.017(24) & 19.516(13) & 19.140(13) & 18.829(13) & 18.898(42) & -- & 0.124(14) \\
OT J010522.2$+$110253 & 20.640(58) & 20.712(23) & 20.563(32) & 20.283(33) & 20.084(87) & -- & 0.105(4) \\
OT J010522.2$+$110253 & 20.833(77) & 20.850(33) & 20.680(40) & 20.451(43) & 19.978(102) & -- & 0.086(3) \\
OT J010522.2$+$110253 & 21.001(95) & 20.914(34) & 20.756(42) & 20.548(52) & 20.215(150) & -- & 0.092(5) \\
OT J010550.1$+$190317 & 19.421(36) & 19.644(16) & 19.818(21) & 19.958(30) & 19.984(99) & -- & 0.060(4) \\
OT J011134.5$+$275922 & 25.312(1930) & 22.292(169) & 21.642(144) & 21.750(207) & 22.566(933) & -- & -- \\
OT J011516.5$+$245530 & 20.759(59) & 20.933(29) & 20.876(36) & 20.786(49) & 20.436(127) & -- & 0.072(2) \\
\end{tabular}
\end{center}
\end{table*}

\addtocounter{table}{-1}
\begin{table*}
\caption{List of dwarf novae (continued)}
\begin{center}
\begin{tabular}{cccccccc}
\hline
Object & $u$ & $g$ & $r$ & $i$ & $z$ & $P_{\rm orb}$ & $P_{\rm est}$\commenta \\
\hline
OT J032839.9$-$010240 & 21.362(158) & 21.220(64) & 21.079(77) & 21.173(115) & 20.947(340) & -- & 0.063(5) \\
OT J032839.9$-$010240 & 21.380(83) & 21.391(42) & 21.145(46) & 21.291(72) & 21.213(232) & -- & 0.062(3) \\
OT J032839.9$-$010240 & 21.394(96) & 21.518(49) & 21.201(49) & 21.423(94) & 21.248(328) & -- & 0.064(5) \\
OT J032839.9$-$010240 & 21.498(145) & 21.563(67) & 21.402(74) & 21.406(104) & 21.534(390) & -- & 0.068(5) \\
OT J032839.9$-$010240 & 21.515(132) & 21.550(67) & 21.246(75) & 21.590(134) & 21.185(328) & -- & 0.072(13) \\
OT J032839.9$-$010240 & 21.554(112) & 21.396(47) & 21.341(75) & 21.526(141) & 21.965(681) & -- & 0.055(7) \\
OT J032839.9$-$010240 & 21.583(151) & 21.305(56) & 21.229(83) & 21.619(156) & 20.863(253) & -- & 0.075(21) \\
\end{tabular}
\end{center}
\end{table*}

\addtocounter{table}{-1}
\begin{table*}
\caption{List of dwarf novae (continued)}
\begin{center}
\begin{tabular}{cccccccc}
\hline
Object & $u$ & $g$ & $r$ & $i$ & $z$ & $P_{\rm orb}$ & $P_{\rm est}$\commenta \\
\hline
OT J051419.9$+$011121 & 19.744(35) & 20.214(21) & 20.209(34) & 19.978(45) & 19.613(106) & -- & 0.082(4) \\
OT J051419.9$+$011121 & 20.249(43) & 20.509(22) & 20.258(25) & 20.135(35) & 19.850(101) & -- & 0.070(2) \\
OT J052033.9$-$000530 & 19.785(36) & 19.566(13) & 19.060(12) & 18.710(11) & 18.466(28) & -- & 0.181(9) \\
OT J052033.9$-$000530 & 20.338(71) & 20.195(24) & 19.505(23) & 19.131(24) & 18.854(65) & -- & 0.286(20) \\
OT J052033.9$-$000530 & 20.579(54) & 20.561(24) & 19.724(18) & 19.304(19) & 18.855(52) & -- & 0.363(50) \\
OT J052033.9$-$000530 & 20.712(71) & 20.473(24) & 19.626(17) & 19.158(17) & 18.826(36) & -- & 0.409(66) \\
OT J052033.9$-$000530 & 20.754(71) & 20.553(30) & 19.714(21) & 19.282(23) & 18.885(62) & -- & 0.433(72) \\
\end{tabular}
\end{center}
\end{table*}

\addtocounter{table}{-1}
\begin{table*}
\caption{List of dwarf novae (continued)}
\begin{center}
\begin{tabular}{cccccccc}
\hline
Object & $u$ & $g$ & $r$ & $i$ & $z$ & $P_{\rm orb}$ & $P_{\rm est}$\commenta \\
\hline
OT J081418.9$-$005022 & 19.084(21) & 19.193(11) & 19.034(12) & 19.015(15) & 18.568(34) & 0.0741\commentb & 0.066(2) \\
OT J081712.3$+$055208 & 21.840(190) & 21.387(55) & 20.857(56) & 20.612(67) & 20.195(192) & -- & 0.488(123) \\
OT J081712.3$+$055208 & 22.024(241) & 21.462(60) & 20.903(55) & 20.971(105) & 20.634(266) & -- & -- \\
OT J081936.1$+$191540 & 20.876(83) & 20.364(24) & 19.938(22) & 19.755(28) & 19.907(116) & -- & 0.264(80) \\
OT J081936.1$+$191540 & 21.482(137) & 20.479(24) & 19.968(24) & 19.796(27) & 19.696(95) & -- & 0.403(149) \\
OT J082019.4$+$474732 & 21.267(94) & 21.361(50) & 21.305(58) & 21.302(84) & 20.894(205) & -- & 0.066(3) \\
OT J082123.7$+$454135 & 19.232(23) & 19.472(12) & 19.004(10) & 18.822(12) & 18.653(33) & -- & 0.126(3) \\
\end{tabular}
\end{center}
\end{table*}

\addtocounter{table}{-1}
\begin{table*}
\caption{List of dwarf novae (continued)}
\begin{center}
\begin{tabular}{cccccccc}
\hline
Object & $u$ & $g$ & $r$ & $i$ & $z$ & $P_{\rm orb}$ & $P_{\rm est}$\commenta \\
\hline
OT J101035.5$+$140239 & 18.457(16) & 17.196(4) & 17.277(5) & 17.359(6) & 17.407(15) & -- & 0.153(32) \\
OT J101035.5$+$140239 & 19.224(22) & 18.051(6) & 17.854(6) & 17.775(7) & 17.713(16) & -- & 0.439(146) \\
OT J103704.6$+$100224 & 23.476(623) & 23.421(274) & 22.812(229) & 22.693(296) & 22.565(656) & -- & -- \\
OT J103704.6$+$100224 & 24.717(895) & 22.857(180) & 22.509(211) & 22.275(274) & 22.686(846) & -- & -- \\
OT J101545.9$+$033312 & 19.811(35) & 20.168(19) & 20.137(24) & 20.130(33) & 19.864(82) & -- & 0.068(2) \\
OT J102146.4$+$234926 & 20.847(64) & 20.725(26) & 20.628(31) & 20.853(58) & 20.436(151) & 0.0554\commentb & 0.063(3) \\
OT J102616.0$+$192045 & 20.108(40) & 20.126(18) & 20.082(22) & 19.853(23) & 19.509(54) & 0.0800\commentb & 0.090(3) \\
\end{tabular}
\end{center}
\end{table*}

\addtocounter{table}{-1}
\begin{table*}
\caption{List of dwarf novae (continued)}
\begin{center}
\begin{tabular}{cccccccc}
\hline
Object & $u$ & $g$ & $r$ & $i$ & $z$ & $P_{\rm orb}$ & $P_{\rm est}$\commenta \\
\hline
OT J153150.8$+$152447 & 22.975(332) & 23.130(157) & 22.686(156) & 22.552(190) & 22.077(429) & -- & 0.156(51) \\
OT J153317.6$+$273428 & 22.338(224) & 21.932(76) & 21.461(74) & 21.146(91) & 21.477(417) & -- & -- \\
OT J153645.2$-$142543 & 22.984(507) & 22.879(150) & 22.913(262) & 22.558(300) & 22.593(785) & -- & 0.092(28) \\
OT J154354.1$-$143745 & 21.219(128) & 21.404(51) & 21.090(55) & 21.001(79) & 20.527(215) & -- & 0.070(3) \\
OT J154428.1$+$335725 & 22.361(276) & 22.108(90) & 21.850(109) & 22.280(251) & 21.613(565) & -- & -- \\
OT J154544.9$+$442830 & 20.850(64) & 20.943(26) & 20.850(34) & 20.718(41) & 20.328(127) & 0.0747\commentb & 0.076(3) \\
OT J155325.7$+$114437 & 22.867(448) & 22.646(142) & 23.474(432) & 22.730(437) & 23.415(863) & -- & -- \\
\end{tabular}
\end{center}
\end{table*}

\addtocounter{table}{-1}
\begin{table*}
\caption{List of dwarf novae (continued)}
\begin{center}
\begin{tabular}{cccccccc}
\hline
Object & $u$ & $g$ & $r$ & $i$ & $z$ & $P_{\rm orb}$ & $P_{\rm est}$\commenta \\
\hline
OT J182142.8$+$212154 & 20.195(53) & 20.260(21) & 20.198(25) & 19.895(27) & 19.438(62) & 0.0794\commentb & 0.089(3) \\
OT J202857.1$-$061803 & 20.767(77) & 20.606(30) & 20.557(36) & 20.597(49) & 20.145(126) & -- & 0.062(2) \\
OT J204001.4$-$144909 & 20.401(66) & 20.563(25) & 20.488(30) & 20.443(38) & 20.301(140) & -- & 0.071(2) \\
OT J204739.4$+$000840 & 23.154(715) & 22.213(131) & 21.997(153) & 21.569(150) & 21.322(512) & -- & -- \\
OT J210034.4$+$055436 & 21.970(189) & 22.113(72) & 22.024(121) & 21.830(141) & 20.948(226) & -- & 0.083(7) \\
OT J210034.4$+$055436 & 22.402(297) & 22.248(90) & 22.160(116) & 21.852(119) & 20.958(198) & -- & 0.094(9) \\
OT J210043.9$-$005212 & 21.178(131) & 21.301(65) & 20.759(52) & 20.643(70) & 20.381(233) & -- & 0.142(18) \\
\end{tabular}
\end{center}
\end{table*}

\addtocounter{table}{-1}
\begin{table*}
\caption{List of dwarf novae (continued)}
\begin{center}
\begin{tabular}{cccccccc}
\hline
Object & $u$ & $g$ & $r$ & $i$ & $z$ & $P_{\rm orb}$ & $P_{\rm est}$\commenta \\
\hline
OT J213829.5$-$001742 & 23.814(723) & 23.531(362) & 22.552(156) & 22.606(245) & 22.941(671) & -- & -- \\
OT J213937.6$-$023913 & 20.145(60) & 20.138(26) & 20.006(31) & 19.791(38) & 19.111(74) & -- & 0.079(2) \\
OT J213937.6$-$023913 & 20.201(68) & 20.010(19) & 19.866(25) & 19.644(31) & 19.206(72) & -- & 0.091(3) \\
OT J213937.6$-$023913 & 20.291(67) & 20.069(19) & 20.019(28) & 19.677(28) & 19.145(61) & -- & 0.099(4) \\
OT J214426.4$+$222024 & 18.909(26) & 17.644(5) & 17.131(5) & 16.984(6) & 16.911(14) & -- & -- \\
OT J214639.9$+$092119 & 22.173(312) & 21.829(71) & 21.710(89) & 21.887(144) & 22.327(740) & -- & 0.065(16) \\
OT J214804.4$+$080951 & 20.543(61) & 20.957(30) & 20.879(38) & 20.686(42) & 20.431(139) & -- & 0.081(3) \\
\end{tabular}
\end{center}
\end{table*}

\addtocounter{table}{-1}
\begin{table*}
\caption{List of dwarf novae (continued)}
\begin{center}
\begin{tabular}{cccccccc}
\hline
Object & $u$ & $g$ & $r$ & $i$ & $z$ & $P_{\rm orb}$ & $P_{\rm est}$\commenta \\
\hline
OT J224505.4$+$011547 & 21.211(135) & 21.463(68) & 21.379(94) & 21.551(182) & 21.406(499) & -- & 0.061(7) \\
OT J224505.4$+$011547 & 21.313(125) & 21.502(62) & 21.554(99) & 21.629(189) & 20.790(333) & -- & 0.073(10) \\
OT J224505.4$+$011547 & 21.438(130) & 21.553(57) & 21.537(72) & 21.665(117) & 21.225(288) & -- & 0.063(5) \\
OT J224753.9$+$235522 & 21.101(142) & 21.188(68) & 20.781(83) & 20.758(130) & 20.278(271) & -- & 0.089(12) \\
OT J224753.9$+$235522 & 21.658(142) & 21.500(52) & 21.282(66) & 21.399(114) & 21.103(305) & -- & 0.068(5) \\
OT J224814.5$+$331224 & 19.390(30) & 19.625(13) & 19.002(12) & 19.105(16) & 19.346(73) & -- & 0.114(20) \\
OT J224814.5$+$331224 & 21.050(86) & 20.938(28) & 20.867(36) & 20.793(52) & 20.583(151) & -- & 0.070(3) \\
\end{tabular}
\end{center}
\end{table*}

\addtocounter{table}{-1}
\begin{table*}
\caption{List of dwarf novae (continued)}
\begin{center}
\begin{tabular}{cccccccc}
\hline
Object & $u$ & $g$ & $r$ & $i$ & $z$ & $P_{\rm orb}$ & $P_{\rm est}$\commenta \\
\hline
ROTSE3 J154041.5$-$002703.2 & 21.057(78) & 21.057(35) & 20.556(34) & 20.276(37) & 19.752(91) & -- & 0.129(10) \\
ROTSE3 J154041.5$-$002703.2 & 21.320(94) & 21.356(44) & 20.976(47) & 20.391(42) & 19.891(107) & -- & 0.144(8) \\
ROTSE3 J154041.5$-$002703.2 & 21.924(201) & 21.824(70) & 21.602(87) & 20.997(79) & 20.317(145) & -- & 0.136(13) \\
ROTSE3 J212313$-$021446.6 & 22.145(223) & 22.117(78) & 21.926(95) & 22.488(232) & 22.126(551) & -- & -- \\
ROTSE3 J214850$-$020622.2 & 23.102(433) & 22.407(91) & 22.293(124) & 22.234(173) & 21.334(296) & -- & -- \\
ROTSE3 J221519.8$-$003257.2 & 20.735(83) & 20.980(37) & 20.859(41) & 20.713(51) & 20.338(131) & -- & 0.074(3) \\
ROTSE3 J221519.8$-$003257.2 & 20.916(118) & 21.192(69) & 20.881(56) & 20.796(68) & 20.335(147) & -- & 0.071(3) \\
\end{tabular}
\end{center}
\end{table*}

\addtocounter{table}{-1}
\begin{table*}
\caption{List of dwarf novae (continued)}
\begin{center}
\begin{tabular}{cccccccc}
\hline
Object & $u$ & $g$ & $r$ & $i$ & $z$ & $P_{\rm orb}$ & $P_{\rm est}$\commenta \\
\hline
SDSS J004335.14$-$003729.8 & 20.265(63) & 19.898(19) & 19.853(25) & 19.954(35) & 20.108(138) & 0.0572 & 0.077(4) \\
SDSS J004335.14$-$003729.8 & 20.269(59) & 19.860(19) & 19.940(26) & 20.032(43) & 19.930(147) & 0.0572 & 0.063(2) \\
SDSS J004335.14$-$003729.8 & 20.276(61) & 19.925(19) & 19.903(28) & 20.020(41) & 20.050(159) & 0.0572 & 0.069(3) \\
SDSS J012940.05$+$384210.4 & 19.680(27) & 19.786(14) & 20.002(20) & 20.179(34) & 20.075(112) & 0.0435\commentb & 0.061(6) \\
SDSS J033449.86$-$071047.8 & 17.569(11) & 17.862(6) & 17.693(6) & 17.577(7) & 17.518(19) & 0.0724\commentb & 0.078(1) \\
SDSS J033710.91$-$065059.4 & 19.621(33) & 19.558(13) & 19.724(19) & 19.969(33) & 20.099(146) & -- & 0.055(2) \\
SDSS J053659.12$+$002215.1 & 19.248(27) & 18.804(9) & 18.953(12) & 19.084(16) & 19.308(75) & -- & 0.053(3) \\
\end{tabular}
\end{center}
\end{table*}

\addtocounter{table}{-1}
\begin{table*}
\caption{List of dwarf novae (continued)}
\begin{center}
\begin{tabular}{cccccccc}
\hline
Object & $u$ & $g$ & $r$ & $i$ & $z$ & $P_{\rm orb}$ & $P_{\rm est}$\commenta \\
\hline
SDSS J090350.73$+$330036.1 & 18.850(18) & 18.829(8) & 18.792(10) & 18.879(12) & 18.619(31) & 0.0591 & 0.060(1) \\
SDSS J090403.48$+$035501.2 & 19.482(30) & 19.218(12) & 19.258(16) & 19.411(24) & 19.439(85) & 0.0597 & 0.058(1) \\
SDSS J090403.48$+$035501.2 & 19.497(28) & 19.304(12) & 19.352(14) & 19.509(22) & 19.388(66) & 0.0597 & 0.055(1) \\
SDSS J090403.48$+$035501.2 & 19.627(34) & 19.335(12) & 19.416(16) & 19.513(25) & 19.412(71) & 0.0597 & 0.059(1) \\
SDSS J090452.09$+$440255.4 & 19.623(33) & 19.381(13) & 19.478(17) & 19.693(28) & 19.774(93) & -- & 0.054(1) \\
SDSS J090628.25$+$052656.9 & 18.824(18) & 18.761(9) & 18.452(8) & 18.096(8) & 17.829(16) & -- & 0.143(4) \\
SDSS J090628.25$+$052656.9 & 19.321(25) & 19.415(12) & 18.833(11) & 18.403(11) & 18.096(25) & -- & 0.223(11) \\
\end{tabular}
\end{center}
\end{table*}

\addtocounter{table}{-1}
\begin{table*}
\caption{List of dwarf novae (continued)}
\begin{center}
\begin{tabular}{cccccccc}
\hline
Object & $u$ & $g$ & $r$ & $i$ & $z$ & $P_{\rm orb}$ & $P_{\rm est}$\commenta \\
\hline
SDSS J124426.26$+$613514.6 & 18.513(17) & 18.753(9) & 18.733(10) & 18.569(12) & 18.365(31) & 0.0992 & 0.084(1) \\
SDSS J124426.26$+$613514.6 & 18.740(18) & 18.718(8) & 18.843(11) & 18.773(13) & 18.431(29) & 0.0992 & 0.072(2) \\
SDSS J124819.36$+$072049.4 & 21.663(141) & 21.326(46) & 21.401(70) & 21.385(89) & 21.285(330) & -- & 0.073(7) \\
SDSS J124819.36$+$072049.4 & 21.786(208) & 21.371(59) & 21.533(94) & 21.329(116) & 21.611(545) & -- & 0.098(17) \\
SDSS J125023.85$+$665525.5 & 18.727(18) & 18.651(8) & 18.653(10) & 18.739(13) & 18.627(39) & 0.0587 & 0.061(1) \\
SDSS J125834.77$+$663551.6 & 20.026(41) & 20.566(25) & 20.234(29) & 20.482(50) & 20.100(132) & -- & 0.065(6) \\
SDSS J130514.73$+$582856.3 & 19.094(22) & 19.283(11) & 19.272(14) & 19.117(17) & 19.037(52) & -- & 0.087(2) \\
\end{tabular}
\end{center}
\end{table*}

\addtocounter{table}{-1}
\begin{table*}
\caption{List of dwarf novae (continued)}
\begin{center}
\begin{tabular}{cccccccc}
\hline
Object & $u$ & $g$ & $r$ & $i$ & $z$ & $P_{\rm orb}$ & $P_{\rm est}$\commenta \\
\hline
SDSS J161909.10$+$135145.5 & 18.897(20) & 18.494(8) & 17.800(6) & 17.429(6) & 17.174(12) & -- & 0.498(111) \\
SDSS J162212.45$+$341147.3 & 19.234(24) & 19.168(10) & 18.648(9) & 18.295(10) & 18.033(25) & -- & 0.266(13) \\
SDSS J162212.45$+$341147.3 & 19.241(26) & 19.119(10) & 18.684(9) & 18.337(9) & 18.042(20) & -- & 0.219(16) \\
SDSS J162212.45$+$341147.3 & 19.245(27) & 19.116(10) & 18.694(10) & 18.335(10) & 18.010(20) & -- & 0.210(16) \\
SDSS J162718.39$+$120435.0 & 20.122(46) & 20.225(18) & 19.786(19) & 19.604(24) & 19.062(49) & 0.1041\commentb & 0.109(6) \\
SDSS J162830.89$+$240259.1 & 19.854(39) & 19.737(16) & 19.775(18) & 19.670(23) & 19.367(61) & -- & 0.073(2) \\
SDSS J162830.89$+$240259.1 & 19.949(33) & 19.780(14) & 19.706(16) & 19.739(21) & 19.525(62) & -- & 0.063(1) \\
\end{tabular}
\end{center}
\end{table*}

\addtocounter{table}{-1}
\begin{table*}
\caption{List of dwarf novae (continued)}
\begin{center}
\begin{tabular}{cccccccc}
\hline
Object & $u$ & $g$ & $r$ & $i$ & $z$ & $P_{\rm orb}$ & $P_{\rm est}$\commenta \\
\hline
SDSS J210449.94$+$010545.8 & 20.259(46) & 20.304(19) & 19.997(22) & 19.869(29) & 19.608(86) & -- & 0.075(3) \\
SDSS J210449.94$+$010545.8 & 20.395(62) & 20.299(24) & 20.116(32) & 19.899(37) & 19.556(91) & -- & 0.080(3) \\
SDSS J210449.94$+$010545.8 & 20.628(183) & 20.965(100) & 20.734(83) & 20.593(90) & 20.368(199) & -- & 0.074(5) \\
SDSS J210449.94$+$010545.8 & 20.717(70) & 20.824(30) & 20.787(42) & 20.605(56) & 20.483(175) & -- & 0.078(4) \\
SDSS J210449.94$+$010545.8 & 20.875(185) & 20.654(53) & 20.617(53) & 20.394(65) & 19.981(156) & -- & 0.080(5) \\
SDSS J210449.94$+$010545.8 & 20.884(78) & 21.020(35) & 20.743(44) & 20.745(68) & 20.451(160) & -- & 0.064(2) \\
SDSS J210449.94$+$010545.8 & 21.079(115) & 21.061(43) & 20.856(54) & 20.730(73) & 20.329(179) & -- & 0.073(4) \\
\end{tabular}
\end{center}
\end{table*}

\addtocounter{table}{-1}
\begin{table*}
\caption{List of dwarf novae (continued)}
\begin{center}
\begin{tabular}{cccccccc}
\hline
Object & $u$ & $g$ & $r$ & $i$ & $z$ & $P_{\rm orb}$ & $P_{\rm est}$\commenta \\
\hline
SDSSp J230351.64$+$010651.0 & 17.433(10) & 18.106(6) & 17.497(5) & 17.379(6) & 17.187(14) & 0.0767 & 0.092(2) \\
SDSSp J230351.64$+$010651.0 & 17.441(10) & 17.975(6) & 17.668(6) & 17.458(7) & 17.315(16) & 0.0767 & 0.084(2) \\
SDSSp J230351.64$+$010651.0 & 17.464(10) & 18.028(6) & 17.636(6) & 17.393(6) & 17.232(14) & 0.0767 & 0.090(2) \\
SDSSp J230351.64$+$010651.0 & 17.512(10) & 17.863(6) & 17.802(6) & 17.575(7) & 17.197(14) & 0.0767 & 0.085(2) \\
SDSSp J230351.64$+$010651.0 & 17.521(10) & 17.917(6) & 17.652(6) & 17.447(6) & 17.061(12) & 0.0767 & 0.078(1) \\
SDSSp J230351.64$+$010651.0 & 17.556(10) & 17.928(6) & 17.610(6) & 17.459(6) & 17.254(14) & 0.0767 & 0.080(1) \\
SDSSp J230351.64$+$010651.0 & 17.615(11) & 17.972(6) & 17.762(6) & 17.516(7) & 17.271(14) & 0.0767 & 0.088(1) \\
\end{tabular}
\end{center}
\end{table*}

\addtocounter{table}{-1}
\begin{table*}
\caption{List of dwarf novae (continued)}
\begin{center}
\begin{tabular}{cccccccc}
\hline
Object & $u$ & $g$ & $r$ & $i$ & $z$ & $P_{\rm orb}$ & $P_{\rm est}$\commenta \\
\hline
SDSS J084108.10$+$102536.2\commentc & 19.818(34) & 20.247(18) & 20.132(21) & 20.015(25) & 20.047(95) & -- & 0.076(4) \\
SDSS J084108.10$+$102536.2\commentc & 20.621(93) & 21.027(43) & 20.808(57) & 20.869(78) & 21.109(376) & -- & 0.064(4) \\
SDSS J084108.10$+$102536.2\commentc & 20.818(124) & 20.822(48) & 20.806(80) & 20.943(112) & 21.158(574) & -- & 0.060(6) \\
SDSS J091242.18$+$620940.1\commentc & 18.729(19) & 18.818(8) & 18.729(10) & 18.442(10) & 17.856(21) & -- & 0.087(2) \\
SDSS J091741.29$+$073647.4\commentc & 21.804(142) & 21.520(52) & 21.396(69) & 21.501(117) & 21.152(304) & -- & 0.081(9) \\
SDSS J091741.29$+$073647.4\commentc & 21.854(149) & 21.435(50) & 21.449(83) & 21.706(141) & 21.096(307) & -- & 0.074(12) \\
SDSS J092620.42$+$034542.3\commentc & 19.455(29) & 19.827(15) & 19.609(16) & 19.619(20) & 19.563(64) & -- & 0.067(1) \\
\end{tabular}
\end{center}
\end{table*}

\addtocounter{table}{-1}
\begin{table*}
\caption{List of dwarf novae (continued)}
\begin{center}
\begin{tabular}{cccccccc}
\hline
Object & $u$ & $g$ & $r$ & $i$ & $z$ & $P_{\rm orb}$ & $P_{\rm est}$\commenta \\
\hline
SDSS J205931.86$-$070516.6\commentc & 20.077(46) & 20.198(22) & 20.178(26) & 19.973(30) & 20.134(118) & -- & 0.095(8) \\
SDSS J205931.86$-$070516.6\commentc & 20.854(100) & 20.755(30) & 21.026(50) & 21.237(87) & 21.556(458) & -- & 0.056(7) \\
SDSS J223854.51$+$053606.8\commentc & 21.441(159) & 21.432(57) & 21.339(92) & 21.314(136) & 20.988(336) & -- & 0.070(5) \\
SDSS J223854.51$+$053606.8\commentc & 21.536(141) & 21.396(55) & 21.243(65) & 21.299(96) & 20.603(220) & -- & 0.068(5) \\
SDSS J223854.51$+$053606.8\commentc & 21.553(118) & 21.548(47) & 21.520(56) & 21.531(84) & 20.847(181) & -- & 0.066(4) \\
\hline
  \multicolumn{8}{l}{\commenta Orbital period estimated with neural network.}\\
  \multicolumn{8}{l}{\commentb Determined from superhump period.}\\
  \multicolumn{8}{l}{\commentc Dwarf novae proposed by \citet{wil10newCVs}.}\\
\end{tabular}
\end{center}
\end{table*}

\begin{table*}
\caption{Estimated extinction neural network classification}\label{tab:dnlist2}
\begin{center}
\begin{tabular}{ccccccccc}
\hline
Object & $g$ & $d$ (pc) & type\commenta & $A_V$ & $P_{\rm orb}$ & ultrashort\commentb & short\commentc & long\commentd \\
\hline
RX And & 14.082(4) & 149 & 2 & 0.079 & 0.2099 & -- & -- & -- \\
BV And & 16.366(4) & 779 & 3 & 0.533 & -- & 0.462(55) & 0.316(47) & 0.222(100) \\
FN And & 18.719(8) & 356 & 3 & 0.144 & -- & 0.000(0) & 0.983(17) & 0.017(17) \\
IZ And & 19.621(13) & 1099 & 3 & 0.226 & -- & 0.000(0) & 0.905(136) & 0.095(136) \\
LL And & 20.073(17) & 550 & 1 & 0.133 & 0.0551 & 0.755(126) & 0.234(125) & 0.011(19) \\
LL And & 20.152(21) & 550 & 1 & 0.133 & 0.0551 & 0.876(199) & 0.124(199) & 0.001(2) \\
LT And & 19.505(11) & 559 & 5 & 0.173 & -- & 0.030(16) & 0.905(85) & 0.065(77) \\
\end{tabular}
\end{center}
\end{table*}

\addtocounter{table}{-1}
\begin{table*}
\caption{Estimated extinction and neural network classification (continued)}
\begin{center}
\begin{tabular}{ccccccccc}
\hline
Object & $g$ & $d$ (pc) & type\commenta & $A_V$ & $P_{\rm orb}$ & ultrashort\commentb & short\commentc & long\commentd \\
\hline
GS Cet & 20.410(21) & 826 & 5 & 0.079 & -- & 0.184(294) & 0.814(297) & 0.002(13) \\
GS Cet & 20.471(23) & 826 & 5 & 0.079 & -- & 0.811(150) & 0.189(151) & 0.001(2) \\
GS Cet & 20.358(23) & 826 & 5 & 0.079 & -- & 0.099(227) & 0.890(253) & 0.011(44) \\
GS Cet & 20.381(19) & 826 & 5 & 0.079 & -- & 0.861(95) & 0.138(96) & 0.001(4) \\
GS Cet & 20.446(23) & 826 & 5 & 0.079 & -- & 0.235(351) & 0.765(351) & 0.000(0) \\
GS Cet & 20.338(20) & 826 & 5 & 0.079 & -- & 0.861(181) & 0.139(181) & 0.000(2) \\
GS Cet & 20.395(23) & 826 & 5 & 0.079 & -- & 0.772(189) & 0.223(189) & 0.005(18) \\
\end{tabular}
\end{center}
\end{table*}

\addtocounter{table}{-1}
\begin{table*}
\caption{Estimated extinction and neural network classification (continued)}
\begin{center}
\begin{tabular}{ccccccccc}
\hline
Object & $g$ & $d$ (pc) & type\commenta & $A_V$ & $P_{\rm orb}$ & ultrashort\commentb & short\commentc & long\commentd \\
\hline
LT Eri & 17.748(5) & 858 & 2 & 0.305 & 0.1702 & 0.000(0) & 0.011(9) & 0.989(9) \\
HQ Gem & 20.386(21) & 3543 & 3 & 1.619 & -- & 0.000(0) & 0.000(1) & 1.000(1) \\
IR Gem & 15.996(5) & 320 & 1 & 0.178 & 0.0684 & 0.025(3) & 0.974(3) & 0.001(0) \\
KZ Gem & 18.666(8) & 1677 & 3 & 0.291 & -- & 0.000(0) & 0.998(1) & 0.002(1) \\
AH Her & -- & 237 & 2 & 0.105 & 0.2581 & -- & -- & -- \\
AH Her & 14.975(10) & 237 & 2 & 0.105 & 0.2581 & 0.000(0) & 0.317(42) & 0.683(42) \\
V478 Her & 17.741(6) & 4244 & 2 & 0.241 & 0.6290 & 0.000(0) & 0.016(15) & 0.984(15) \\
\end{tabular}
\end{center}
\end{table*}

\addtocounter{table}{-1}
\begin{table*}
\caption{Estimated extinction and neural network classification (continued)}
\begin{center}
\begin{tabular}{ccccccccc}
\hline
Object & $g$ & $d$ (pc) & type\commenta & $A_V$ & $P_{\rm orb}$ & ultrashort\commentb & short\commentc & long\commentd \\
\hline
V405 Peg & 16.100(3) & 1251 & 2 & 0.637 & 0.1776 & 0.000(0) & 0.042(13) & 0.958(13) \\
V405 Peg & 16.792(5) & 1251 & 2 & 0.637 & 0.1776 & -- & -- & -- \\
FO Per & 16.859(4) & 271 & 2 & 0.119 & 0.1719 & 0.000(0) & 0.198(59) & 0.802(59) \\
GK Per & -- & 455 & 1 & 0.645 & 1.9968 & -- & -- & -- \\
KT Per & 15.316(3) & 207 & 2 & 0.343 & 0.1627 & 0.000(0) & 0.000(0) & 1.000(0) \\
QY Per & 20.313(21) & 550 & 1 & 0.382 & 0.0760\commente & 0.014(11) & 0.705(308) & 0.282(310) \\
V336 Per & 19.703(15) & 1286 & 3 & 0.555 & -- & 0.000(0) & 0.000(0) & 1.000(0) \\
\end{tabular}
\end{center}
\end{table*}

\addtocounter{table}{-1}
\begin{table*}
\caption{Estimated extinction and neural network classification (continued)}
\begin{center}
\begin{tabular}{ccccccccc}
\hline
Object & $g$ & $d$ (pc) & type\commenta & $A_V$ & $P_{\rm orb}$ & ultrashort\commentb & short\commentc & long\commentd \\
\hline
KS UMa & 17.418(5) & 360 & 1 & 0.023 & 0.0680 & 0.008(3) & 0.990(3) & 0.002(1) \\
HV Vir & 19.181(11) & 300 & 1 & 0.103 & 0.0571 & 0.802(62) & 0.196(62) & 0.002(3) \\
OU Vir & 18.564(7) & 550 & 1 & 0.124 & 0.0727 & 0.344(40) & 0.622(46) & 0.034(21) \\
QZ Vir & 14.864(3) & 101 & 1 & 0.042 & 0.0588 & 0.147(7) & 0.842(10) & 0.011(5) \\
VW Vul & 15.772(3) & 430 & 2 & 0.395 & 0.1687 & 0.001(0) & 0.197(23) & 0.802(23) \\
1502$+$09 & 18.995(9) & 426 & 5 & 0.104 & -- & 0.038(16) & 0.962(16) & 0.000(0) \\
1502$+$09 & 19.111(10) & 426 & 5 & 0.104 & -- & 0.031(15) & 0.969(15) & 0.000(0) \\
\end{tabular}
\end{center}
\end{table*}

\addtocounter{table}{-1}
\begin{table*}
\caption{Estimated extinction and neural network classification (continued)}
\begin{center}
\begin{tabular}{ccccccccc}
\hline
Object & $g$ & $d$ (pc) & type\commenta & $A_V$ & $P_{\rm orb}$ & ultrashort\commentb & short\commentc & long\commentd \\
\hline
NSV 18230 & 20.429(22) & 1224 & 3 & 0.042 & -- & 0.015(18) & 0.963(93) & 0.022(94) \\
NSV 19466 & 18.290(7) & 1748 & 3 & 0.097 & -- & 0.123(29) & 0.876(30) & 0.001(1) \\
NSV 19466 & 17.686(6) & 1748 & 3 & 0.097 & -- & 0.024(4) & 0.974(4) & 0.003(1) \\
NSV 20657 & 21.200(38) & 1414 & 3 & 0.070 & -- & 0.178(226) & 0.667(310) & 0.155(288) \\
OT J000024.7$+$332543 & 20.760(31) & 1149 & 3 & 0.159 & -- & 0.010(12) & 0.953(113) & 0.037(109) \\
OT J000130.5$+$050624 & 20.967(34) & 1106 & 3 & 0.113 & -- & 0.000(0) & 0.331(393) & 0.669(393) \\
OT J000659.6$+$192818 & 21.231(56) & 3082 & 3 & 0.143 & -- & 0.030(54) & 0.940(127) & 0.030(116) \\
\end{tabular}
\end{center}
\end{table*}

\addtocounter{table}{-1}
\begin{table*}
\caption{Estimated extinction and neural network classification (continued)}
\begin{center}
\begin{tabular}{ccccccccc}
\hline
Object & $g$ & $d$ (pc) & type\commenta & $A_V$ & $P_{\rm orb}$ & ultrashort\commentb & short\commentc & long\commentd \\
\hline
OT J032651.7$+$011513 & 22.091(122) & 2768 & 6 & 0.339 & -- & 0.000(0) & 0.039(161) & 0.961(161) \\
OT J032651.7$+$011513 & 22.407(123) & 2768 & 6 & 0.339 & -- & 0.000(0) & 0.003(23) & 0.997(23) \\
OT J032651.7$+$011513 & 22.350(95) & 2768 & 6 & 0.339 & -- & 0.000(0) & 0.025(113) & 0.975(113) \\
OT J032651.7$+$011513 & 22.397(126) & 2768 & 6 & 0.339 & -- & 0.000(0) & 0.010(82) & 0.990(82) \\
OT J032651.7$+$011513 & 22.495(108) & 2768 & 6 & 0.339 & -- & 0.000(0) & 0.022(109) & 0.978(109) \\
OT J032651.7$+$011513 & 22.235(98) & 2768 & 6 & 0.339 & -- & 0.000(0) & 0.030(125) & 0.970(125) \\
OT J032651.7$+$011513 & 22.486(98) & 2768 & 6 & 0.339 & -- & 0.000(0) & 0.006(42) & 0.994(42) \\
\end{tabular}
\end{center}
\end{table*}

\addtocounter{table}{-1}
\begin{table*}
\caption{Estimated extinction and neural network classification (continued)}
\begin{center}
\begin{tabular}{ccccccccc}
\hline
Object & $g$ & $d$ (pc) & type\commenta & $A_V$ & $P_{\rm orb}$ & ultrashort\commentb & short\commentc & long\commentd \\
\hline
OT J032839.9$-$010240 & 21.774(82) & 592 & 3 & 0.350 & -- & 0.141(292) & 0.842(305) & 0.016(85) \\
OT J032839.9$-$010240 & 21.525(60) & 592 & 3 & 0.350 & -- & 0.292(385) & 0.683(400) & 0.025(145) \\
OT J032839.9$-$010240 & 21.584(55) & 592 & 3 & 0.350 & -- & 0.035(162) & 0.546(468) & 0.418(471) \\
OT J032839.9$-$010240 & 21.448(50) & 592 & 3 & 0.350 & -- & 0.381(398) & 0.265(368) & 0.354(431) \\
OT J032902.0$+$060047 & 21.625(64) & 1789 & 3 & 1.130 & -- & 0.001(3) & 0.521(449) & 0.479(450) \\
OT J032902.0$+$060047 & 21.274(42) & 1789 & 3 & 1.130 & -- & 0.011(19) & 0.553(360) & 0.436(365) \\
OT J032902.0$+$060047 & 22.037(72) & 1789 & 3 & 1.130 & -- & 0.001(3) & 0.277(383) & 0.723(384) \\
\end{tabular}
\end{center}
\end{table*}

\addtocounter{table}{-1}
\begin{table*}
\caption{Estimated extinction and neural network classification (continued)}
\begin{center}
\begin{tabular}{ccccccccc}
\hline
Object & $g$ & $d$ (pc) & type\commenta & $A_V$ & $P_{\rm orb}$ & ultrashort\commentb & short\commentc & long\commentd \\
\hline
OT J055730.1$+$001514 & 22.920(137) & 1377 & 6 & 2.156 & -- & 0.001(4) & 0.652(457) & 0.347(457) \\
OT J055842.8$+$000626 & 20.329(20) & 714 & 3 & 1.721 & -- & 0.000(0) & 0.001(3) & 0.999(3) \\
OT J055842.8$+$000626 & 19.803(15) & 714 & 3 & 1.721 & -- & 0.000(0) & 0.000(0) & 1.000(0) \\
OT J055842.8$+$000626 & 20.434(22) & 714 & 3 & 1.721 & -- & 0.000(0) & 0.001(9) & 0.999(9) \\
OT J055842.8$+$000626 & 20.732(29) & 714 & 3 & 1.721 & -- & 0.000(0) & 0.000(0) & 1.000(0) \\
OT J073055.5$+$425636 & 22.761(150) & 1185 & 6 & 0.182 & -- & 0.536(456) & 0.251(394) & 0.213(374) \\
OT J073339.3$+$212201 & 20.447(27) & 2068 & 6 & 0.172 & -- & 0.000(0) & 0.000(0) & 1.000(0) \\
\end{tabular}
\end{center}
\end{table*}

\addtocounter{table}{-1}
\begin{table*}
\caption{Estimated extinction and neural network classification (continued)}
\begin{center}
\begin{tabular}{ccccccccc}
\hline
Object & $g$ & $d$ (pc) & type\commenta & $A_V$ & $P_{\rm orb}$ & ultrashort\commentb & short\commentc & long\commentd \\
\hline
OT J081712.3$+$055208 & 21.462(60) & 7044 & 3 & 0.082 & -- & 0.000(0) & 0.287(419) & 0.713(419) \\
OT J081936.1$+$191540 & 20.364(24) & 4275 & 3 & 0.162 & -- & 0.000(0) & 0.338(409) & 0.662(409) \\
OT J081936.1$+$191540 & 20.479(24) & 4275 & 3 & 0.162 & -- & 0.000(0) & 0.902(236) & 0.098(236) \\
OT J082019.4$+$474732 & 21.361(50) & 2322 & 3 & 0.139 & -- & 0.200(223) & 0.763(250) & 0.037(152) \\
OT J082123.7$+$454135 & 19.472(12) & 2214 & 3 & 0.141 & -- & 0.000(0) & 0.004(11) & 0.996(11) \\
OT J082123.7$+$454135 & 19.923(19) & 2214 & 3 & 0.141 & -- & 0.000(0) & 0.815(115) & 0.184(115) \\
OT J082123.7$+$454135 & 19.987(20) & 2214 & 3 & 0.141 & -- & 0.001(1) & 0.952(78) & 0.047(78) \\
\end{tabular}
\end{center}
\end{table*}

\addtocounter{table}{-1}
\begin{table*}
\caption{Estimated extinction and neural network classification (continued)}
\begin{center}
\begin{tabular}{ccccccccc}
\hline
Object & $g$ & $d$ (pc) & type\commenta & $A_V$ & $P_{\rm orb}$ & ultrashort\commentb & short\commentc & long\commentd \\
\hline
OT J092839.3$+$005944 & 21.236(136) & 2474 & 3 & 0.232 & -- & 0.020(86) & 0.381(408) & 0.599(425) \\
OT J092839.3$+$005944 & 21.775(76) & 2474 & 3 & 0.232 & -- & 0.011(65) & 0.752(360) & 0.237(360) \\
OT J092839.3$+$005944 & 21.800(79) & 2474 & 3 & 0.232 & -- & 0.000(0) & 0.562(396) & 0.438(396) \\
OT J092839.3$+$005944 & 21.901(71) & 2474 & 3 & 0.232 & -- & 0.000(1) & 0.572(410) & 0.428(410) \\
OT J101035.5$+$140239 & 17.196(4) & 6408 & 3 & 0.140 & -- & 0.000(0) & 1.000(0) & 0.000(0) \\
OT J101035.5$+$140239 & 18.051(6) & 6408 & 3 & 0.140 & -- & 0.000(0) & 1.000(0) & 0.000(0) \\
OT J103704.6$+$100224 & 23.421(274) & 3921 & 6 & 0.083 & -- & 0.012(90) & 0.429(473) & 0.558(476) \\
\end{tabular}
\end{center}
\end{table*}

\addtocounter{table}{-1}
\begin{table*}
\caption{Estimated extinction and neural network classification (continued)}
\begin{center}
\begin{tabular}{ccccccccc}
\hline
Object & $g$ & $d$ (pc) & type\commenta & $A_V$ & $P_{\rm orb}$ & ultrashort\commentb & short\commentc & long\commentd \\
\hline
OT J144316.5$-$010222 & 22.417(106) & 1963 & 3 & 0.140 & -- & 0.190(328) & 0.469(441) & 0.341(441) \\
OT J144316.5$-$010222 & 22.236(87) & 1963 & 3 & 0.140 & -- & 0.008(29) & 0.830(278) & 0.162(278) \\
OT J145502.2$+$143815 & 20.117(21) & 1435 & 3 & 0.119 & -- & 0.000(0) & 0.004(9) & 0.996(9) \\
OT J145502.2$+$143815 & 20.485(22) & 1435 & 3 & 0.119 & -- & 0.000(0) & 0.007(46) & 0.993(46) \\
OT J145502.2$+$143815 & 20.282(23) & 1435 & 3 & 0.119 & -- & 0.000(0) & 0.001(6) & 0.999(6) \\
OT J145921.8$+$354806 & 21.544(39) & 1110 & 2 & 0.040 & 0.0822\commente & 0.000(0) & 0.165(304) & 0.835(304) \\
OT J151020.7$+$182303 & 21.424(43) & 11218 & 3 & 0.093 & -- & 0.000(0) & 0.003(17) & 0.997(17) \\
\end{tabular}
\end{center}
\end{table*}

\addtocounter{table}{-1}
\begin{table*}
\caption{Estimated extinction and neural network classification (continued)}
\begin{center}
\begin{tabular}{ccccccccc}
\hline
Object & $g$ & $d$ (pc) & type\commenta & $A_V$ & $P_{\rm orb}$ & ultrashort\commentb & short\commentc & long\commentd \\
\hline
OT J165002.8$+$435616 & 22.758(164) & 4749 & 3 & 0.056 & -- & 0.000(1) & 0.521(447) & 0.479(447) \\
OT J165002.8$+$435616 & 22.376(113) & 4749 & 3 & 0.056 & -- & 0.028(131) & 0.749(396) & 0.223(388) \\
OT J165002.8$+$435616 & 22.855(161) & 4749 & 3 & 0.056 & -- & 0.050(186) & 0.368(443) & 0.582(464) \\
OT J170115.8$-$024159 & 22.977(203) & 2061 & 6 & 1.080 & -- & 0.022(138) & 0.687(443) & 0.291(436) \\
OT J170151.6$+$132131 & 21.317(76) & 4327 & 3 & 0.248 & -- & 0.000(0) & 0.100(262) & 0.900(262) \\
OT J170606.1$+$255153 & 21.479(66) & 1768 & 3 & 0.112 & -- & 0.048(94) & 0.836(257) & 0.116(258) \\
OT J170609.7$+$143452 & 18.231(7) & 642 & 3 & 0.458 & -- & 0.009(2) & 0.991(2) & 0.000(0) \\
\end{tabular}
\end{center}
\end{table*}

\addtocounter{table}{-1}
\begin{table*}
\caption{Estimated extinction and neural network classification (continued)}
\begin{center}
\begin{tabular}{ccccccccc}
\hline
Object & $g$ & $d$ (pc) & type\commenta & $A_V$ & $P_{\rm orb}$ & ultrashort\commentb & short\commentc & long\commentd \\
\hline
OT J211550.9$-$000716 & 22.795(566) & 3279 & 3 & 0.233 & -- & 0.002(18) & 0.588(481) & 0.410(482) \\
OT J211550.9$-$000716 & 23.313(183) & 3279 & 3 & 0.233 & -- & 0.004(43) & 0.099(284) & 0.897(293) \\
OT J211550.9$-$000716 & 23.120(198) & 3279 & 3 & 0.233 & -- & 0.001(6) & 0.172(360) & 0.828(361) \\
OT J212025.1$+$194157 & 21.819(64) & 1578 & 6 & 0.259 & -- & 0.029(123) & 0.949(183) & 0.022(135) \\
OT J212555.1$-$032406 & 21.977(88) & 2928 & 6 & 0.217 & -- & 0.438(472) & 0.022(139) & 0.539(474) \\
OT J212555.1$-$032406 & 22.000(179) & 2928 & 6 & 0.217 & -- & 0.554(472) & 0.101(289) & 0.345(446) \\
OT J212633.3$+$085459 & 20.757(26) & 2077 & 3 & 0.182 & -- & 0.018(31) & 0.945(73) & 0.037(69) \\
\end{tabular}
\end{center}
\end{table*}

\addtocounter{table}{-1}
\begin{table*}
\caption{Estimated extinction and neural network classification (continued)}
\begin{center}
\begin{tabular}{ccccccccc}
\hline
Object & $g$ & $d$ (pc) & type\commenta & $A_V$ & $P_{\rm orb}$ & ultrashort\commentb & short\commentc & long\commentd \\
\hline
OT J220449.7$+$054852 & 20.035(19) & 886 & 3 & 0.233 & -- & 0.001(0) & 0.720(295) & 0.279(295) \\
OT J220449.7$+$054852 & 20.940(32) & 886 & 3 & 0.233 & -- & 0.000(0) & 0.033(85) & 0.967(85) \\
OT J221128.7$-$030516 & 19.555(13) & 1229 & 3 & 0.305 & -- & 0.224(74) & 0.776(74) & 0.000(0) \\
OT J221128.7$-$030516 & 19.718(17) & 1229 & 3 & 0.305 & -- & 0.108(61) & 0.890(65) & 0.002(10) \\
OT J221128.7$-$030516 & 19.864(17) & 1229 & 3 & 0.305 & -- & 0.124(82) & 0.875(83) & 0.001(2) \\
OT J221128.7$-$030516 & 20.177(24) & 1229 & 3 & 0.305 & -- & 0.010(9) & 0.985(31) & 0.005(31) \\
OT J221128.7$-$030516 & 20.597(25) & 1229 & 3 & 0.305 & -- & 0.441(254) & 0.559(254) & 0.000(0) \\
\end{tabular}
\end{center}
\end{table*}

\addtocounter{table}{-1}
\begin{table*}
\caption{Estimated extinction and neural network classification (continued)}
\begin{center}
\begin{tabular}{ccccccccc}
\hline
Object & $g$ & $d$ (pc) & type\commenta & $A_V$ & $P_{\rm orb}$ & ultrashort\commentb & short\commentc & long\commentd \\
\hline
OT J232551.5$-$014024 & 18.853(9) & 1226 & 3 & 0.196 & -- & 0.000(0) & 0.000(0) & 1.000(0) \\
OT J232619.4$+$282650 & 20.893(27) & 1720 & 3 & 0.369 & -- & 0.006(12) & 0.833(277) & 0.161(280) \\
OT J232619.4$+$282650 & 20.933(39) & 1720 & 3 & 0.369 & -- & 0.012(21) & 0.703(323) & 0.285(325) \\
OT J232619.4$+$282650 & 21.074(32) & 1720 & 3 & 0.369 & -- & 0.018(21) & 0.491(301) & 0.491(307) \\
OT J233938.7$-$053305 & 18.996(10) & 4296 & 3 & 0.110 & -- & 0.000(0) & 0.004(7) & 0.996(7) \\
OT J234440.5$-$001206 & 20.308(30) & 600 & 2 & 0.122 & 0.0745\commente & 0.005(7) & 0.970(117) & 0.025(118) \\
OT J234440.5$-$001206 & 20.358(23) & 600 & 2 & 0.122 & 0.0745\commente & 0.012(9) & 0.988(10) & 0.000(1) \\
\end{tabular}
\end{center}
\end{table*}

\addtocounter{table}{-1}
\begin{table*}
\caption{Estimated extinction and neural network classification (continued)}
\begin{center}
\begin{tabular}{ccccccccc}
\hline
Object & $g$ & $d$ (pc) & type\commenta & $A_V$ & $P_{\rm orb}$ & ultrashort\commentb & short\commentc & long\commentd \\
\hline
ROTSE3 J221519.8$-$003257.2 & 21.182(59) & 1757 & 3 & 0.317 & -- & 0.092(189) & 0.723(371) & 0.185(348) \\
ROTSE3 J221519.8$-$003257.2 & 21.455(51) & 1757 & 3 & 0.317 & -- & 0.003(5) & 0.862(289) & 0.135(290) \\
ROTSE3 J221519.8$-$003257.2 & 21.679(59) & 1757 & 3 & 0.317 & -- & 0.012(45) & 0.937(180) & 0.051(176) \\
ROTSE3 J221519.8$-$003257.2 & 21.348(84) & 1757 & 3 & 0.317 & -- & 0.168(273) & 0.554(390) & 0.278(380) \\
ROTSE3 J221519.8$-$003257.2 & 21.598(50) & 1757 & 3 & 0.317 & -- & 0.083(138) & 0.850(223) & 0.067(187) \\
RX J1715.6$+$6856 & 18.633(8) & 1235 & 2 & 0.112 & 0.0683 & 0.157(30) & 0.840(30) & 0.004(3) \\
RX J1831.7$+$6511 & 17.037(4) & 661 & 2 & 0.149 & 0.167 & 0.008(1) & 0.973(14) & 0.018(13) \\
\end{tabular}
\end{center}
\end{table*}

\addtocounter{table}{-1}
\begin{table*}
\caption{Estimated extinction and neural network classification (continued)}
\begin{center}
\begin{tabular}{ccccccccc}
\hline
Object & $g$ & $d$ (pc) & type\commenta & $A_V$ & $P_{\rm orb}$ & ultrashort\commentb & short\commentc & long\commentd \\
\hline
SDSS J053659.12$+$002215.1 & 18.849(9) & 274 & 5 & 0.777 & -- & 0.999(0) & 0.001(0) & 0.000(0) \\
SDSS J053659.12$+$002215.1 & 18.776(9) & 274 & 5 & 0.777 & -- & 0.977(70) & 0.023(70) & 0.000(0) \\
SDSS J074355.56$+$183834.8 & 20.089(18) & 1095 & 5 & 0.120 & -- & 0.000(0) & 0.000(0) & 1.000(0) \\
SDSS J074355.56$+$183834.8 & 19.785(15) & 1095 & 5 & 0.120 & -- & 0.000(0) & 0.000(0) & 1.000(0) \\
SDSS J074531.92$+$453829.6 & 19.041(9) & 300 & 1 & 0.154 & 0.0528 & 0.619(125) & 0.381(125) & 0.000(0) \\
SDSS J074640.62$+$173412.8 & 19.767(13) & 1062 & 2 & 0.116 & 0.0649\commente & 0.065(31) & 0.903(39) & 0.032(25) \\
SDSS J074640.62$+$173412.8 & 19.807(16) & 1062 & 2 & 0.116 & 0.0649\commente & 0.001(1) & 0.115(158) & 0.884(159) \\
\end{tabular}
\end{center}
\end{table*}

\addtocounter{table}{-1}
\begin{table*}
\caption{Estimated extinction and neural network classification (continued)}
\begin{center}
\begin{tabular}{ccccccccc}
\hline
Object & $g$ & $d$ (pc) & type\commenta & $A_V$ & $P_{\rm orb}$ & ultrashort\commentb & short\commentc & long\commentd \\
\hline
SDSS J090452.09$+$440255.4 & 19.381(13) & 498 & 5 & 0.051 & -- & 0.920(62) & 0.080(62) & 0.000(0) \\
SDSS J090628.25$+$052656.9 & 18.761(9) & 1448 & 3 & 0.148 & -- & 0.000(0) & 0.017(13) & 0.983(13) \\
SDSS J090628.25$+$052656.9 & 19.415(12) & 1448 & 3 & 0.148 & -- & 0.000(0) & 0.000(0) & 1.000(0) \\
SDSS J091001.63$+$164820.0 & 18.832(9) & 434 & 5 & 0.106 & -- & 0.010(7) & 0.989(7) & 0.000(0) \\
SDSS J091001.63$+$164820.0 & 18.781(9) & 434 & 5 & 0.106 & -- & 0.006(1) & 0.994(1) & 0.000(0) \\
SDSS J091001.63$+$164820.0 & 18.967(9) & 434 & 5 & 0.106 & -- & 0.002(1) & 0.995(4) & 0.002(3) \\
SDSS J091001.63$+$164820.0 & 19.252(11) & 434 & 5 & 0.106 & -- & 0.003(1) & 0.994(6) & 0.003(5) \\
\end{tabular}
\end{center}
\end{table*}

\addtocounter{table}{-1}
\begin{table*}
\caption{Estimated extinction and neural network classification (continued)}
\begin{center}
\begin{tabular}{ccccccccc}
\hline
Object & $g$ & $d$ (pc) & type\commenta & $A_V$ & $P_{\rm orb}$ & ultrashort\commentb & short\commentc & long\commentd \\
\hline
SDSS J124058.03$-$015919.2 & 19.545(13) & 379 & 2 & 0.087 & 0.0259 & 0.962(78) & 0.038(78) & 0.000(0) \\
SDSS J124417.89$+$300401.0 & 19.571(12) & 871 & 3 & 0.054 & -- & 0.040(21) & 0.960(21) & 0.000(0) \\
SDSS J124426.26$+$613514.6 & 18.753(9) & 3067 & 2 & 0.050 & 0.0992 & 0.024(7) & 0.867(68) & 0.110(68) \\
SDSS J124426.26$+$613514.6 & 18.718(8) & 3067 & 2 & 0.050 & 0.0992 & 0.043(10) & 0.956(10) & 0.001(1) \\
SDSS J124819.36$+$072049.4 & 21.326(46) & 1887 & 3 & 0.097 & -- & 0.321(316) & 0.462(349) & 0.217(339) \\
SDSS J124819.36$+$072049.4 & 21.371(59) & 1887 & 3 & 0.097 & -- & 0.053(152) & 0.684(403) & 0.264(391) \\
SDSS J125023.85$+$665525.5 & 18.651(8) & 320 & 1 & 0.050 & 0.0587 & 0.534(67) & 0.465(68) & 0.000(1) \\
\end{tabular}
\end{center}
\end{table*}

\addtocounter{table}{-1}
\begin{table*}
\caption{Estimated extinction and neural network classification (continued)}
\begin{center}
\begin{tabular}{ccccccccc}
\hline
Object & $g$ & $d$ (pc) & type\commenta & $A_V$ & $P_{\rm orb}$ & ultrashort\commentb & short\commentc & long\commentd \\
\hline
SDSS J160111.53$+$091712.7 & 20.603(26) & 872 & 5 & 0.131 & -- & 0.000(1) & 0.571(395) & 0.429(395) \\
SDSS J160419.02$+$161548.5 & 19.082(10) & 1592 & 3 & 0.119 & -- & 0.063(13) & 0.933(14) & 0.004(2) \\
SDSS J160501.35$+$203056.9 & 19.361(11) & 199 & 2 & 0.191 & 0.0567 & 0.896(54) & 0.104(54) & 0.000(0) \\
SDSS J160501.35$+$203056.9 & 19.883(15) & 199 & 2 & 0.191 & 0.0567 & 0.756(167) & 0.243(167) & 0.002(5) \\
SDSS J160501.35$+$203056.9 & 19.891(15) & 199 & 2 & 0.191 & 0.0567 & 0.907(74) & 0.092(74) & 0.001(1) \\
SDSS J160501.35$+$203056.9 & 19.926(16) & 199 & 2 & 0.191 & 0.0567 & 0.916(81) & 0.083(81) & 0.000(0) \\
SDSS J161030.35$+$445901.7 & 19.753(15) & 585 & 5 & 0.049 & -- & 0.966(74) & 0.032(62) & 0.002(17) \\
\end{tabular}
\end{center}
\end{table*}

\addtocounter{table}{-1}
\begin{table*}
\caption{Estimated extinction and neural network classification (continued)}
\begin{center}
\begin{tabular}{ccccccccc}
\hline
Object & $g$ & $d$ (pc) & type\commenta & $A_V$ & $P_{\rm orb}$ & ultrashort\commentb & short\commentc & long\commentd \\
\hline
SDSS J210014.12$+$004446.0 & 18.380(7) & 769 & 2 & 0.249 & 0.0840\commente & 0.108(26) & 0.872(41) & 0.020(22) \\
SDSS J210014.12$+$004446.0 & 18.366(7) & 769 & 2 & 0.249 & 0.0840\commente & 0.295(68) & 0.639(95) & 0.066(63) \\
SDSS J210014.12$+$004446.0 & 18.453(7) & 769 & 2 & 0.249 & 0.0840\commente & 0.046(10) & 0.638(95) & 0.315(98) \\
SDSS J210014.12$+$004446.0 & 18.574(8) & 769 & 2 & 0.249 & 0.0840\commente & 0.058(15) & 0.885(32) & 0.056(33) \\
SDSS J210014.12$+$004446.0 & 18.407(12) & 769 & 2 & 0.249 & 0.0840\commente & 0.242(52) & 0.594(85) & 0.164(105) \\
SDSS J210014.12$+$004446.0 & 18.630(8) & 769 & 2 & 0.249 & 0.0840\commente & 0.093(18) & 0.844(24) & 0.063(23) \\
SDSS J210014.12$+$004446.0 & 18.731(8) & 769 & 2 & 0.249 & 0.0840\commente & 0.165(33) & 0.824(33) & 0.011(8) \\
\end{tabular}
\end{center}
\end{table*}

\addtocounter{table}{-1}
\begin{table*}
\caption{Estimated extinction and neural network classification (continued)}
\begin{center}
\begin{tabular}{ccccccccc}
\hline
Object & $g$ & $d$ (pc) & type\commenta & $A_V$ & $P_{\rm orb}$ & ultrashort\commentb & short\commentc & long\commentd \\
\hline
SDSS J225831.18$-$094931.7 & 15.587(3) & 410 & 1 & 0.124 & 0.0830\commente & 0.086(5) & 0.914(5) & 0.000(0) \\
SDSS J233325.92$+$152222.2 & 18.521(7) & 355 & 4 & 0.206 & 0.0577 & 0.411(55) & 0.588(55) & 0.000(0) \\
SDSS J233325.92$+$152222.2 & 18.745(9) & 355 & 4 & 0.206 & 0.0577 & 0.069(35) & 0.922(31) & 0.010(9) \\
SDSS J233325.92$+$152222.2 & 18.761(8) & 355 & 4 & 0.206 & 0.0577 & 0.429(65) & 0.532(60) & 0.039(19) \\
SDSSp J081321.91$+$452809.4 & 18.259(7) & 1933 & 2 & 0.159 & 0.289 & 0.000(0) & 0.000(0) & 1.000(0) \\
SDSSp J081321.91$+$452809.4 & 18.517(8) & 1933 & 2 & 0.159 & 0.289 & 0.000(0) & 0.000(0) & 1.000(0) \\
SDSSp J081610.84$+$453010.2 & 20.056(17) & 1454 & 3 & 0.195 & -- & 0.000(0) & 0.334(258) & 0.666(258) \\
\end{tabular}
\end{center}
\end{table*}

\addtocounter{table}{-1}
\begin{table*}
\caption{Estimated extinction and neural network classification (continued)}
\begin{center}
\begin{tabular}{ccccccccc}
\hline
Object & $g$ & $d$ (pc) & type\commenta & $A_V$ & $P_{\rm orb}$ & ultrashort\commentb & short\commentc & long\commentd \\
\hline
SDSS J073758.55$+$205544.5\commentf & 19.984(18) & 195 & 5 & 0.075 & -- & 0.941(61) & 0.059(61) & 0.000(0) \\
SDSS J074500.58$+$332859.6\commentf & 22.185(82) & 510 & 5 & 0.132 & -- & 0.000(1) & 0.189(332) & 0.810(332) \\
SDSS J074500.58$+$332859.6\commentf & 22.235(97) & 510 & 5 & 0.132 & -- & 0.000(0) & 0.121(304) & 0.879(304) \\
SDSS J074859.55$+$312512.6\commentf & 17.768(5) & 97 & 5 & 0.070 & -- & 0.804(25) & 0.162(17) & 0.034(13) \\
SDSS J075107.50$+$300628.4\commentf & 20.213(27) & 866 & 3 & 0.161 & -- & 0.008(8) & 0.992(8) & 0.000(1) \\
SDSS J075107.50$+$300628.4\commentf & 19.777(15) & 866 & 3 & 0.161 & -- & 0.554(128) & 0.443(128) & 0.003(4) \\
SDSS J075117.00$+$100016.2\commentf & 18.488(7) & 692 & 3 & 0.058 & -- & 0.193(38) & 0.807(38) & 0.000(0) \\
\end{tabular}
\end{center}
\end{table*}

\addtocounter{table}{-1}
\begin{table*}
\caption{Estimated extinction and neural network classification (continued)}
\begin{center}
\begin{tabular}{ccccccccc}
\hline
Object & $g$ & $d$ (pc) & type\commenta & $A_V$ & $P_{\rm orb}$ & ultrashort\commentb & short\commentc & long\commentd \\
\hline
SDSS J142953.56$+$073231.2\commentf & 21.208(45) & 204 & 5 & 0.073 & -- & 0.000(0) & 0.429(374) & 0.571(374) \\
SDSS J142953.56$+$073231.2\commentf & 21.150(53) & 204 & 5 & 0.073 & -- & 0.000(0) & 0.055(155) & 0.945(155) \\
SDSS J142953.56$+$073231.2\commentf & 21.175(39) & 204 & 5 & 0.073 & -- & 0.000(0) & 0.041(134) & 0.959(134) \\
SDSS J151109.79$+$574100.3\commentf & 21.239(35) & 2788 & 3 & 0.048 & -- & 0.040(64) & 0.935(90) & 0.025(71) \\
SDSS J151109.79$+$574100.3\commentf & 21.514(51) & 2788 & 3 & 0.048 & -- & 0.001(2) & 0.916(168) & 0.084(168) \\
SDSS J151109.79$+$574100.3\commentf & 21.361(39) & 2788 & 3 & 0.048 & -- & 0.001(1) & 0.722(373) & 0.278(373) \\
SDSS J152124.38$+$112551.9\commentf & 21.615(51) & 505 & 5 & 0.118 & -- & 0.274(352) & 0.469(424) & 0.257(384) \\
\end{tabular}
\end{center}
\end{table*}

\section{Distribution of Orbital Periods}\label{sec:porbdist}

   The present study provides a sample of DNe whose quiescent states
were recorded by SDSS and whose outbursts were detected with CRTS.
Since most of non-magnetic CVs below the period gap are DNe,
this sample can be used to explore the distribution of orbital
periods of CVs toward the shortest end.  This sample has
an advantage of being uniformly sampled, and less biased
than any known previous samples, including SDSS CVs, which were
selected initially by colors and then by spectroscopy.  Since there
is a chance that on the SDSS color criterion might have missed a significant
part of CVs, particularly WZ Sge-type dwarf novae, inside
the color exclusion zones (cf. subsection \ref{sec:bouncers}),
the present outburst-selected sample would be
an excellent alternative.\footnote{
   It would be worth noting that both OT J074727.6$+$065050 and
   OT J012059.6$+$325545 mentioned in subsection \ref{sec:bouncers} were
   discovered by amateur observers and not by the CRTS.  There might
   have been a difference in detection policies between the CRTS
   and amateur observers, and it might affect the homogeneity of
   the CRTS transients.  This aspect needs to be investigated further.
}

   In discussing the parent population of an outburst-selected
sample, we need to incorporate the effect of the detection
probability.  \citet{uem10shortPCV} proposed a method to
estimate the intrinsic $P_{\rm orb}$-distribution of DNe using a
Bayesian approach.  In this subsection, we applied this method to the
present sample of the CRTS transients.  

   The CRTS-SDSS sample was selected in the following way: 
first, we selected DNe having $P_{\rm est}=70$--$130$~min, since by 
the Bayesian model we used
objects only in this region; second, we excluded objects whose 
$P_{\rm est}$ had a large error of $>40$~\% of $P_{\rm est}$.
We obtained the remaining 123 SDSS objects whose outbursts were
detected by CRTS.  We used weighted mean $P_{\rm est}$ in the case
that multiple SDSS observations provided multiple $P_{\rm est}$ for
the same object.  Table~\ref{tab:crtsotlist} lists their names,
$P_{\rm est}$, their errors, and the number of recorded outbursts
in the CRTS data.  Most of them are new sources discovered by
CRTS (namely ``OT" sources), while CRTS detected outbursts of several
known sources, which are also included in the table.  The inclusion
of these already-known sources in the sample is important for avoiding 
a bias resulting from exclusion of already-known, 
i.e. frequently outbursting, relatively long-$P_{\rm orb}$ objects.

\begin{figure}
  \begin{center}
    \FigureFile(88mm,88mm){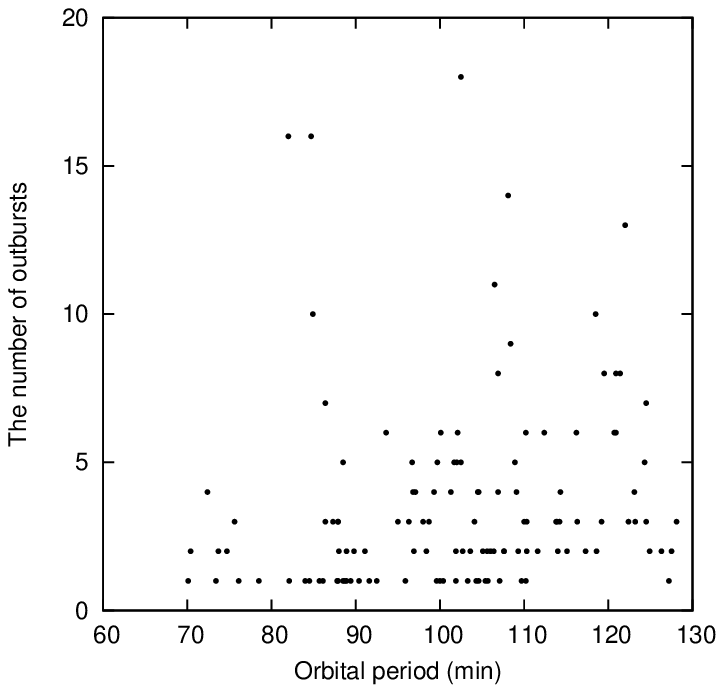}
  \end{center}
  \caption{The number of outbursts of the CRTS-SDSS sample. The
    abscissa denotes the orbital period in min.}\label{fig:ts} 
\end{figure}

   \citet{uem10shortPCV} assumed that the detection probability
($=1/T_{\rm s}$, where $T_{\rm s}$ is the typical recurrence time
of superoutbursts) is independent of $P_{\rm orb}$ in a long $P_{\rm orb}$ 
region of $P_{\rm orb}\gtrsim 85$~min, while it is lower in a shorter 
$P_{\rm orb}$ region at $P_{\rm orb}\lesssim 85$~min.  This assumption
can be tested with the present CRTS-SDSS sample.   Figure~\ref{fig:ts}
shows the relation between the number of recorded outbursts and
$P_{\rm est}$ of the sample.  As can be seen in the figure,
the CRTS-SDSS sample follows the trend assumed in \citet{uem10shortPCV}:
the number of high activity sources is small in a short period region of
$P_{\rm est}\lesssim 85$~min.  

\begin{figure}
  \begin{center}
    \FigureFile(88mm,88mm){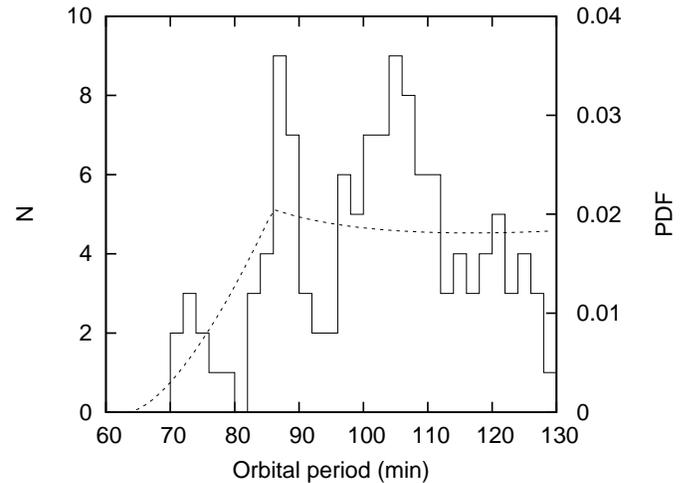}
  \end{center}
  \caption{Distribution of $P_{\rm est}$ of the CRTS-SDSS sample. 
    The dashed line indicates the probability density function (PDF) 
    of the best model derived from our analysis (for details
    see the text).}\label{fig:pdistr}
\end{figure}

   The $P_{\rm est}$ distribution of our CRTS-SDSS sample is shown in 
figure~\ref{fig:pdistr}.  In \citet{uem10shortPCV}, two samples, 
the ``ASAS" and ``mixed" samples, were analyzed.  The ASAS sample 
included 42 DNe, whose outbursts were detected with the ASAS survey 
from 2003 to 2007.  The mixed sample included 146 DNe, whose outbursts
were reported to VSNET.  We applied the Kolmogorov-Smirnov (KS) test to
evaluate whether the CRTS-SDSS sample differs from those samples.  The
KS probabilities were calculated to be 0.18 and 0.08 for the ASAS and
mixed samples, respectively.  Hence, this test indicates that the
CRTS-SDSS sample has a different $P_{\rm orb}$ distribution from the
samples in \citet{uem10shortPCV}.  Their $P_{\rm orb}$ distribution 
apparently has a peak at $P_{\rm orb}\sim 86$~min.  
In the CRTS-SDSS sample, the
distribution is rather flat in a region of $P_{\rm orb}\gtrsim
86$~min.  The relatively flat distribution may be partly due to
a smearing effect of the $P_{\rm orb}$ distribution resulting
from errors in estimating $P_{\rm est}$.
As shown in table \ref{tab:crtsotlist}, some of $P_{\rm est}$ have
large uncertainties, which imply that the real distribution was
significantly smeared out.

\begin{figure}
  \begin{center}
    \FigureFile(88mm,88mm){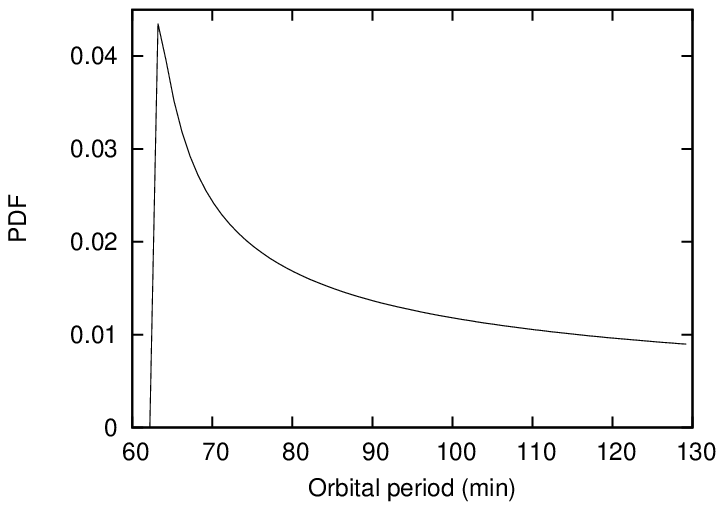}
  \end{center}
  \caption{Probability density function (PDF) of the intrinsic 
   $P_{\rm orb}$ distribution of DNe estimated from the CRTS-SDSS sample.}
  \label{fig:intr}
\end{figure}

   We estimated the intrinsic $P_{\rm orb}$ distribution using the 
CRTS-SDSS sample.  The intrinsic $P_{\rm orb}$ distribution is 
described in the model is as follows:
\begin{eqnarray}
I(p)= \left \{
\begin{array}{ll}
p^{-\alpha} e^{-\alpha/p}/A_I & (p \ge 1) \\
0 & (p < 1)
\end{array} \right. \\
p = P_{\rm orb}-P_{\rm min}+1\; ({\rm min}),
\end{eqnarray}
where $\alpha$ is a parameter for the profile of the $P_{\rm orb}$ 
distribution, $P_{\rm min}$ the minimum $P_{\rm orb}$, and  $A_I$
the normalization factor.
A larger $\alpha$ yields a distribution with a stronger  
spike near $P_{\rm min}$.  The distribution is flat in the case of 
$\alpha=0$.  Using the same model as in \citet{uem10shortPCV} and
adopting the derived $n=2.0$ for the dependence of the detection
probability on $P_{\rm orb}$, we calculated the posterior probability
distribution of $\alpha$ and $P_{\rm min}$ from the CRTS-SDSS sample.
The medians and 68.3\% confidence intervals were derived from
the estimated posterior distribution, which are listed in
table~\ref{tab:post}.  The table also includes the results in
\citet{uem10shortPCV}.  The model for the observed distribution is
indicated by the dashed line in figure~\ref{fig:pdistr}.
The intrinsic $P_{\rm orb}$ distribution,
which is obtained from the parameters, is depicted in
figure~\ref{fig:intr}.  As can be seen in this figure, a period spike
feature appears just above $P_{\rm min}$.  It is consistent with the
results in \citet{uem10shortPCV}.

   On the other hand, the parameter $\alpha$ is significantly smaller
than those in \citet{uem10shortPCV}, as shown in
table~\ref{tab:post}.   As a result, the spike feature is less
prominent.  As mentioned above, the samples used in
\citet{uem10shortPCV} have a peak at $P_{\rm orb}\sim 86$~min in their
$P_{\rm orb}$ distribution.   Such a profile tends to give a higher
$\alpha$ in our model.  The small $\alpha$ for the CRTS-SDSS sample is
probably due to the relatively flat distribution in a region of
$P_{\rm orb}\gtrsim 86$~min; in other words, it is a  potential smearing
effect of the distribution in $P_{\rm est}$.   The parameter
$P_{\rm min}$ is estimated to be smaller than those in
\citet{uem10shortPCV}.  This parameter is highly dependent on
the number of objects near $P_{\rm min}$.   In the CRTS-SDSS sample,
there are larger number of objects in the region of
$P_{\rm orb}=70$--$80$, compared to the samples in \citet{uem10shortPCV}.
This may also be due to the smearing effect and the period minimum
determined from $P_{\rm est}$ could be shorter than the real $P_{\rm min}$.
The systematic trends of shortening of the $P_{\rm min}$ and
smaller $\alpha$ have been confirmed by artificially introducing ``zitter''
errors to the samples in \citet{uem10shortPCV}.
We thus could not find very convincing difference in the distribution
of $P_{\rm orb}$ between CRTS-SDSS and past outburst-selected samples.
Most objects with a short $P_{\rm est}$,
however, do not have a measured $P_{\rm orb}$ (either spectroscopically
or photometrically), and there remains a possibility that these
objects may be genuine ultrashort-period CVs.  Characterizations
of these objects by further observations are desired.

\begin{table}
\caption{CRTS sources for the estimation of the intrinsic $P_{\rm orb}$
  distribution.}\label{tab:crtsotlist}
\begin{tabular}{cccl}
\hline
$P_{\rm est}$\commenta & error\commenta & $N$\commentb & Object \\
\hline
 70.1 & 14.0 & 1 & OT J004902.0$+$074726 \\ 
 70.4 &  7.2 & 2 & OT J151037.4$+$084104 \\ 
 72.4 & 18.2 & 4:& OT J164748.0$+$433845 \\ 
 73.4 &  2.4 & 1 & EL UMa \\ 
 73.7 &  4.1 & 2 & OT J073758.5$+$205545 \\ 
 74.7 &  4.1 & 2 & OT J010550.1$+$190317 \\ 
 75.6 &  2.5 & 3 & OT J035003.4$+$370052 \\ 
 76.1 &  3.0 & 1 & OT J084555.1$+$033930 \\ 
 78.5 &  2.6 & 1 & OT J104411.4$+$211307 \\ 
 82.0 &  2.1 &16 & OT J010329.0$+$331822 \\ 
 82.1 & 29.4 & 1 & OT J230425.8$+$062546 \\ 
 84.0 & 15.0 & 1 & OT J223606.3$+$050517 \\ 
 84.5 &  2.2 & 1 & OT J112253.3$-$111037 \\ 
 84.7 & 22.8 &16 & OT J215630.5$-$031957 \\ 
 84.9 &  1.5 &10:& OT J141002.2$-$124809 \\ 
 85.7 &  2.5 & 1 & OT J005824.6$+$283304 \\ 
 86.1 &  4.5 & 1 & OT J084041.5$+$000520 \\ 
 86.4 &  1.9 & 7:& OT J074419.7$+$325448 \\ 
 86.4 &  8.6 & 3 & OT J223136.0$+$180747 \\ 
 87.3 & 20.3 & 3 & OT J102937.7$+$414046 \\ 
 87.8 &  1.3 & 1 & OT J032839.9$-$010240 \\ 
 87.9 &  3.4 & 1 & OT J102637.0$+$475426 \\ 
 87.9 &  4.4 & 3 & OT J075414.5$+$313216 \\ 
 87.9 &  4.6 & 3:& OT J163311.3$-$011132 \\ 
 88.0 &  4.0 & 2 & OT J163120.9$+$103134 \\ 
 88.5 &  4.8 & 5:& OT J224505.4$+$011547 \\ 
 88.5 &  4.8 & 1 & OT J102146.4$+$234926 \\ 
 88.7 & 45.3 & 1 & OT J091453.6$+$113402 \\ 
 88.9 &  1.9 & 1 & OT J130030.3$+$115101 \\ 
 88.9 &  1.9 & 2 & OT J132536.0$+$210037 \\ 
 89.4 &  5.7 & 1 & OT J222824.1$+$134944 \\ 
 89.8 &  9.0 & 2 & OT J082654.7$-$000733 \\ 
 90.4 &  2.6 & 1 & OT J202857.1$-$061803 \\ 
 91.1 & 13.0 & 2 & OT J004807.2$+$264621 \\ 
 91.6 &  5.6 & 1 & OT J223058.3$+$210147 \\ 
 92.5 &  2.2 & 1 & OT J221128.7$-$030516 \\ 
 93.6 &  2.2 & 6 & OT J085603.8$+$322109 \\ 
 95.0 &  1.7 & 3 & OT J124417.9$+$300401 \\ 
 95.9 &  2.3 & 1 & OT J014150.4$+$090822 \\ 
 96.3 &  4.5 & 3:& OT J084413.7$-$012807 \\ 
 96.7 &  0.1 & 5 & ASAS J224349$+$0809.5 \\ 
 96.8 &  2.8 & 4 & OT J162806.2$+$065316 \\ 
 96.9 & 23.3 & 2 & OT J214639.9$+$092119 \\ 
 97.1 &  4.5 & 4:& OT J082019.4$+$474732 \\ 
 98.0 &  4.6 & 3:& OT J043517.8$+$002941 \\ 
 98.4 &  2.9 & 2 & OT J101545.9$+$033312 \\ 
 98.7 &  2.9 & 3 & OT J040659.8$+$005244 \\ 
 99.3 &  3.2 & 4:& OT J210954.1$+$163052 \\ 
 99.6 &  8.1 & 1 & OT J020056.0$+$195727 \\ 
 99.7 &  4.8 & 5 & OT J075648.0$+$305805 \\ 
100.0 &  1.9 & 1:& OT J011516.5$+$245530 \\ 
100.1 &  4.7 & 6:& OT J154354.1$-$143745 \\ 
100.4 &  1.5 & 1:& OT J221344.0$+$173252 \\ 
101.3 &  2.0 & 4 & OT J164950.4$+$035835 \\ 
\hline
  \multicolumn{4}{l}{\commenta Unit min.} \\
  \multicolumn{4}{l}{\commentb Number of outbursts recorded by CRTS.} \\
\end{tabular}
\end{table}

\addtocounter{table}{-1}
\begin{table}
\caption{CRTS sources for the estimation of the intrinsic $P_{\rm orb}$
  distribution (continued).}
\begin{tabular}{cccl}
\hline
$P_{\rm est}$\commenta & error\commenta & $N$\commentb & Object \\
\hline
101.7 &  5.1 & 5 & OT J162619.8$-$125557 \\ 
101.9 &  2.8 & 2 & OT J001538.3$+$263657 \\ 
101.9 &  2.4 & 1 & OT J215815.3$+$094709 \\ 
102.0 & 35.4 & 5:& OT J162012.0$+$115257 \\ 
102.1 &  3.8 & 6:& OT J224814.5$+$331224 \\ 
102.5 &  3.3 & 5:& OT J204001.4$-$144909 \\ 
102.5 &  1.2 &18:& V1032 Oph\\ 
102.7 &  0.8 & 2 & OT J170609.7$+$143452 \\ 
103.3 &  5.2 & 1 & OT J075332.0$+$375801 \\ 
103.6 &  4.0 & 2 & OT J105835.1$+$054706 \\ 
104.1 & 66.9 & 3:& OT J091634.6$+$130358 \\ 
104.3 &  3.3 & 1 & OT J210650.6$+$110250 \\ 
104.5 &  2.6 & 4 & OT J103317.3$+$072119 \\ 
104.6 & 19.9 & 1 & OT J214959.9$+$124529 \\ 
104.6 &  3.0 & 4 & OT J004500.3$+$222708 \\ 
105.1 &  9.1 & 2 & OT J224753.9$+$235522 \\ 
105.4 &  5.6 & 1 & OT J074928.0$+$190452 \\ 
105.6 &  1.9 & 2 & OT J043829.1$+$004016 \\ 
105.7 &  2.3 & 1 & SDSS J152419.33$+$220920.0 \\ 
106.0 &  5.9 & 2 & OT J081418.9$-$005022 \\ 
106.4 &  5.0 & 2 & OT J043546.9$+$090837 \\ 
106.5 & 41.1 &11:& OT J090852.2$+$071640 \\ 
106.9 &  6.2 & 4 & OT J224823.7$-$092059 \\ 
106.9 &  6.1 & 8 & V844 Her\\ 
107.1 &  2.0 & 1 & OT J051419.9$+$011121 \\ 
107.6 &  3.7 & 2 & OT J112509.7$+$231036 \\ 
107.6 &  8.0 & 2 & OT J082603.7$+$113821 \\ 
108.1 &  2.1 &14:& OT J080853.7$+$355053 \\ 
108.4 &  1.7 & 9 & QW Ser \\ 
108.9 &  8.1 & 5 & OT J124819.4$+$072050 \\ 
109.1 & 11.5 & 4 & OT J160844.8$+$220610 \\ 
109.3 &  2.5 & 2 & OT J112332.0$+$431718 \\ 
109.7 & 33.3 & 1 & UW Tri\\ 
110.0 &  9.7 & 3 & OT J231142.8$+$204036 \\ 
110.2 &  4.1 & 1 & OT J154544.9$+$442830 \\ 
110.2 &  3.1 & 6 & OT J003304.0$+$380106 \\ 
110.3 & 31.6 & 2 & OT J090516.1$+$120451 \\ 
110.3 & 15.8 & 3 & OT J144316.5$-$010222 \\ 
111.6 &  1.5 & 2 & HS 1016$+$3412 \\ 
112.4 &  2.5 & 6 & OT J021110.2$+$171624 \\ 
113.8 &  2.9 & 3 & SDSS J100515.39$+$191108.0 \\ 
113.9 &  3.9 & 3 & OT J011543.2$+$333724 \\ 
114.0 &  2.6 & 2 & TT Boo\\ 
114.2 &  5.4 & 3 & OT J170702.5$+$165339 \\ 
114.3 &  4.1 & 4:& OT J043742.1$+$003048 \\ 
115.1 &  4.5 & 2:& OT J214804.4$+$080951 \\ 
116.2 &  0.4 & 6 & NSV04838 \\ 
116.3 &  4.4 & 3 & OT J000024.7$+$332543 \\ 
117.3 & 29.5 & 2 & OT J082821.8$+$105344 \\ 
118.5 & 10.6 & 10& OT J163942.7$+$122414 \\ 
118.6 &  3.9 & 2 & OT J141712.0$-$180328 \\ 
119.2 &  4.7 & 3:& OT J212633.3$+$085459 \\ 
119.5 &  3.5 & 8 & OT J084358.1$+$425037 \\ 
120.7 & 12.7 & 6 & OT J085113.4$+$344449 \\ 
\hline
  \multicolumn{4}{l}{\commenta Unit min.} \\
  \multicolumn{4}{l}{\commentb Number of outbursts recorded by CRTS.} \\
\end{tabular}
\end{table}

\addtocounter{table}{-1}
\begin{table}
\caption{CRTS sources for the estimation of the intrinsic $P_{\rm orb}$
  distribution (continued).}
\begin{tabular}{cccl}
\hline
$P_{\rm est}$\commenta & error\commenta & $N$\commentb & Object \\
\hline
120.9 & 27.1 & 8 & OT J155430.6$+$365043 \\ 
120.9 &  5.1 & 6 & OT J213937.6$-$023913 \\ 
121.4 &  4.2 & 8:& OT J224253.4$+$172538 \\ 
122.0 &  9.2 &13:& OT J232619.4$+$282650 \\ 
122.4 &  4.0 & 3 & OT J085409.4$+$201339 \\ 
123.1 & 13.5 & 4 & OT J001158.3$+$315544 \\ 
123.2 & 11.8 & 3 & OT J152037.9$+$040948 \\ 
124.3 &  2.9 & 5 & OT J000659.6$+$192818 \\ 
124.5 &  3.3 & 3 & EG Aqr\\ 
124.5 &  8.7 & 7 & OT J041636.9$+$292806 \\ 
124.9 & 35.7 & 2 & OT J211550.9$-$000716 \\ 
126.3 &  3.1 & 2:& OT J223909.8$+$250331 \\ 
127.2 & 23.2 & 1 & OT J214842.5$-$000723 \\ 
127.5 & 10.3 & 2 & OT J215636.3$+$193242 \\ 
128.1 &  3.7 & 3 & OT J102616.0$+$192045 \\ 
\hline
  \multicolumn{4}{l}{\commenta Unit min.} \\
  \multicolumn{4}{l}{\commentb Number of outbursts recorded by CRTS.} \\
\end{tabular}
\end{table}

\begin{table}
  \caption{Parameters estimated from the Bayesian
  analysis.}\label{tab:post} 
  \begin{center}
    \begin{tabular}{ccc}
     \hline 
     Sample & $\alpha$   & $P_{\rm min}$ (min)\\
     \hline
     CRTS-SDSS  & $0.49^{+0.22}_{-0.24}$ & $63.2^{+2.7}_{-3.8}$ \\
     \hline
     ASAS$^*$      & $1.00^{+0.28}_{-0.30}$ & $71.9^{+2.4}_{-3.8}$ \\
     all RKcat$^*$ & $0.98^{+0.16}_{-0.16}$ & $72.5^{+1.3}_{-1.6}$ \\
     \hline
     \multicolumn{3}{l}{\footnotesize{$^*$\citet{uem10shortPCV}}}
    \end{tabular}
  \end{center}
\end{table}

\section{Conclusion}

   We investigated de-reddened SDSS colors of known dwarf novae,
and found correlations between orbital periods and colors.
The $u-g$ color is a particularly good indicator for systems with
the shortest orbital periods and can be used to distinguish
WZ Sge-type candidates.
We have also developed a method for estimating orbital periods of
dwarf novae from SDSS colors in quiescence using an artificial
neural network.
For typical objects below the period gap with considerable 
photometric accuracy,
we could estimate orbital periods to a 1 $\sigma$ error of 22 \%.
The error in estimation is worse for longer period systems.
We also developed a neural-network-based method for a categorical
classification.  This method has proven to be efficient in classifying
objects into three categories (WZ Sge type, SU UMa type and SS Cyg/Z Cam type)
and works for very faint objects to a limit of $g$=21.
Using these methods, we investigated the distribution of orbital
periods of dwarf novae from a modern transient survey (CRTS).
We confirmed that the number of detected outbursts were smaller
in systems with orbital periods shorter than 85 min.  The estimated
parent population was different from those of earlier outburst-selected
samples investigated by \citet{uem10shortPCV}, in that the present sample
tends to give a flatter distribution toward the shortest period and
a shorter estimate of the period minimum.  Although it is likely this was
a result of uncertainties resulting from neural-network analysis
and photometric errors, it is necessary to confirm of the real nature of
objects having the shortest estimated periods.
We also provide estimated orbital periods, estimated classifications and
supplementary information for known dwarf novae with quiescent SDSS
photometry.
We have also shown that there are a significant number of transients
whose quiescent colors are inside the WD exclusion box of SDSS
spectroscopic follow-up, and suggested that the fraction of
shortrst-$P_{\rm orb}$ objects and period bouncers were significantly
biased in the SDSS CVs.

\section*{Note added in proof:}

The following objects have been named in \citet{NameList80b}
during the proofreading period: 
2QZ J142701.6$-$012310 = V558 Vir,
GSC 847.1021 = IU Leo,
HS 1016$+$3412 = AC LMi,
HS 1340$+$1524 = HW Boo,
OT J074727.6$+$065050 = DY CMi,
OT J080714.2$+$113812 = KK Cnc,
OT J084555.1$+$033930 = V498 Hya,
OT J102146.4$+$234926 = IK Leo,
ROTSE3 J151453.6$+$020934.2 = V418 Ser,
SDSS J005050.88$+$000912.6 = GS Cet,
SDSS J013132.39$-$090122.3 = GY Cet,
SDSS J040714.78$-$064425.1 = LT Eri,
SDSS J074531.92$+$453829.6 = EQ Lyn,
SDSS J080434.20$+$510349.2 = EZ Lyn,
SDSS J084400.10$+$023919.3 = V495 Hya,
SDSS J090103.93$+$480911.1 = PU UMa,
SDSS J090452.09$+$440255.4 = FV Lyn,
SDSS J123813.73$-$033933.0 = V406 Vir,
SDSS J124426.26$+$613514.6 = V351 UMa,
SDSS J125023.85$+$665525.5 = OV Dra,
SDSS J133941.11$+$484727.5 = V355 UMa,
SDSS J150240.98$+$333423.9 = NZ Boo,
SDSS J151413.72$+$454911.9 = PP Boo,
SDSS J155644.24$-$000950.2 = V493 Ser,
SDSSp J081321.91$+$452809.4 = FH Lyn,
SDSSp J082409.73$+$493124.4 = FV Lyn.

\subsection*{V1047 Aql}

According to Rod Stubbings, the observation by Greg Bolt during
the 2005 August outburst detected superhumps, and the superhump period
was about 0.074 d.  This result is consistent with our categorical
classification.

\vskip 3mm

This work was supported by the Grant-in-Aid for the Global COE Program
``The Next Generation of Physics, Spun from Universality and Emergence"
from the Ministry of Education, Culture, Sports, Science and Technology
(MEXT) of Japan.
This work was partly supported by the Grant-in-Aid for Young Scientists (B)
No.11020523 (HM) from the MEXT of Japan.
We are grateful to the Catalina Real-time Transient Survey
team for making their real-time detection of transient objects available to
the public.
Funding for the SDSS has been provided by the Alfred P. Sloan Foundation,
the Participating Institutions, the National Aeronautics and
Space Administration, the National Science Foundation, the US Department
of Energy, the Japanese MEXT and the Max Planck Society.
The SDSS web site is $<$http://www.sdss.org$>$.
Simbad's VizieR service, the NASA Astrophysics Data System
and the AAVSO VSX service have been important resources of information
for this research.
We are grateful to Elena Pavlenko for making historical materials
of NSV 00895 available to us and to Tomohito Ohshima for helping us
in astrometry of SX LMi.
We are also grateful to many numbers of observers including the VSNET
Collaboration, AAVSO, CVNET, BAA VSS alert and AVSON networks
who have greatly contributed to the detections of outbursts of dwarf novae
and determinations of superhump or orbital periods.

\appendix
\section{Notes on Individual Objects}\label{sec:individual}

   In addition to statistics of dwarf novae, the neural network analysis
has clarified a wealth of previously unknown properties of individual
dwarf novae.  We discuss on new findings of individual objects.
We also present some negative results, or suggestible identifications
on objects whose reality or identification have been ambiguous.

\subsection*{FN And}

   Our analysis favors an orbital period above the period gap.
Our unpublished photometry during a long outburst in 1994 July did not
detect superhumps.  The object is likely an SS Cyg-type dwarf nova.
The relatively rapid fading rate (1.2 mag d$^{-1}$) during the terminal
stage of the outburst suggests a relatively short orbital period,
consistent with the present analysis.  The object appears to be
similar to a short-$P_{\rm orb}$ SS Cyg-type dwarf nova AR And
(\cite{szk84AAVSO}; \cite{tay96arandamcaspyper}).

\subsection*{IZ And}

   The current analysis is consistent with the lack of superhumps
and the relatively slow fading rate \citep{kat01izand}.

\subsection*{LL And}

   \citet{kat04lland} suggested that this object may have a massive
secondary for its $P_{\rm orb}$.  The present analysis did not show
a very strong contribution from the secondary, as has been observed
in QZ Ser.

\subsection*{PT And}

   This object was originally considered to be a nova in M31
\citep{gru58ptand}.  Additional detections of the outbursts led to
a classification of a possible dwarf nova \citep{sha89ptand}.
\citet{alk00ptand} presented a finding chart and summary of past
observations.  The object underwent another outburst in 2010,
and was confirmed to be a dwarf nova (T. Ohshima et al., in preparation).
Although the SDSS image is present, there is no photometric entry.
A visual inspection of the image suggests a 22 mag quiescent counterpart.

\subsection*{V496 Aur}

   Although the period was not meaningfully estimated with the 
neural-network analysis due to its faintness, the colors tend to be 
in the region of WZ Sge-type dwarf novae (subsection \ref{sec:twocolor}).  
There were five outbursts recorded by CRTS;the behavior is unlikely to be
that of a WZ Sge-type dwarf nova.  The object may be related to ``borderline''
WZ Sge-type dwarf novae.

\subsection*{T Boo}

   The existence of this object has long been debated (see e.g.
\cite{due87novaatlas}), and has also been suspected to be a dwarf nova.
The candidates 5 in \citet{dow89oldnova} has a blue color
(SDSS J141356.53$+$190328.6, classified as an F2-9 type according to
\cite{rin96oldnovaspec}), while candidate 2 is a red object.
The invisibility of the candidate 2 on a $U$ image of \citet{dow89oldnova}
can be understood as being a result of a large $u-g$ color index
(type K5-M4 according to \cite{rin96oldnovaspec}.
Candidates 1, 4, 8 and 9 are galaxies.
The present result is consistent with \citet{sha90oldnova}.

\subsection*{LM Cas}

   \citet{liu00CVspec3} suggested that the fainter component of a close
pair is the CV.  Our measurement of the fainter component is consistent
with the dwarf-nova classification.

\subsection*{GZ Cet}

   The estimated orbital period represented a markedly large departure
from the real period, which is expected for a dwarf nova with an unusually
luminous secondary.  Among samples with known $P_{\rm orb}$, GZ Cet
showed the widest deviation among dwarf novae below the period gap,
and no systems with intermediate deviations were present in our sample.
This suggests that such systems are relatively rare.

\subsection*{EU CMa}

   Our analysis favors an orbital period above the period gap.
Our unpublished photometry during a long outburst in 2003 February did not
detect superhumps.  The object is likely to be an SS Cyg-type dwarf nova.

\subsection*{DE Cnc}

   This object was originally detected during the course of a search
for flare stars \citep{got77decnc}.  The object was caught in outburst
on 1977 February 21 at $U$=15.8.  An examination of archival plates
detected another outburst on 1931 January 11 at $B$=14.6.  Other fainter
outbursts were also suspected.  \citet{got79decnc} further studied
the object and suggested it to be a dwarf nova.  \citet{mun98CVspec5}
observed the counterpart suggested by \citet{DownesCVatlas2} and obtained
a featureless continuum resembling field G--K stars.
\citet{liu00CVspec3} obtained a similar result.
The present analysis has confirmed that the color is also unlikely that of
a quiescent dwarf nova.  Since the position marked on the finding chart
in \citep{got77decnc} appears to be slightly south of the position
suggested by \citet{DownesCVatlas2}, the true quiescent counterpart
may be fainter.  An inspection of the SDSS image was unable to detect
a promising quiescent counterpart.

\subsection*{GZ Cnc}

   The $u-g$ and $g-z$ colors for this objects are approximately consistent
with those of WZ Sge-type dwarf novae.  The neural-network analysis, 
however, distinguished
this object from WZ Sge-type dwarf novae.  A combination of color
criteria and neural network would be helpful in identifying certain
classes of objects.

\subsection*{HH Cnc}

   Our analysis favors an orbital period above the period gap.
The result is consistent with the lack of superhumps \citep{kat01tmzv36},
and our unpublished photometry during the 2003 January outburst.
In the spectrum presented by \citet{szk07SDSSCV6}, there was no evidence
for the secondary.

\subsection*{FU Com}

   Although \citet{DownesCVatlas3} proposed the counterpart, citing
\citet{how90faintCV3} as a reference, \citet{how90faintCV3} were unable to
identify the variable object in their field.  The proposed counterpart
did not show a color resembling a CV (listed in table \ref{tab:dnlist}).
After examining the position, we found a likely X-ray counterpart
1RXS J123054.8$+$270759, which has a likely blue optical counterpart
SDSS J123053.11$+$270822.0, \timeform{32''} distant from the X-ray
position.  This object is a quasar ($z$=0.472, SDSS spectroscopy).
Its classification appears to be consistent with the low-amplitude
variation of FU Com reported in \citet{luy65fucom}.

\subsection*{FY Com}

   Due to the lack of the finding chart, it is almost impossible to identify 
the object unambiguously.  Although \citet{DownesCVatlas3} proposed
the counterpart, this object is not particularly blue.
A relatively blue object SDSS J124925.74$+$272511.9 ($g$ = 21.0) is
present within the error of the original report, although its $u-g$ color
is unlikely for a CV.

\subsection*{GP Com}

   This is a helium CV permanently (as of now) in quiescence
(\cite{sma75gpcom}; \cite{tsu97amcvn}).  It is included in this survey 
in order to examine the response of the neural network to quiescent helium
dwarf novae.

\subsection*{IM Com}

   IM Com was discovered as an eruptive object by \citet{zac82ctbooimcom}.
The object exhibited a 0.7 mag brightening in 6 d in 1981 March
(see figure 4 in \cite{zac82ctbooimcom}) and already faded after 30 d.
Although GCVS classified the object as a possible dwarf nova, the original
author suggested either nova like (NL)-type variable or a dwarf nova.
The identification in \citet{DownesCVatlas3} is correct [note that the
chart in \citet{zac82ctbooimcom} was intentionally reversed].
Although \citet{mun98CVspec5} reported a featureless continuum resembling
those og the field G--K stars, the SDSS $g-r$ color is much bluer than 
in the assumed candidate for DE Cnc with a similar description.  
The large $u-g = +1.0$,
however, is rare in CVs.  The object should deserve further research.

\subsection*{MT Com}

   This object was proposed to be the optical counterpart of the EUV
transient RE J1255$+$266 (\cite{dah95j1255}; \cite{wat96j1255};
\cite{whe00j1255}).  \citet{pat05j1255} proposed a possible orbital
period of 0.0829 d and suggested it a candidate period bouncer.

\subsection*{V1081 Cyg}

   Although the orbital period was not directly estimated due to a
lack of the $r$-band data, we can obtain a period of 0.2--0.7 d
assuming an $r$ magnitude of 16.8--17.3.  The object is likely a
long-$P_{\rm orb}$ dwarf nova.  The result is consistent with
a strong red continuum reported in \citet{bru92CVspec2}.

\subsection*{V1089 Cyg}

   \citet{liu99CVspec1} reported a reddish continuum in the spectrum,
although no clear signature of the secondary was detected.
The present analysis is in agreement with the spectroscopic result.

\subsection*{V1363 Cyg}

   The object is renowned for its long-term variability, unlike most of
other dwarf novae (\cite{mil71cygvar}; \cite{pin72v1454cyg}).
The object was classified either as a Z Cam-type dwarf nova \citep{GCVS}
or as a VY Scl-type NL variable \citep{bru92CVspec2}.
\citet{sch90DNspec} also reported the presence of a secondary that is directly
visible in the spectrum.  This spectrum was apparently taken during its
faint state.  The object recently underwent major outbursts in
2007--2008 and 2011, superimposed on a gradually fading bright state
lasting since 2005.  The 2011 outburst was well observed, and was composed
of a slow rise by $\sim$3.5 mag in 30 d, and a slow fade lasting more
than 40 d.  The overall behavior of the outburst resembled those of two
famous long-$P_{\rm orb}$ DNe, V630 Cas (\cite{whi73v630cas};
\cite{war94v630cas}; \cite{oro01v630cas}) and V1017 Sgr \citep{sek92v1017sgr}.
The present analysis strongly favor a long-$P_{\rm orb}$
dwarf nova.  The inferred period is close to the 0.4 d periodicity
detected during the outburst (vsnet-alert 13326), although both values
have large uncertainties.  The unusual behavior of this object seems
to be understood as if the system generally has low, highly variable
mass-transfer rate, and the natural consequence of a long orbital period
which de-stabilize the outer accretion disk due to the thermal disk
instability.

\subsection*{V1449 Cyg}

   The color is unusually red.
This may have been resulted from an unresolved companion due to the
crowded field, or a contribution from a luminous secondary.
The classification appears to be secure from the recorded light curve
\citep{pin72v1454cyg}.
This object has been confirmed to show H$\alpha$ in emission
\citep{IPHAS}.

\subsection*{V1697 Cyg}

   This is a faint red star in SDSS, and unlikely a CV.

\subsection*{V1711 Cyg}

   The suggested object \citep{DownesCVatlas3} is not particularly blue,
and it is less likely a CV.  No photometric data were available.

\subsection*{V2466 Cyg}

   A very faint object is present in SDSS image.  No photometric data
were available.

\subsection*{DO Dra}

   Although this object is better understood to be an outbursting
intermediate polar (\cite{pat92dodra}; \cite{pat93dodraXray}), we included
this object in the current study because the accretion disk in this
system is mostly in cold state, as in quiescent dwarf novae.

\subsection*{HQ Gem}

   This object has been confirmed to show H$\alpha$ in emission
\citep{IPHAS}.

\subsection*{KZ Gem}

   The classification appears to be secure from the multiple outbursts
reported to VSNET \citep{VSNET}.  The brightest maximum reached a visual
magnitudes of 14.0.

\subsection*{MV Gem}

   This variable was discovered by \citet{hof68an290277}.
\citet{zwi96CVspec} reported a mid-F dwarf spectrum, and it is not usually
considered as a CV.  An inspection of the SDSS image, the object is
a close pair, and the fainter object ($g$=20.4) has a color similar to
CVs.  The object (not included in the table) apparently needs to be examined.
According to the observations by M. Iida, the object (in combined light)
brightened more than 2 mag in 7 d.

\subsection*{V610 Her}

   The object listed (\timeform{16h 43m 39.09s}, \timeform{+22D 31' 25.2''})
is \timeform{10.5''} distant from the nominal GCVS position, but has
a more CV-like color.  This degree of uncertainty in the original
identification can be reconciled with the quality of the
hand-written chart (\cite{hof68an290277}; \cite{kat99var6}).
The suggested counterpart in \citet{DownesCVatlas1} was correct.

\subsection*{DO Leo}

   Although \citet{GCVS}, \citet{DownesCVatlas3} and \citet{RKCat}
classified this object as NL, the object is actually an eclipsing
Z Cam-type dwarf nova.\footnote{
cf. $<$http://nesssi.cacr.caltech.edu/catalina/20010514/
105141150564100059p.html$>$.
}  The SDSS observations was recorded during its quiescent state.

\subsection*{HM Leo}

   Outbursts fading more rapidly than what is expected from the orbital
period have been reported (e.g. 0.9 mag d$^{-1}$ for the 2009 March outburst).
This object might be somewhat similar to DO Dra in its outburst
characteristics.

\subsection*{SS LMi}

   The true position is \timeform{10h 34m 05.85s}, \timeform{+31D 07' 59.6''},
correctly listed in \citet{GCVSelectronic2011} but several catalogs
including VizieR give incorrect coordinates.  The color is consistent
with the WZ Sge-type classification.

\subsection*{SX LMi}

   This object is a close double.  Although the presence of a companion was
already mentioned in \citet{wag98sxlmi}, it was not perfectly clear
whether \citet{wag98sxlmi} measured only the bright component in obtaining
the quiescent light curve.  The astrometry of our CCD image during the
outburst in 1994 December favored a position close to the brighter
component.  The result has confirmed the unusually low outburst amplitude
claimed by \citet{nog97sxlmi}.  It would be worth noting that the fainter
component is also a blue object in SDSS, contrary to the suggestion
in \citet{wag98sxlmi}.  Although the object underwent frequent outburst
in the 1990's (\cite{nog97sxlmi}; \cite{kat01sxlmi}), the frequency of
outbursts (particularly superoutbursts) has remarkably reduced since 2007.

\subsection*{V699 Oph}

   This object is an an unresolved companion to a 16 mag star.
These stars remained unresolved in SDSS, and we did not include this
object in the sample.

\subsection*{GR Ori}

   The object was discovered as a possible nova in 1916 \citep{thi16grori}.
It was recorded at a photographic magnitude of 11.5 on January 30
and gradually faded to 12.5 on February 7, and further faded to 13.0
on February 30.  The object has long been considered as a distant nova
(\cite{mcl45novadist}; \cite{due85halonova}), although there was a
suggestion of a dwarf nova \citep{bru57DNatlas}.
During a ``reconnaissance program'' of old suspected novae,
\citet{rob00oldnova} identified GR Ori as a 22.8 mag blue object.
The present identification by the SDSS data has confirmed this
identification.  Although the result of neural-network analysis was
inconclusive due to faintness of the object, the color (especially
the $u-g$ color) resembles those of WZ Sge-type dwarf novae.
The relatively rapid decline, especially at the late epoch of the outburst,
seems to strengthen this classification.

\subsection*{V370 Peg}

   This object is a transient object reported in \citet{mod98v370pegiauc}.
The true quiescent counterpart is a blue object east to the cataloged
object (red star in SDSS) in \citet{GCVSelectronic2011}.
The blue object well matches the finding chart (object in outburst)
available at $<$http://astro.berkeley.edu/bait/public\_html/cv.gif$>$.
The object is not in the SDSS photometric catalog.
The object was again recorded in outburst in 2010 October
(Eddy Muyllaert, cf. vsnet-alert 12304), reaching a magnitude of 16.7
(unfiltered CCD).  This outburst lasted at least for 9 d.

\subsection*{GK Per}

   The object was included because this old nova exhibits dwarf nova-type
outbursts (e.g. \cite{bia86gkper}; \cite{can86gkper}).

\subsection*{QY Per}

   The object is a well-confirmed SU UMa-type dwarf nova with long
outburst intervals for its $P_{\rm orb}$ (\cite{kat00qyperiauc}; \cite{Pdot}).
The variation of the superhump amplitudes was also unusual \citep{Pdot3}.

\subsection*{V336 Per}

   This object is renowned for its low outburst frequency \citep{bus79VS17}
and was suspected to be a WZ Sge-type dwarf nova \citep{kat01hvvir}.\footnote{
See also the description $<$http://hea.iki.rssi.ru/\~{}denis/
CVmonitoring.html$>$.
}
\citet{liu00CVspec3} also suggested the low mass-transfer rate from
its spectroscopy and inferred a short $P_{\rm orb}$.  In 2009 August,
a long-awaited outburst was reported (cf. vsnet-alert 11447).
A lack of superhumps in time-series observations by Ian Miller,
Jeremy Shears and David Boyd was remarkable (vsnet-alert 11452).
The negative result for an SU UMa (or WZ Sge) type dwarf nova can be
reconciled with the present neural-network analysis.

\subsection*{V372 Per}

   The object has a low outburst frequency \citep{rom76DN}.
Photometry during the state $\sim$3 mag above quiescence indicated
a relatively red color \citep{rom76DN}, which is in agreement with
the present study.

\subsection*{XY Psc}

   The object was discovered as a candidate supernova near UGC 729 in 1972
\citep{ros72xypsciauc1}.  \citet{ros72xypsciauc2} later suggested
the object to be a dwarf nova with a large outburst amplitude.
Although identification of the object remained ambiguous
for a long time, \citet{hen01xypsc} identified a counterpart based on
original discovery photographs.  \citet{kat01hvvir} suggested that the
object is a very likely candidate for a WZ Sge-type dwarf nova based on
the lack of outbursts for a long time [note, however, there was a
photographic record of an outburst on 1985 January 12
(vsnet-chat 432)\footnote{
$<$http://www.kusastro.kyoto-u.ac.jp/vsnet/Mail/vsnet-chat/msg00432.html$>$
}, which still needs to be confirmed].
The present neural-network analysis seems to strengthen the SU UMa-type
nature of this object, although the color is not so striking as those
of extreme WZ Sge-type dwarf novae.

\subsection*{AH Psc}

   This is a missing BD star, which has been suspected to be some kind
of an eruptive object (cf. \cite{due87novaatlas}).  There is no good
candidate in SDSS.

\subsection*{AS Psc}

   This object is a dwarf nova in the field of M33
(see \cite{kat01uvgemfsandaspsc} for its history and references).
\citet{tho02aspsc} further identified the quiescent counterpart.
This object was also detected in SDSS to be a blue object.
The color seems to suggest a more usual SU UMa-type dwarf nova rather
than an extreme WZ Sge-type dwarf nova (see also discussions in
\cite{kat01uvgemfsandaspsc} and \cite{tho02aspsc}).
There was another outburst detected by CRTS at a magnitude of 15.8
(unfiltered CCD) in 2008 January.\footnote{
$<$http://nesssi.cacr.caltech.edu/catalina/20010316/
103161320074100046p.html$>$
}

\subsection*{X Ser}

   This object is an old nova showing dwarf nova-type outbursts similar
to GK Per (\cite{hon98nloutburst}; \cite{hon01nloutburst}).  The most
recent outburst was detected by CRTS in 2009 August and reached $V$=14.2.
The fading stage of the outburst lasted for $\sim$30 d and bore a
similarity with outbursts of GK Per and V630 Cas (see the note for
V1363 Cyg for the references).
The orbital parameters are also similar to those of GK Per and V630 Cas
\citep{tho00v533herv446herxser}.

\subsection*{QZ Ser}

   \citet{GCVSelectronic2011} gives the position of the brighter component
of a close pair.  The true CV is the fainter component.

\subsection*{XX Tau}

   Although this is a classical nova, \citet{sch05oldnova} recently reported
that the spectrum of this nova is similar to those of dwarf novae,
and suggested a low mass-transfer rate.  We have confirmed the SDSS
color similar to those of dwarf novae.  Although it is not certain whether
the present neural-network analysis applies to such a system, the color
favors a system below the period gap.  \citet{rod05oldnova} reported
a possible photometric periodicity of 0.14 d.
Further detailed research is certainly needed.

\subsection*{TX Tri}

   This object was discovered as a dwarf nova by \citep{kur78twtritxtri}.
\citet{ric79m33blueobjectaspsc} independently detected this variable
and suggested it to be a quasar based on its color.
Further examination by \citet{ric81txtri} indicated that the object
is a dwarf nova.  Based on the presence of He\textsc{II} and
C\textsc{III}/N\textsc{III} lines, the authors proposed that the
object is similar to WZ Sge.  \citet{ric89txtri} measured the cycle
length of outbursts to be 90 d, while \citet{kur78twtritxtri}
suggested a period of 70 d.  The spectroscopic observation by
\citet{liu00CVspec3} was taken during an outburst, and the contribution
by the secondary was not detected.  Our analysis seems to preclude
the possibility of a WZ Sge-type object, and the object is likely
a usual SS Cyg-type dwarf nova above the period gap.

\subsection*{UZ Tri}

   The object was detected in outburst only on two nights:
1980 November 1 (14.2 mag, photographic) and November 27 (15.5 mag).
The object was invisible on 1980 September 7 and 1981 January 23
\citep{mei86uztri}.  According to \citet{mei86uztri}, there is
a colorless star and a galaxy on the Palomer Sky Survey at the location
of UZ Tri, and either of them underwent an outburst.
\citet{DownesCVatlas3} proposed that the former is the candidate quiescent
counterpart.  We provide SDSS data for the candidate object.
The object appears to be slightly too red for a CV.

\subsection*{NSV 00895}

   The object was discovered by T. Hvan as a possible dwarf nova
\citep{mes40nsv895}.  \citet{kho53nsv895} studied this object and
suggested that it may be a dwarf nova with a very long interval
($>$ 1000 d) of outbursts, or a nova.  \citet{kho53nsv895} presented
a reliable finding chart and description in relation to the UGC 2172
(referred to as a nebula).  The information is sufficient to search
for a counterpart in SDSS.  There was no promising candidate for
a quiescent dwarf nova.  \citet{wen90nsv00895} reported another
positive measurement and its light curve.  \citet{wen90nsv00895}
concluded that it is not possible to determine whether NSV 00895 is
a nova or a supernova.  According to \citet{tre09ngc1023},
UGC 2172 is a dwarf irregular galaxy of the NGC 1023 group
at a distance of 1 Mpc.  The lack of a promising quiescent candidate
and the proximity of the galaxy strongly favor a supernova 
of $M_{\rm pg} = -19$

\subsection*{NSV 02853}

   None of the objects around the star marked in \citet{DownesCVatlas3}
show a CV-like color.

\subsection*{NSV 04394}

   The object was originally discovered by \citet{wil72nsv4394}
and was suspected to be a dwarf nova or a supernova.
There was no promising candidate according to \citet{DownesCVatlas3}.
There are two blue SDSS objects in this field:
SDSS J090938.24$+$444633.2 ($g=21.9$), \timeform{1''} distant from
the nominal position of NSV 04394 and SDSS J090939.78$+$444638.2 ($g=21.2$)
\timeform{17''} distant.  We adopted the first candidate in the table.
If this is indeed a dwarf nova, the amplitude strongly suggests
a WZ Sge-type dwarf nova.

\subsection*{NSV 04498}

   This object was discovered by \citet{luy68faintBS}.  \citet{hae96faintCV}
were unable to identify a variable ozbject in the field.  The blue object
SDSS J092849.78$+$231114.2 ($g$=19.9, \timeform{68''} distant from
the reported approximate coordinates) might be a viable candidate.

\subsection*{NSV 05031}

   The coordinates are \timeform{10h57m49.81s}, \timeform{-21D56'58.0''}.
The object is identical with CTCV J1057$-$2156 \citep{aug10CTCVCV2}
whose orbital period is 1.68 hr.  The existence of a past outburst
and the quiescent spectrum suggest an SU UMa-type dwarf nova.

\subsection*{NSV 05543}

   This object was recorded by \citet{rue33nsv5543}.  \citet{DownesCVatlas3}
selected a very blue object as the possible quiescent counterpart.
The spectrum of this object (SDSS J121829.75$+$400522.3) in the SDSS
archive is dominated by Balmer absorption lines (classified as A0 in the
SDSS archive) and lacks emission lines.
The identified object is unlikely to be a dwarf nova in quiescence;
\timeform{33''} distant from this object, there is another relatively
blue object (SDSS J121831.75$+$400546.1) with $g$=21.0.  The $u-g$ color,
however, is atypical for a CV.

\subsection*{NSV 09445}

   There was no very promising candidate within the circle shown in
\citet{DownesCVatlas3}.  The blue object (less likely related to NSV 09445)
marked in \citet{DownesCVatlas3}
(=SDSS J173930.43$+$210851.4, $g=18.8$) has a blue color in SDSS.
The likely association with a ROSAT source 1RXS J173930.9$+$210847
might suggest a QSO.

\subsection*{NSV 14681}

   The fainter component of a pair.

\subsection*{NSV 18230}

   This object (=KUV 09313$+$4052) was reported to be a likely dwarf nova
\citep{nog79nsv18230}.  CRTS recorded additional two outbursts.
The brightest outburst reached an unfiltered CCD magnitude of 16.9.

\subsection*{NSV 18704 (SN 1985J), RS Psc}

   Although both objects were discovered as supernovae, there are
suspicions that they are possible foreground variables.
Although both objects have SDSS images, the galaxies were too bright to
search for possible candidate stellar images.

\subsection*{NSV 19466}

   The object is suspected to be either a dwarf nova or a VY Scl-type
NL object \citep{szk03SDSSCV2}.  The spectrum suggests a dwarf nova.

\subsection*{1502$+$09}

   This object was reported in \citet{fil85150209} which was caught in
outburst during a spectroscopic survey.  The quiescent spectra presented
in \citet{fil85150209} and taken by SDSS, are sufficient to classify
the object as a dwarf nova.  Our neural-network analysis strongly
favors a short orbital period and this object is a good candidate for
an SU UMa-type dwarf nova.  No further outburst has been recorded.

\subsection*{2QZ J142701.6$-$012310, OT J204739.4$+$000840,
SDSS J012940.05$+$384210.4, SDSS J124058.03$-$015919.2,
SDSS J204739.40$+$000840.3}

   These objects are helium dwarf novae.

\subsection*{OT J162322.2$+$121334}

   During this survey, we noticed that this object (=CSS080606:162322$+$121334,
not included in tables) showed multiple deep fadings similar to
VY Scl-type CVs (\cite{war95book}; \cite{lea99vyscl}; \cite{hon04vyscl}).
The object has a blue SDSS counterpart, and fortunately, there were its
observations both in high and low states.
Assuming that the color of VY Scl-type CVs in very low state is similar
to dwarf novae, we estimated an estimated $P_{\rm orb}$ of 0.13(3) d using
the present neural network.  The estimated period is compatible with
the expected $P_{\rm orb}$ for a VY Scl-type CV.  The range of variability
based on SDSS data is $g$=17.0--21.9.  Further spectroscopic confirmation
is desired.

\subsection*{OT J170606.1$+$255153}

   This is an anonymous object near UGC 10700 \citep{gra03j1706iauc8153}.
The CV-type nature has been established \citep{fil03j1706iauc8158}.
The SDSS counterpart suggests a large ($\sim$ 6 mag) outburst amplitude
and the object is a good candidate for an SU UMa-type dwarf nova.

\subsection*{OT J213806.6$+$261957}

   This WZ Sge-type dwarf nova \citep{Pdot2} is a companion to a brighter
star.  Although these stars are barely resolved in SDSS image,
photometric data are given for the combined magnitude.  We did not
include this object in the sample.

\subsection*{OT J215815.3$+$094709}

   The detection of superhumps was reported in \citet{Pdot2}.  Prior to
this CRTS discovery, two bright superoutbursts were recorded by ASAS-3
(2003 August, $V$=12.75 and 2004 October, $V$=12.99).  The large amplitudes
of superhumps reported in \citet{Pdot2} may have been somewhat anomalous
(cf. \cite{Pdot3}) considering that the 2010 observation was obtained
when the object was much fainter (around 13.4 mag).  The present 
neural-network analysis showed no particular anomaly.

\subsection*{OT J220449.7$+$054852, ROTSE3 J004626$+$410714}

   The blue component of a close pair is selected.

\subsection*{OT J223958.2$+$231837}

   The object was spectroscopically confirmed as a CV (CRTS page).
The light curve is rather unusual for a typical dwarf nova.

\subsection*{SDSS J075939.79$+$191417.3}

   The object was classified either a dwarf nova or a polar
\citep{szk06SDSSCV5}.  We classified it a dwarf nova based on the general
appearance of the spectrum.  The object has recently been shown to be
eclipsing \citep{dra10eclipsingWD}.  No outburst has been recorded.

\subsection*{SDSS J204448.92$-$045928.8}

   The long orbital period of 1.68(1) d was reported
from a radial-velocity study \citep{pet05CVlongP}.  The detection of
a long-duration outburst \citep{wil09j2044} also supports a long
$P_{\rm orb}$.  The failure to reproduce the $P_{\rm orb}$ by the
neural-network analysis
is not very surprising since our method is insensitive for characterizing
objects with $P_{\rm orb}$ longer than 1 d (another long-$P_{\rm orb}$
system X Ser did not yield a meaningful value; in this case it may
have been a result of the contribution from the luminous
inner accretion disk).  It is noteworthy that ($V-J$) color of this object
is similar to that of another long-$P_{\rm orb}$ system GK Per, and bluer than
most of secondary-dominated SS Cyg-type dwarf novae.
A neural-network analysis using categories provides better results for
these systems: all objects have a high probability for the {\it long} category.

\subsection*{SDSS J233325.92$+$152222.2}

   The object is an intermediate polar with a long spin period
\citep{sou07j2333}.  We included it in this sample because the object
is not dominated by the disk and system resembles dwarf novae,
as in EX Hya (\cite{hel89exhya}; \cite{hel00exhyaoutburst}).

\subsection*{SN 1964O}

   The object was suspected to be a dwarf nova rather than a supernova
in \citet{DownesCVatlas2}.  Although two candidates were listed in
\citet{DownesCVatlas3}, both stars are K stars.  We identified a more
likely counterpart (SDSS J150849.85$+$552807.6) between these two stars.
The magnitudes listed in the table refers to this object.
If this is the true quiescent counterpart, the amplitude of the 1964
outburst was at least 6 mag.

\section{Names of Optical Transients}

   In the main text of this paper, we have used coordinated-based names for
optical transients (OTs), as in the series of papers starting with
\citet{Pdot}.  This measure has been taken for two reasons: consistency
of the names between this paper and a series of our ones (e.g. \cite{Pdot}),
and the reader's convenience with unique identifications of the objects 
by coordinates
(CRTS identifiers lack the accuracy in the right ascension for unique
identifications).  We list original IDs for CRTS (CSS, MLS and SSS objects)
and references for other transients (table \ref{tab:otlist}).

\begin{table*}
\caption{Identification list of optical transients}\label{tab:otlist}
\begin{center}
\begin{tabular}{cc|cc}
\hline
Object & ID or discoverer & Object & ID or discoverer \\
\hline
OT J000024.7$+$332543 & CSS100910:000025$+$332543 & OT J073921.2$+$222454 & CSS080303:073921$+$222454 \\
OT J000130.5$+$050624 & CSS101127:000130$+$050624 & OT J074222.5$+$172807 & MLS110309:074223$+$172807 \\
OT J000659.6$+$192818 & CSS100602:000700$+$192818 & OT J074419.7$+$325448 & CSS091029:074420$+$325448 \\
OT J001158.3$+$315544 & CSS101111:001158$+$315544 & OT J074727.6$+$065050 & Itagaki \citep{yam08j0747cbet1216} \\
OT J001340.0$+$332124 & CSS101111:001340$+$332124 & OT J074820.0$+$245759 & CSS101115:074820$+$245759 \\
OT J001538.3$+$263657 & CSS090918:001538$+$263657 & OT J074928.0$+$190452 & MLS091117:074928$+$190452 \\
OT J002500.2$+$073349 & CSS081123:002500$+$073350 & OT J075332.0$+$375801 & CSS100407:075332$+$375801 \\
OT J002656.6$+$284933 & CSS101212:002657$+$284933 & OT J075414.5$+$313216 & CSS110414:075414$+$313216 \\
OT J003203.6$+$314510 & CSS091220:003204$+$314510 & OT J075648.0$+$305805 & CSS080406:075648$+$305805 \\
OT J003304.0$+$380106 & CSS110115:003304$+$380106 & OT J080428.4$+$363104 & CSS091116:080428$+$363104 \\
OT J003500.0$+$273620 & CSS081031:003500$+$273620 & OT J080714.2$+$113812 & Itagaki \citep{yam08j0747cbet1216} \\
OT J004500.3$+$222708 & CSS081102:004500$+$222708 & OT J080729.7$+$153442 & CSS090317:080730$+$153442 \\
OT J004518.4$+$185350 & CSS081024:004518$+$185349 & OT J080853.7$+$355053 & CSS080416:080854$+$355053 \\
OT J004606.7$+$052100 & CSS100911:004607$+$052060 & OT J081030.6$+$002429 & CSS100108:081031$+$002429 \\
OT J004807.2$+$264621 & CSS090925:004807$+$264621 & OT J081414.9$+$080450 & CSS080202:081415$+$080450 \\
OT J004902.0$+$074726 & CSS090204:004902$+$074726 & OT J081418.9$-$005022 & CSS080409:081419$-$005022 \\
OT J005152.9$+$204017 & CSS091026:005153$+$204017 & OT J081712.3$+$055208 & CSS091215:081712$+$055208 \\
OT J005824.6$+$283304 & CSS101009:005825$+$283304 & OT J081936.1$+$191540 & CSS100202:081936$+$191540 \\
OT J010329.0$+$331822 & CSS110116:010329$+$331822 & OT J082019.4$+$474732 & CSS091213:082019$+$474732 \\
OT J010411.6$-$031341 & CSS091009:010412$-$031341 & OT J082123.7$+$454135 & CSS090224:082124$+$454135 \\
OT J010522.2$+$110253 & CSS080921:010522$+$110253 & OT J082603.7$+$113821 & CSS110124:082604$+$113821 \\
OT J010550.1$+$190317 & CSS091016:010550$+$190317 & OT J082654.7$-$000733 & CSS080306:082655$-$000733 \\
OT J011134.5$+$275922 & CSS081024:011134$+$275922 & OT J082821.8$+$105344 & CSS071231:082822$+$105344 \\
OT J011516.5$+$245530 & CSS101008:011517$+$245530 & OT J082908.4$+$482639 & CSS091120:082908$+$482639 \\
OT J011543.2$+$333724 & CSS081205:011543$+$333724 & OT J084041.5$+$000520 & CSS090331:084041$+$000520 \\
OT J011613.8$+$092216 & MLS101127:011614$+$092216 & OT J084127.4$+$210053 & MLS110522:084127$+$210054 \\
OT J012059.6$+$325545 & Itagaki (vsnet-alert 12431) & OT J084358.1$+$425037 & CSS080309:084358$+$425037 \\
OT J014150.4$+$090822 & MLS101213:014150$+$090822 & OT J084413.7$-$012807 & CSS090331:084414$-$012807 \\
OT J020056.0$+$195727 & CSS090117:020056$+$195727 & OT J084555.1$+$033930 & Itagaki \citep{yam08j0845cbet1225} \\
OT J021110.2$+$171624 & CSS080130:021110$+$171624 & OT J085113.4$+$344449 & CSS080401:085113$+$344449 \\
OT J021308.0$+$184416 & CSS101214:021308$+$184416 & OT J085409.4$+$201339 & CSS080506:085409$+$201339 \\
OT J023211.7$+$303636 & CSS101029:023212$+$303636 & OT J085603.8$+$322109 & CSS100508:085604$+$322109 \\
OT J025615.0$+$191611 & CSS090128:025615$+$191611 & OT J085822.9$-$003729 & CSS071112:085823$-$003729 \\
OT J032651.7$+$011513 & CSS100219:032652$+$011513 & OT J090016.7$+$343928 & CSS110512:090017$+$343928 \\
OT J032839.9$-$010240 & CSS101116:032840$-$010240 & OT J090239.7$+$052501 & CSS080304:090240$+$052501 \\
OT J032902.0$+$060047 & CSS101202:032902$+$060047 & OT J090516.1$+$120451 & CSS091022:090516$+$120451 \\
OT J033104.4$+$172540 & MLS100911:033104$+$172540 & OT J090852.2$+$071640 & CSS110125:090852$+$071640 \\
OT J035003.4$+$370052 & CSS100113:035003$+$370052 & OT J091453.6$+$113402 & CSS081107:091454$+$113402 \\
OT J040659.8$+$005244 & Itagaki \citep{yam08j0406cbet1463} & OT J091534.9$+$081356 & \citet{don08j0915} \\
OT J041636.9$+$292806 & CSS100317:041637$+$292806 & OT J091634.6$+$130358 & MLS100315:091635$+$130358 \\
OT J041734.6$-$061357 & CSS081202:041735$-$061357 & OT J092839.3$+$005944 & CSS110212:092839$+$005945 \\
OT J042142.1$+$340329 & CSS090403:042142$+$340328 & OT J101035.5$+$140239 & MLS110305:101036$+$140239 \\
OT J042229.3$+$161430 & CSS091024:042229$+$161430 & OT J101545.9$+$033312 & CSS090429:101546$+$033312 \\
OT J042434.2$+$001419 & CSS081030:042434$+$001419 & OT J102146.4$+$234926 & \citet{chr06j1021cbet746} \\
OT J043020.0$+$095318 & CSS080929:043020$+$095318 & OT J102616.0$+$192045 & Itagaki \citep{yam09j1026cbet1644} \\
OT J043517.8$+$002941 & CSS090219:043518$+$002941 & OT J102637.0$+$475426 & Itagaki \citep{yam09j1026cbet1644} \\
OT J043546.9$+$090837 & CSS091123:043547$+$090837 & OT J102937.7$+$414046 & CSS080305:102938$+$414046 \\
OT J043742.1$+$003048 & CSS081203:043742$+$003048 & OT J103317.3$+$072119 & CSS080208:103317$+$072119 \\
OT J043829.1$+$004016 & CSS100218:043829$+$004016 & OT J103704.6$+$100224 & MLS101214:103705$+$100224 \\
OT J044216.0$-$002334 & CSS071115:044216$-$002334 & OT J103738.7$+$124250 & CSS090516:103739$+$124250 \\
OT J051419.9$+$011121 & CSS091109:051420$+$011121 & OT J104411.4$+$211307 & CSS100217:104411$+$211307 \\
OT J052033.9$-$000530 & CSS080207:052034$-$000530 & OT J105550.1$+$095621 & CSS080130:105550$+$095621 \\
OT J055730.1$+$001514 & CSS080307:055730$+$001514 & OT J105835.1$+$054706 & CSS081025:105835$+$054706 \\
OT J055842.8$+$000626 & CSS100114:055843$+$000626 & OT J112112.0$-$130843 & CSS110301:112112$-$130842 \\
OT J073055.5$+$425636 & CSS090911:073056$+$425636 & OT J112253.3$-$111037 & CSS100603:112253$-$111037 \\
OT J073339.3$+$212201 & CSS091111:073339$+$212201 & OT J112332.0$+$431718 & CSS090602:112332$+$431717 \\
OT J073559.9$+$220132 & MLS110305:073600$+$220132 & OT J112509.7$+$231036 & CSS110226:112510$+$231036 \\
OT J073758.5$+$205545 & CSS110301:073759$+$205545 & OT J112634.0$-$100210 & CSS080227:112634$-$100210 \\
\hline
\end{tabular}
\end{center}
\end{table*}

\addtocounter{table}{-1}
\begin{table*}
\caption{Identification list of optical transients (continued)}
\begin{center}
\begin{tabular}{cc|cc}
\hline
Object & ID or discoverer & Object & ID or discoverer \\
\hline
OT J115330.2$+$315836 & CSS080201:115330$+$315836 & OT J173307.9$+$300635 & CSS090820:173308$+$300635 \\
OT J122756.8$+$622935 & CSS110503:122757$+$622934 & OT J175901.1$+$395551 & CSS100510:175901$+$395551 \\
OT J123833.7$+$031854 & CSS090615:123834$+$031854 & OT J182142.8$+$212154 & Itagaki (vsnet-alert 11952) \\
OT J124027.4$-$150558 & CSS071218:124027$-$150558 & OT J202857.1$-$061803 & CSS110621:202857$-$061803 \\
OT J124417.9$+$300401 & CSS080427:124418$+$300401 & OT J204001.4$-$144909 & CSS101116:204001$-$144908 \\
OT J124819.4$+$072050 & CSS090516:124819$+$072050 & OT J204739.4$+$000840 & \citet{and08amcvn} \\
OT J130030.3$+$115101 & CSS080702:130030$+$115101 & OT J210034.4$+$055436 & CSS101009:210034$+$055436 \\
OT J132536.0$+$210037 & CSS090102:132536$+$210037 & OT J210043.9$-$005212 & CSS090615:210044$-$005212 \\
OT J134052.1$+$151341 & CSS100531:134052$+$151341 & OT J210205.7$+$025834 & CSS091017:210206$+$025834 \\
OT J135219.0$+$280917 & CSS090705:135219$+$280917 & OT J210650.6$+$110250 & CSS090924:210651$+$110250 \\
OT J135336.0$-$022043 & CSS100405:135336$-$022043 & OT J210704.5$+$014416 & CSS090428:210705$+$014416 \\
OT J135716.8$-$093239 & CSS110208:135717$-$093238 & OT J210846.4$-$035031 & CSS110513:210846$-$035031 \\
OT J141002.2$-$124809 & CSS080502:141002$-$124809 & OT J210954.1$+$163052 & CSS100404:210954$+$163052 \\
OT J141712.0$-$180328 & CSS080425:141712$-$180328 & OT J211550.9$-$000716 & CSS090828:211551$-$000716 \\
OT J142548.1$+$151502 & CSS110628:142548$+$151502 & OT J212025.1$+$194157 & CSS090817:212025$+$194157 \\
OT J144011.0$+$494734 & CSS090530:144011$+$494734 & OT J212555.1$-$032406 & CSS080511:212555$-$032406 \\
OT J144316.5$-$010222 & CSS100115:144317$-$010222 & OT J212633.3$+$085459 & CSS100509:212633$+$085459 \\
OT J145502.2$+$143815 & CSS100407:145502$+$143815 & OT J213122.4$-$003937 & Itagaki \citep{yam08j2131cbet1631} \\
OT J145921.8$+$354806 & CSS110613:145922$+$354806 & OT J213309.4$+$155004 & CSS080404:213309$+$155004 \\
OT J151020.7$+$182303 & CSS080514:151021$+$182303 & OT J213432.3$-$012040 & CSS081120:213432$-$012040 \\
OT J151037.4$+$084104 & CSS110306:151037$+$084104 & OT J213701.8$+$071446 & Itagaki (vsnet-alert 10670) \\
OT J152037.9$+$040948 & CSS110301:152038$+$040948 & OT J213829.5$-$001742 & CSS100524:213830$-$001742 \\
OT J152501.8$-$013021 & CSS100506:152502$-$013021 & OT J213937.6$-$023913 & CSS091119:213938$-$023913 \\
OT J153150.8$+$152447 & CSS080401:153151$+$152447 & OT J214426.4$+$222024 & CSS100520:214426$+$222024 \\
OT J153317.6$+$273428 & CSS110611:153318$+$273428 & OT J214639.9$+$092119 & CSS110613:214640$+$092119 \\
OT J153645.2$-$142543 & SSS110721:153645$-$142543 & OT J214804.4$+$080951 & CSS080505:214804$+$080951 \\
OT J154354.1$-$143745 & CSS100531:154354$-$143745 & OT J214842.5$-$000723 & CSS071116:214843$-$000723 \\
OT J154428.1$+$335725 & CSS090322:154428$+$335725 & OT J214959.9$+$124529 & CSS090722:215000$+$124529 \\
OT J154544.9$+$442830 & CSS110428:154545$+$442830 & OT J215344.7$+$123524 & CSS090526:215345$+$123524 \\
OT J155325.7$+$114437 & CSS080424:155326$+$114437 & OT J215630.5$-$031957 & CSS090728:215630$-$031956 \\
OT J155430.6$+$365043 & CSS081009:155431$+$365043 & OT J215636.3$+$193242 & CSS090622:215636$+$193242 \\
OT J155748.0$+$070543 & CSS100507:155748$+$070543 & OT J215815.3$+$094709 & CSS100615:215815$+$094709 \\
OT J160204.8$+$031632 & CSS080331:160205$+$031632 & OT J220031.2$+$033431 & CSS100624:220031$+$033431 \\
OT J160232.2$+$161733 & CSS080424:160232$+$161732 & OT J220449.7$+$054852 & CSS091019:220450$+$054852 \\
OT J160524.1$+$060816 & CSS080428:160524$+$060816 & OT J221128.7$-$030516 & CSS100519:221129$-$030516 \\
OT J160844.8$+$220610 & CSS080302:160845$+$220610 & OT J221232.0$+$160140 & CSS090911:221232$+$160140 \\
OT J162012.0$+$115257 & CSS080415:162012$+$115257 & OT J221344.0$+$173252 & CSS090917:221344$+$173252 \\
OT J162235.7$+$035247 & CSS090601:162236$+$035247 & OT J222002.3$+$113825 & CSS081230:222002$+$113825 \\
OT J162605.7$+$225044 & CSS080514:162606$+$225044 & OT J222548.1$+$252511 & CSS091027:222548$+$252511 \\
OT J162619.8$-$125557 & CSS090419:162620$-$125557 & OT J222724.5$+$284404 & CSS090531:222724$+$284404 \\
OT J162656.8$-$002549 & CSS080426:162657$-$002549 & OT J222824.1$+$134944 & CSS100615:222824$+$134944 \\
OT J162806.2$+$065316 & CSS110611:162806$+$065316 & OT J222853.7$+$295115 & CSS091110:222854$+$295114 \\
OT J163120.9$+$103134 & CSS080505:163121$+$103134 & OT J223018.8$+$292849 & CSS101010:223019$+$292849 \\
OT J163239.3$+$351108 & CSS110507:163239$+$351108 & OT J223058.3$+$210147 & CSS080501:223058$+$210147 \\
OT J163311.3$-$011132 & CSS100601:163311$-$011132 & OT J223136.0$+$180747 & CSS090911:223136$+$180747 \\
OT J163942.7$+$122414 & CSS080131:163943$+$122414 & OT J223235.4$+$304105 & CSS081107:223235$+$304105 \\
OT J164146.8$+$121026 & CSS080606:164147$+$121026 & OT J223418.5$-$035530 & CSS090910:223418$-$035530 \\
OT J164624.8$+$180808 & CSS100616:164625$+$180808 & OT J223606.3$+$050517 & CSS080611:223606$+$050517 \\
OT J164748.0$+$433845 & CSS100513:164748$+$433845 & OT J223909.8$+$250331 & CSS100521:223910$+$250331 \\
OT J164950.4$+$035835 & CSS100707:164950$+$035835 & OT J223958.2$+$231837 & CSS090826:223958$+$231837 \\
OT J165002.8$+$435616 & CSS090930:165003$+$435616 & OT J223958.4$+$342306 & CSS091016:223958$+$342305 \\
OT J170115.8$-$024159 & CSS090612:170116$-$024158 & OT J224253.4$+$172538 & CSS090622:224253$+$172538 \\
OT J170151.6$+$132131 & CSS110426:170152$+$132131 & OT J224505.4$+$011547 & CSS100624:224505$+$011547 \\
OT J170606.1$+$255153 & \citet{fil03j1706iauc8158} & OT J224753.9$+$235522 & CSS100521:224754$+$235522 \\
OT J170609.7$+$143452 & CSS090205:170610$+$143452 & OT J224814.5$+$331224 & CSS091110:224814$+$331224 \\
OT J170702.5$+$165339 & CSS090818:170702$+$165339 & OT J224823.7$-$092059 & CSS081029:224824$-$092059 \\
OT J171223.1$+$362516 & CSS090516:171223$+$362516 & OT J225749.6$-$082228 & Itagaki (vsnet-outburst 10891) \\
OT J172515.5$+$073249 & CSS100706:172516$+$073249 & OT J230115.4$+$224111 & CSS080923:230115$+$224111 \\
\hline
\end{tabular}
\end{center}
\end{table*}

\addtocounter{table}{-1}
\begin{table*}
\caption{Identification list of optical transients (continued)}
\begin{center}
\begin{tabular}{cc|cc}
\hline
Object & ID or discoverer & Object & ID or discoverer \\
\hline
OT J230131.1$+$040417 & CSS080907:230131$+$040417 & OT J231552.3$+$271037 & CSS100610:231552$+$271037 \\
OT J230425.8$+$062546 & Nishimura \citep{nak11j2304cbet2616} & OT J232551.5$-$014024 & CSS091116:232551$-$014024 \\
OT J230711.3$+$294011 & CSS090926:230711$+$294010 & OT J232619.4$+$282650 & CSS080930:232619$+$282650 \\
OT J231110.9$+$013003 & Itagaki (vsnet-outburst 8239) & OT J233938.7$-$053305 & MLS100902:233939$-$053305 \\
OT J231142.8$+$204036 & CSS091108:231143$+$204036 & OT J234440.5$-$001206 & MLS100904:234441$-$001206 \\
OT J231308.1$+$233702 & Itagaki (TCP J23130812$+$2337018) & -- & -- \\
\hline
\end{tabular}
\end{center}
\end{table*}

\end{document}